\documentclass[%
 reprint,
 superscriptaddress,
 amsmath,amssymb,
 aps,
]{revtex4-2}

\usepackage{graphicx}
\usepackage{dcolumn}
\usepackage{bm}

\usepackage{xcolor}
\usepackage{hanging}
\usepackage{array}
\usepackage{multirow}

\usepackage{soul}
\usepackage{ragged2e} 
\usepackage{afterpage}

\begin{document}


\preprint{CTLearningGoalFramework_Bott2020}

\title{Developing a learning goal framework for computational thinking \\ in computationally integrated physics classrooms}

\author{Daniel P. Weller}
\affiliation{Department of Physics \& Astronomy, Michigan State University, East Lansing, MI, 48824}

\author{Theodore E. Bott}
\affiliation{Department of Physics \& Astronomy, Michigan State University, East Lansing, MI, 48824}

\author{Marcos D. Caballero}
\affiliation{Department of Physics \& Astronomy, Michigan State University, East Lansing, MI, 48824}
\affiliation{Department of Computational Mathematics, Science, \& Engineering and CREATE for STEM Institute, Michigan State University, East Lansing, MI, 48824}
\affiliation{Department of Physics \& Center for Computing in Science Education, University of Oslo, 0316 Oslo, Norway}

\author{Paul W. Irving}
\affiliation{Department of Physics \& Astronomy, Michigan State University, East Lansing, MI, 48824}

\date{\today}

\begin{abstract}
Computational thinking has been a recent focus of education research within the sciences. However, there is a dearth of scholarly literature on how best to teach and to assess this topic, especially in disciplinary science courses. Physics classes with computation integrated into the curriculum are a fitting setting for investigating computational thinking. In this paper, we lay the foundation for exploring computational thinking in introductory physics courses. First, we review relevant literature to synthesize a set of potential learning goals that students could engage in when working with computation. The computational thinking framework that we have developed features 14 practices contained within 6 different categories. We use in-class video data as existence proofs of the computational thinking practices proposed in our framework. In doing this work, we hope to provide ways for teachers to assess their students' development of computational thinking, while also giving physics education researchers some guidance on how to study this topic in greater depth.
\end{abstract}

\maketitle

\section{Introduction}\label{Introduction}

In recent history, computation has transformed from being a tool that assisted scientific research to being fundamental to the very fabric of doing science. Consequently, computational thinking (CT) is an important emerging set of skills for students in the 21st century~\cite{Mohaghegh2016}. Computational thinking is the underlying set of skills and practices that scaffold the ``doing'' of computation. The topic of CT was first promulgated in an influential 2006 article by Jeanette Wing who described CT as ``thinking like a computer scientist''~\cite{Wing2006}. This seminal work expounded the idea that computer science education should be thought of as more than just programming; it is a conceptualizing of the fundamental skills required to solve complex problems. Wing proposed that other disciplines, especially mathematics and engineering, could easily complement computer science and that learning about CT is an important experience for everyone, not just computer scientists. More recently, CT has been defined as ``the thought processes involved in formulating a problem and expressing its solution(s) in such a way that a computer -- human or machine -- can effectively carry out''~\cite{Wing2017}. Since then, global education efforts have focused on providing students experience with CT~\cite{Bocconi2016,Brackmann2016,So2020}. Thus, the prevalence of computational thinking in modern educational and professional spaces motivates our interest in this subject.

Although computational thinking was initially proposed by the computer science community, it has since been incorporated into disciplinary science education standards~\cite{Barr2011a,Sneider2014}. For example, the Next Generation Science Standards (NGSS), which have been adopted by a majority of states in the United States, emphasize using mathematics and computational thinking as one of the eight major scientific practices that all students should encounter in their \mbox{K-12} education~\cite{K12Framework2012}. The expectation being outlined in these frameworks is that computational thinking becomes a practice of doing science across all STEM based disciplines. For example, through grades \mbox{9-12}, an example of CT is when students ``create/revise a computational model or simulation of a phenomenon.'' This is a broad expectation that could be applied to multiple disciplines. However, expectations have also been formalized for specific disciplines. For instance, in the case of high school physical science, one performance expectation states, ``Students who demonstrate understanding can create a computational model to calculate the change in the energy of one component in a system when the change in energy of the other component(s) and energy flows in and out of the system are known''~\cite{K12Framework2012}. This benchmark provides one example of how CT could manifest in high school physics classrooms. The increased expectancy around both general and specific CT based learning outcomes indicates a need to study best practices of supporting the development of and assessing CT in introductory physics education.

Despite the emphasis being placed on CT, language around this subject is often vague and can lead to significantly different learning outcomes depending on how it is implemented in the classroom~\cite{Grover2013}. Notice that over the years, CT has been framed in a variety ways (e.g., ``skills'' versus ``practices'' versus ``thought processes''). While many prominent studies agree that CT does not need to be done within a programming context, the subject has, nevertheless, largely been studied in classrooms with some form of coding. In this way, CT concepts are often conflated with the disciplinary core ideas of computer science. Wing calls computer science ``the study of computation,'' and she differentiates CT by focusing on it as a fundamental and conceptual problem-solving skill~\cite{Wing2006}. Computational thinking describes the practices involved in how a computer scientist thinks, rather than the rote skills required to carry out plans and ideas with a computer. Since CT does not require the use of a computer, these ideas can be applied in a computational or non-computational (i.e., analytical) learning environment. The ``CS Unplugged'' collection of teaching materials is one example of efforts made to teach students about CT without the use of a computer~\cite{CSUnplugged}. With that being said, it is expected that natural differences in the manifestation of CT will emerge based on different classroom implementation strategies (e.g., learning activities with a computer versus learning activities without a computer). Hence, careful work must be done to clarify what one means when discussing CT, how it changes depending on pedagogical approach, and how CT influences the design and implementation of computationally integrated activities.

Computational modeling activities have been proposed to be a rich area for exploring CT in introductory physics~\cite{Aiken2013,Caballero2015,Caballero2014,Caballero2012,Aksit2020,Dwyer2013,Eidin2020,Farris2014,Lee2020}. As an example, the repository of exercise sets from the Partnership for Integration of Computation in Undergraduate Physics (PICUP) features a plethora of computational modeling activities that incorporate CT practices~\cite{PICUP}. Physics courses that utilize computational modeling activities are often termed ``integrated'' courses because they do not focus on programming as the primary objective. Instead, computationally integrated courses infuse computational ideas within existing physics course content. Preliminary work has been completed to study the affordances and constraints of integrating computation within physics classes at the university, high school, and middle school levels~\cite{Weller2018,Weller2019,Leary2018,Oleynik2019,Bott2019,Hutchins2020,Caballero2012a,Kortemeyer2020}. For instance, a theoretical framework was proposed by Sengupta et al.\ in 2013, which described a modeling cycle with CT ideas underlying the entire sequence~\cite{Sengupta2013}. Their CT-infused modeling cycle includes steps like scientific inquiry (i.e., developing understanding of scientific phenomena and modeling practices), algorithm design (i.e., developing understanding of programming techniques), and engineering (i.e., iteratively refining a model). This cycle is underscored by CT because it is iterative and involves developing computational procedures to represent physical models. More recently, Orban and Teeling-Smith discussed the manifestation of CT in introductory physics classes~\cite{Orban2020}. Clearly, computational modeling and computationally integrated physics courses can engage students with CT concepts and practices. Hence, computational modeling activities are apposite settings for investigating CT at high school and introductory physics level.

Preliminary research has addressed assessment techniques for CT in introductory science classes, but there is still a marked lack of agreed upon assessment strategies for examining CT in introductory physics~\cite{Basu2018, Caballero2012, Weintrop2014}. For example, some work has been done by Wilensky and colleagues to develop an assessment framework for CT in secondary science classrooms~\cite{Swanson2019}. However, in an attempt to be broad and all-encompassing, this framework omits the nuanced details required to explore CT in-depth for each specific discipline, such as physics versus chemistry or biology. We claim that the universality of one CT framework is problematic when applying that framework to a different context. Thus, one of the primary goals of this manuscript is to develop a framework specifically aimed at investigating CT in introductory physics classrooms. From a research perspective, we are interested in providing a framework that will allow for the study of how CT manifests in what the students are doing in the classroom so that research questions around activity design and integration approach can be answered. At this stage in CT research it is useful to understand how design decisions around activities can influence the CT practices that occur rather than simply assessing students' development on a pre-post test.

In this work, we lay the foundation for a computational thinking framework specifically aimed at physics educators and researchers interested in teaching and studying CT when utilizing computational modeling learning activities. For researchers, we give a detailed description of the practices in our context and highlight some preliminary results from its initial implementation. For physics educators, we aim to demystify to some extent what CT practices can look like in the classroom to guide future activity design and assessment. In doing this work, we attempt to satisfy calls for integrating computation with introductory physics curricula~\cite{Chonacky2008,AAPT2002}. To address the aforementioned calls regarding the importance of CT in K-12 education, this study focuses on computation integrated with physics at the high school and introductory level, ranging from physical science to advanced placement (AP) physics to intro physics classes at the college level. This report will review previous literature to develop a framework specific to computationally integrated introductory physics courses, and then we will provide video data from in-class interactions between high school students to serve as existence proofs for these practices in our context. In Section \ref{LitReview}, we will review previous CT frameworks. Section \ref{Context} will provide details of our particular context. Then, Section \ref{Methods} will outline our research methods and Section \ref{Framework} will provide details about practices contained within our framework. Section \ref{Results} will  summarize the results from our application of this framework to two different classroom contexts. Lastly, Sections \ref{Discussion} and \ref{Conclusions} will be our discussion and conclusions from this work. Overall, this research may act as the initial exploration of what broader CT frameworks look like when filtered through a specific contextual lens. Ultimately, we hope to highlight the importance of context (i.e., discipline, class level, pedagogical strategies, computational platform, etc.) when determining the intended learning outcomes for computationally integrated physics courses.

\section{Review of Previous Computational Thinking Frameworks}\label{LitReview}

We reviewed a number of primary research studies in the field of CT. An important point of emphasis is that the literature included here is not meant to fully encompass all research pertaining to CT. Rather, it seeks to present a contextually diverse subset of CT frameworks. Throughout the review process, we stress the importance of context by taking information from pre-existing knowledge sources and including emergent ideas from empirical data sources. Our approach allows for the emergence and transformation of CT skills from other dissimilar contexts to our particular case. We will discuss our interpretations of specific practices in Section \ref{Framework}. In this section, we will provide a general background of each framework included in the synthesis of our framework. We follow with an analysis of how the other frameworks coincide and differ with regards to context and to thematic content.

\subsection{Overview of Frameworks}

Before we can compare and contrast across the various frameworks, we must first look at each study individually. Each framework illustrates a unique manifestation of CT for a specific context. This subsection provides an abridgment of the frameworks involved in our study before presenting any analysis across the frameworks. The following paragraphs offers a background of each study included in Table \ref{ShortenedLitReview_Table}. We selected the most comprehensive and relevant CT frameworks, and then adapted them to our context. After reviewing the key details of each study, the following subsections discuss how each study's context might have influenced the respective framework and analyze recurring themes within the frameworks.

\begin{table*}[t]
  \caption{Table summarizing information from previous major CT frameworks.
  \label{ShortenedLitReview_Table}}
  \begin{ruledtabular}
    \begin{tabular}{c c c c}
      \textbf{Framework} & \textbf{Level} & \textbf{Discipline} & \textbf{CT Categories} \\
      \hline

      Barr \& Stephenson, 2011 & K-12 & Computer Science & Core CT Concepts and Capabilities \\
      \hline

      Berland \& Lee, 2011 & Undergraduate & Interdisciplinary & Categories of CT \\
      \hline

      \multirow{3}{*}{Brennan \& Resnick, 2012} & \multirow{3}{*}{K-12} & \multirow{3}{*}{Interdisciplinary} & CT Concepts \\
      & & & CT Practices \\
      & & & CT Perspectives \\
      \hline

      \multirow{4}{*}{Weintrop et al., 2016} & \multirow{4}{*}{High School} & \multirow{4}{*}{Math \& Science} & Data Practices \\
      & & & Modeling \& Simulation Practices \\
      & & & Computational Problem Solving Practices \\
      & & & Systems Thinking Practices \\
      \hline

      \multirow{2}{*}{AAPT, 2016} & \multirow{2}{*}{Undergraduate} & \multirow{2}{*}{Physics} & Technical Computing Skills \\
      & & & Computational Physics Skills \\
      \hline

      Shute et al., 2017 & K-12 & Interdisciplinary & CT Facets \\
      \hline

      Rich et al., 2020 & Elementary & Math \& Science & CT Practices \\

    \end{tabular}
  \end{ruledtabular}
\end{table*}

\subsubsection*{``Bringing Computational Thinking to K-12: What is Involved and What is the Role of the Computer Science Education Community?'' -- Barr \& Stephenson (2011)}

Barr and Stephenson provide a general commentary on integrating computational thinking broadly into K-12 education. A key takeaway of the study is that CT integration is complex, and requires ``systemic change, teacher engagement, and development of significant resources''~\cite{Barr2011}. The paper emphasizes that it is no longer sufficient to wait  until the collegiate level to integrate CT into curriculum. More difficulty arises because there is no widely agreed upon definition of computational thinking. The authors' acknowledge the dilemma of defining CT in the broad context of K-12 education, which they say is a ``highly complex, highly politicized environment where multiple competing priorities, ideologies, pedagogies, and ontologies all vie for dominance''~\cite{Barr2011}. They discuss a 2009 collaboration between the Computer Science Teachers Association (CSTA) and the International Society for Technology in Education (ISTE), in which the two organizations sought to develop ``a shared vision and set of strategies for embedding computational thinking across the K-12 curriculum, most especially in the STEM subject areas''~\cite{Barr2011}. A key distinction is made, which states that the purpose of the collaboration was not to define CT, but rather to reach a consensus of what CT means in K-12. A major result of this collaboration was establishing the notion that disconnecting CT from CS would better accommodate the diverse set of disciplines in K-12 education. In discussing how CT manifests itself in the classroom, the point is made that CT ``is a problem solving methodology that can be automated and transferred and applied across subjects''~\cite{Barr2011}.

The items in the framework are called ``Core Computational Thinking Concepts and Capabilities'' \cite{Barr2011}. A key influence on the results of this framework was that definitions should be coupled with examples that demonstrate how computational thinking can be incorporated in the classroom, and so an example from five different disciplines -- CS, Math, Science, Social Studies, and Language Arts -- is provided for each concept/capability. Many of these concepts/capabilities are evident in other frameworks, namely problem decomposition, abstraction, algorithms and procedures, and simulation. Other concepts/capabilities, such as Automation, are less common in other frameworks. Furthermore, the collaboration resulted in agreement of various strategies or characteristics that could be considered broadly beneficial to any learning experience. Some examples of these strategies/characteristics include increased use of computational vocabulary, acceptance of failed solution attempts, and teamwork by students. The paper concludes by recommending strategies to achieve the necessary systemic change for CT integration, most notably that ``the larger CS community can help by providing suitable materials and taking advantage of opportunities to work with K-12 administrators'' \cite{Barr2011}.

\subsubsection*{``Collaborative Strategic Board Games as a Site for Distributed Computational Thinking'' -- Berland \& Lee (2012)}

Berland and Lee conducted a study that explores how computational thinking manifests itself in a distinctive context: playing a strategy-based board game entitled ``Pandemic'' \cite{Leacock2007}. Uniquely, the framework established by this study is better described as a coding mechanism, which is applied to data in the form of observations of students playing the board game together. In addition, this code book stems directly from real examples of students collaborating in a ``socially distributed way,'' and is therefore meant to be viewed through a lens of ``distributed computation,'' which specifically refers to team or group-based computational activities: ``When games are collaborative – that is, a game requires that players work in joint pursuit of a shared goal – the computational thinking is easily observed as distributed across several participants'' \cite{Berland2011}.

The framework consists of five core aspects of computational thinking, which come from Wing's 2006 article ``Computational Thinking'' \cite{Wing2006}, as shown in Table \ref{ShortenedLitReview_Table}. The authors explicitly state that these five aspects of CT ``do not encompass the full range of cognitive capabilities or processes that are involved in thinking computationally'' \cite{Berland2011}, but are rather a subset of CT that is satisfactory for their research purposes. The paper gives a comprehensive explanation of the ``Pandemic'' board game before beginning a discussion around how examples of CT appeared during gameplay. Unsurprisingly, the authors indicate that ``any effort to tie player behaviors to computational thinking is complicated by the lack of a concrete definition of computational thinking'' \cite{Berland2011}, a point of discussion for nearly every framework reviewed in this study. Thus, the coding framework is ``imperfect (at best), and perhaps their most serious problem is that they are overlapping and mutually dependent'' \cite{Berland2011}. Each category/code is presented in a table that includes descriptions, rationale behind inclusion, and examples of student dialogue that was coded in each respective category.

\subsubsection*{``New frameworks for studying and assessing the development of computational thinking'' -- Brennan \& Resnick (2012)}

Broadly, the main research goal of Brennan and Resnick's study is to explore how ``design-based learning activities – in particular, programming interactive media – support the development of computational thinking in young people'' \cite{Brennan2012}. The study focuses on the use of Scratch, which is defined as ``a programming environment that enables young people to create their own interactive stories, games, and simulations, and then share those creations in an online community with other young programmers from around the world'' \cite{Brennan2012}. The paper first defines a CT framework for the Scratch context, before ultimately moving into a discussion around assessment of young people who engage in programming. This framework is unique, as it is organized according to three dimensions: computational concepts, computational practices, and computational perspectives, the content of which can be seen in Table \ref{ShortenedLitReview_Table}. Definitions and examples within the Scratch context are given for each concept, practice, or perspective.

Each category within the framework appears to have a different function. The seven concepts within the computational concepts category resonate specifically with programming, yet are said to ``transfer to other programming (and non-programming) contexts'' \cite{Brennan2012}. Many of these concepts are not found in other frameworks, most likely because they are most evident in a strictly CS-programming context. Some concepts however, such as Parallelism and Data, are discussed thoroughly in other frameworks as notable elements of computational thinking. Interestingly, Brennan and Resnick's computational practices section is heavily focused on approaches to learning: ``Computational practices focus on the process of thinking and learning, moving beyond \textit{what} you are learning to \textit{how} you are learning'' \cite{Brennan2012}. Finally, the computational perspectives category is said to stem from Scratchers' descriptions of ``evolving understandings of themselves, their relationships to others, and the technological world around them'' \cite{Brennan2012}. The practices within focus on expressing oneself creatively through computation, enriching a computational experience through interaction with others, and questioning the complicated and often taken-for-granted technologies in the modern world.

\subsubsection*{``Defining Computational Thinking for Mathematics and Science Classrooms'' -- Weintrop et al. (2016)}

Weintrop et al.'s main research goal behind this study, which was part of a larger effort to ``infuse computational thinking into high school science and mathematics curricular materials,'' was to ``propose a definition of computational thinking for mathematics and science,'' specifically at the high school level \cite{Weintrop2016}. The definition proposed by Weintrop and colleagues is a four-category taxonomy of computational thinking practices, which is used to argue ``for the approach of embedding computational thinking in mathematics and science contexts,'' and facilitate discussion of how ``to bring current educational efforts in line with the increasingly computational nature of modern science and mathematics'' \cite{Weintrop2016}. The paper argues for the inclusion of CT-based curricula in math and science courses. The three main points from this argument highlight (1) the fact that these disciplines are seeing a rapid increase in computational influences,  (2) the pedagogic importance of utilizing computational tools and skillsets, and (3) the increasing underrepresentation of women and minorities in computational fields.

The taxonomy was developed via a clearly defined step-by-step process. The authors began by reviewing existing literature around CT, including two National Research Council publications \cite{Council2010,Council2010a}. Next, they reviewed and classified various activities designed to introduce computational thinking in math and science classes. They then took a working taxonomy to independent researchers, which ``undertook another round of revisions and categorized the individual practices into distinct categories''~\cite{Weintrop2016}. This taxonomy was then used in a professional development setting, which yielded mostly positive feedback with a few suggestions (such as the transition from ``skills'' to more actionable ``practices''~\cite{Weintrop2016}. Finally, numerous interviews with STEM professionals who regularly use computation helped inform the construction of the final draft of the framework.

The four overarching categories within the taxonomy include Data Practices, Modeling \& Simulating Practices, Computational Problem-Solving Practices, and Systems Thinking Practices \cite{Weintrop2016}, each of which contains five to seven sub-practices. The taxonomy is comprehensive; it contains a total of 22 practices, which is far more than any other recent framework. What is omitted from the framework is also notable, namely the fact there are no affect-based ideas (such as teamwork, creativity, pedagogical influences, etc.). The paper then highlights various high school-appropriate lesson plans that incorporate CT, and illustrates how the taxonomy is applicable for the activities. Some examples of these activities Video Games, DNA Sequencing, and Gas Laws~\cite{Weintrop2016}. The framework contains an above-average amount of practices, and each practice is described thoroughly with definitions and examples. 

\subsubsection*{``AAPT Recommendations for Computational Physics in the Undergraduate Physics Curriculum'' -- AAPT (2016)}

The recommendations for computational physics from the American Association of Physics Teachers (AAPT) ~\cite{AAPT2016} was written expressly for the purpose of encouraging an increased emphasis on computation within undergraduate physics curricula. The main motivation for this work lies around the increased recognition of computing as a particularly useful skill for physics majors in their post-baccalaureate degrees: ``Skills necessary for using a computer to calculate answers to physics problems, i.e., computational physics skills, are highly valued by research, industry, and many other employment sectors'' \cite{AAPT2016}. It is stated that the development of computational skills begins with fundamental tools and skills, on which students can build ``technical computing skills'' and ``computational physics skills'' \cite{AAPT2016}. Some of the basis skills include spreadsheets, integrated mathematical computing packages, general-purpose programming languages, and special-purpose software \cite{AAPT2016}. The authors discuss that computational tools have learning curves, and it takes time to develop proficiency as a student. After providing extensive detail of the specific technical computing skills and computational physics skills, the paper then highlights specific challenges to integrating computation into an undergraduate physics curriculum. These challenges revolve around time management, instructor and student experience, and software and hardware problems~\cite{Leary2018}. Finally, in the appendix, numerous examples of each skill are given to provide more insight into how these skills manifest themselves in a typical undergraduate physics classroom.

As mentioned, the paper establishes a set of skills, which are organized into either technical computing skills or computational physics skills, and can be seen in Table \ref{ShortenedLitReview_Table}. There are only three technical computing skills: processing data, representing data visually, and preparing documents and presentations that are authentic to the discipline. Interestingly, the computational physics skills are said to be characteristic of ``computational physics thinking,'' which the authors refer to as ``a synthesis of physics principles and algorithmic thinking'' \cite{AAPT2016}.  There are not any direct connections to computational thinking, but there is a lot of overlap between existing CT literature and the content within this paper. Additionally, AAPT seems to emphasize modeling and working with code instead of problem-solving or solution construction (this aligns with our goals as well; we will discuss our intended research outcomes at a later point). One major point from this paper is that these skills are directly linked to an assessment scheme. The skills are things students should be able to do after some experience, which is a further reminder of the general nature of this framework as an guideline for evaluating students in a physics classroom. A few of these skills are significantly different from those included in other frameworks, such as ``translating a model into code'' and ``choosing scales \& units.'' However, the majority of the other skills, including ``subdividing a model into a set of computational tasks'' (i.e., decomposing) and ``debugging, testing, and validating code'' (i.e., debugging) matched well with existing CT literature \cite{AAPT2016}. Our goal was to be inclusive with the breadth of CT practices discovered in the literature review.  As a result, this source provided multiple contributions to our framework, especially because it included practices explicitly related to physics.

\subsubsection*{``Demystifying computational thinking'' -- Shute et al. (2017)}

The paper from Shute et al. aims to examine past literature around CT to resolve conclusively the lack-of-definition issue that has plagued computational thinking researchers since its conception. They began by collecting and analyzing ``approximately 70 papers'' that were associated with CT, especially in early education (below collegiate level). Of these 70 papers, 25 were ultimately omitted due to various effects (small sample sizes, lack of direct focus on CT, etc.), which the authors felt would sway the results too heavily. Shute et al.\ then shift to a thorough summary of the literature review, which highlights the characteristics of computational thinking, interventions to develop computational thinking, assessment of computational thinking, and various computational thinking models \cite{Shute2017}.

The paper defines CT as: ``The conceptual foundation required to solve problems effectively and efficiently (i.e., algorithmically, with or without the assistance of computers) with solutions that are reusable in different contexts'' \cite{Shute2017}. The last piece of this definition implies a contextual reliance, which resonates with our findings. The authors point out that their definition illustrates how CT is portrayed through actions, which are the basis for performance-based assessments \cite{Shute2017}. In addition to defining CT, a framework of CT practices is established, and this framework can be seen in Table \ref{ShortenedLitReview_Table}. It is stated that the four most common components of CT from the literature are abstraction, decomposition, algorithms, and debugging, and they used these four practices as the basis for their framework structure. Other practices that are included in the framework are either standalone practices alongside these four (iteration and generalization) or they can be considered one element of the primary four (modeling is a sub-component under abstraction) \cite{Shute2017}.


\subsubsection*{``Teacher implementation profiles for integrating computational thinking into elementary mathematics and science instruction'' -- Rich et al. (2020)}

The most recent framework comes from Rich and colleagues~\cite{Rich2020}. The central research question of this study was ``How do elementary school teachers create opportunities for their students to engage in CT practices during mathematics and science lessons?'' \cite{Rich2020}. The study used eight teachers from five different schools, each of whom participated in a series of professional development workshops that (1) steadily introduced computational thinking in the contexts of math and science courses, (2) formally introduced to the four selected practices as seen in Table \ref{ShortenedLitReview_Table}, and (3) allowed teachers to evaluate activities and assessment strategies with peers and workshop administrators. After evaluating teachers' implementations of CT-styled activities in their respective classrooms, the researchers established a concrete set of strategies used by teachers to create CT opportunities. Examples of such strategies include framing, prompting, and inviting reflection \cite{Rich2020}. When coding in-class data, three categories were taken into account: examples of CT practices, whether these examples were explicit or implicit, and strategy-implementation from teachers. The results walk through four profiles: CT as problem-solving strategies, CT to structure lessons, CT through prompting, and CT for teacher planning. Results indicate that the first profile, CT as a Problem-Solving Strategy, was the most commonly witnessed profile~\cite{Rich2020}. Rich et al.~discuss how these results could impact professional development in the future.

Their framework only includes four practices: abstraction, decomposition, debugging, and patterns. It is stated early in the paper that the focus is to ``work with teachers on understanding how they implement CT -- and not on defining and redefining CT...'' \cite{Rich2020}. This is a strong context for what we are doing. We want to see how CT practices look in our context, like Rich did. Therefore, it is important that we recognize that this framework is not meant to be completely comprehensive, and thus we should not consider omissions from the framework as significant influences on our framework. Rather, it will be important to focus strictly on the practices that were included for specific reasons discussed by the authors. For example, the authors state that ``Decomposition and Patterns were selected because ``interviews with our participating teachers suggested they saw many connections between these practices and their mathematics and science teaching'' \cite{Rich2020}. Debugging was emphasized because ``a focus on finding and correcting mistakes appealed to them \cite{Rich2020}. Lastly, abstraction was chosen because ``it has been identified as a particularly important CT practice by several CS education researchers'' \cite{Rich2020}. 
The framework containing these four practices was established to purposely capture only major aspects of CT that are heavily emphasized in research or were associated with desirable learning goals for the teachers involved in the study.



\subsection{Contextual Differences Between Frameworks}

As we review the contextual differences between the frameworks included in our literature survey, we are implicitly highlighting the fact the context is an influential factor on how CT should be evaluated and assessed in the classroom. While there is a generalizability to these frameworks, we are interested in the certain pieces that are more or less applicable for different contexts. The following paragraphs highlight the major contextual differences between the studies above, which include the frameworks' intended audiences, disciplines, and research goals. We argue why these differences may have an effect on each respective framework.


CT frameworks that are meant to apply to a wide variety of grade-levels, such as all of K-12, tend to generalize their descriptions of CT practices. One would expect for CT practices, and other general problem solving skills, to appear inherently different as students move through grade levels, especially from primary to secondary schooling, due to a natural progression in schoolwork. There should be distinctions between the level of engagement with CT practices because the type of learning activities change, as detailed by NGSS's benchmark recommendations around computational thinking \cite{K12Framework2012}. This is one example that supports our position on the importance of context when applying a CT framework to one specific classroom. While a generalized CT framework may appear comprehensive and applicable to virtually all STEM classes, in practice, the application of a framework is much more detail-oriented. For example, if a high school instructor was looking to evaluate how their students were decomposing problems, key indicators of decomposition (perhaps detailed in a rubric or assessment layout of some sort) would look different than those of an elementary school teacher. The high school teacher might emphasize higher level decomposition of more complicated problems, while the elementary school teacher might only emphasize simpler forms of decomposition. Therefore, if a framework is meant to applicable to an inclusive array of contexts (such as K-12 STEM), then it must detail the distinctions between higher and lower level application of CT practices. At different levels, students encounter different concepts. Currently, it is not well understood what the generalizability of CT practices means for different pedagogies, education levels, or concepts. The inclusivity of previous frameworks is not understood based on context and different student levels. You can create a broad, generalized framework, but how applicable are they? How can we know if they are useful without considering pedagogical details. Each set of CT practices is contextually specific, and when developing such a specialized set of practices, one cannot ignore the details of their intended context. If a framework promotes a generalized applicability without providing any details on how the framework is applied across differing education levels then the framework will struggle to be more than an introductory guide to a researcher of CT or an integrator of CT into the classroom.

In addition to differences in the particular education level, CT studies often differ in their intended disciplines. For instance, reports from Barr \& Stephenson~\cite{Barr2011} and Grover \& Pea~\cite{Grover2013} are situated in the realm of computer science, while the framework from Weintrop et al.~\cite{Weintrop2016} focuses on math and science classrooms. Other frameworks, like Brennan and Resnick's, aim at application across diversity of different disciplines. If the framework was intended for computer science classrooms, then the practices will likely differ from those of a framework for computationally-integrated physics courses. Frameworks that are meant to apply to multiple disciplines may have less specificity than frameworks intended for one specific discipline. In particular, we focus on CT integrated within physics classes. These types of courses are different than computer-science or coding courses, because they often do not explicitly attempt to teach students how to code. Rather, the purpose of integrating computation into physics class is often to provide an alternate avenue for learning the physics material, as well as to promote the utility of computation as a tool for doing physics \cite{Caballero2012a}. This distinction impacts how CT is evaluated in the classroom; if it is a computer science course, then CT will likely be evaluated differently than in a computationally-integrated physics course. For this reason, the AAPT framework \cite{AAPT2016} has more applicability than other frameworks for our context, because it is intended specifically for the the discipline at hand: computationally-integrated physics. Frameworks that are intended for dissimilar disciplines might not capture the unique characteristics of a physics class, and this needs to be taken into account when assessing each studies' impact on our framework.

It is clear that CT studies are often conducted with different research goals in mind. Of course, some researchers aimed at synthesizing universally applicable CT frameworks. At the same time, others sought to formulate a framework for the purpose of creating attitudinal surveys~\cite{Korkmaz2017} or exploring how CT ideas affect students working in groups~\cite{Berland2011}. An example of this would be the framework from Berland \& Lee, which applies an original framework to investigate the use of CT when students played strategic board games~\cite{Berland2011}. In the end, they proposed that board games could serve as an important foundation for further exploration of CT in the context of games. In such a case where the framework was created with a specific research goal in mind, the practices included in that framework were highly context-specific. In the case of Berland \& Lee, the practice of ``Distributed Computation'' is geared toward the specific type of dialogue-based data that was produced in the study. One of our intended research outcomes is to produce a comprehensive framework to describe what CT might look like for our context, so that high school teachers might be provided with guidelines that will influence their curriculum design and assessment strategies around computation. Studies that share a focus on assessment of computational activities in their research outcomes might resonate with our work. Thus, the alignment of various studies' research goals with ours has an impact on their relevance for our framework.

Many of the differences between CT frameworks are attributed to contextual distinctions, including intended audience, discipline of origin, and research goal. Some of these distinctions are aligned or misaligned with our context, and we argue that it is important to consider one's context when applying a CT framework to one particular classroom. If the context of a previous study closely matches our context, it is more likely that the corresponding framework will have a greater impact on the development of our framework. Likewise, if a study's context is largely different from ours, then it is unlikely that it will have a significant influence on our view of CT in high school physics. The next section will provide more specific details around the thematic content of each study, as well as highlight some key decisions about the structure and organization of our framework. 

\subsection{Comparing \& Contrasting Recurring Themes}

The following subsections provide a commentary on the thematic differences between each of the frameworks, as well as our decisions of how these themes influenced our framework. Themes discussed here include practice grouping and tiering, idea-based CT vs. action-based CT, terminology, and affectual elements of CT. These frameworks are organized differently, not only in structure and grouping of practice, but in overall language as well. We provide a background of how different frameworks approach each specific theme, followed by our approach with our framework.

\subsubsection*{Framework Structure, Practice Grouping, and Grain Sizes}

One key distinction that separates these frameworks is if and how they group their framework elements. Some studies, such as those from Weintrop et al. and Brennan \& Resnick, group every practice together in various categories (For Weintrop et al., it's Modeling \& Simulating, Computational Problem-Solving, Data, and Systems Thinking practices, and for Brennan \& Resnick, it's concepts, practices, and perspectives). However, we see a different approach from studies like Barr \& Stephenson. Interestingly, none of the concepts/capabilities in this framework are tiered or grouped together in any way, even if similarities between them are present (i.e., Data collection, Data analysis, and Data representation are separated, rather than falling into one ``Data'' category). Shute et al. fall in the middle, with a hybridized approach (some categories with practices within, and other standalone practices).

The ways in which practices are organized has implications about how they are defined. Related practices are typically considered so because they occur at the same stage of a problem-solving or modeling process, or they fall under a similar theme. Weintrop et al.'s Data Practices category contains five practices, all associated with handling a dataset. Brennan \& Resnick's ``Perspectives'' category contains three perspectives (Expressing, Connecting, and Questioning) that are each associated with students' ``evolving understandings of themselves, their relationships to others, and the technological world around them'' \cite{Brennan2012}. We feel that categorizing our CT framework based on the general structure/order of practices as they occur in the classroom is useful for our context. When teachers are evaluating students' development of CT skills, it may be useful to have a sense of which skills pertain to which specific stage of the model development. We have attempted to categorize each practice in this way, but some of our practices are present throughout the entirety of the problem-solving process. Thus, we ultimately have a hybridized approach like Shute et al. \cite{Shute2017}, where each category corresponds to a step/stage of the student's approach to solving the problem/modeling the physics, and each standalone practice corresponds to a practice that is present throughout the process, and can happen at any stage.

The grain sizes of various CT practices have implications for both the authors' value for the given practice, as well as how it is defined. Where the practice grouping idea above discusses how practices relate to each other in parallel (from one piece of the framework to another), the hierarchical tiering of practices in various frameworks highlights which practices are said to be encompassed by others. Rather than being grouped together as in the same category, they are tiered, with the lower practice being described as one example/manifestation of the larger practice. This tiering of practices can be influenced by a number of factors. For example, inside of Weintrop et al.'s ``Modeling \& Simulating practices'' (where modeling itself can be described as a CT practice), there is the sub-practice of ``creating computational abstractions''~\cite{Weintrop2014}. However, in Shute et al.'s framework, the abstraction category contains the practice of modeling. The two frameworks have a flipped tiering. Weintrop et al. highlights creating abstractions as a form of modeling, and Shute et al. highlight modeling as a form of abstraction. Reasons for this choice of tiering are not readily available in all frameworks, and so we can not always know why one framework values a certain practice over another. We argue that tiering practices matters, and there must be a reason why a certain practice falls under a larger umbrella practice. We feel that tiering practices is necessary, because the large-scale outer practices can often be vague or highly generalizable. Classifying numerous practices as examples/manifestations of a larger practice helps reduce this vagueness.

\subsubsection*{Language of ``ideas'' versus ``actions''}

Another subtle detail to consider is the specific language around the items in a framework. Some frameworks discuss their elements of CT as ideas or concepts (nouns), and others identify them as actions (verbs). For example, Weintrop et al. refer to each of the items in their taxonomy as ``CT practices,'' which all end in -ing (Collecting Data, Assessing models, creating abstractions, etc.) \cite{Weintrop2016}. Additionally, the AAPT framework describes a set of ``skills'' around computational physics, which are also described in an action-based form \cite{AAPT2016}. By contrast, some frameworks opt for broad definitions of abstract ideas. Barr \& Stephenson's framework is represented by a table of ``concepts and capabilities,'' each of which are not action-based \cite{Barr2011}. Likewise, Berland \& Lee's study includes ``examples of CT'' in the form of idea-based concepts that encompass a number of actions \cite{Berland2011}. We ultimately chose to organize our framework components in the form of actionable practices, because we felt that our framework (which is designed with teachers of all levels of computational experience in mind) should be able to serve as an evaluation tool for teachers and not only be used by researchers to study a context. Because the framework is made up of actionable items, then we expect that it will be easier for teachers to classify behavior in the classroom with the framework, allowing for more straightforward and effective assessment. We also hope that it will be easier for researchers to identify occurrences of CT in the classroom so that they can answer research questions around integration or activity design.

\subsubsection*{Vocabulary of practice titles versus definitions/descriptions}

In addition to the action-based language of various framework components, we also need to consider the similarities and differences between the terminology used across the literature. Two frameworks could have included the same practice/idea, but named each respective idea uniquely. Literary support for the practice of utilizing generalization as an element of CT came from Brennan \& Resnick (``Reusing and Remixing'') \cite{Brennan2012}, Rich et al. (``Patterns'') \cite{Rich2020}, and Shute et al. (``Generalization'') \cite{Shute2017}. Including group work in our framework stemmed from Berland \& Lee's ``Distributed Computation'' \cite{Berland2011}, Korkmaz et al.'s ``Cooperativity'' \cite{Korkmaz2017}, and Brennan \& Resnick's ``Connecting'' \cite{Brennan2012}. Additional examples can be seen in the language around debugging and algorithm building, which both have a number of justifying sources with unique terminology.

Although these studies often seek to justify the same practices, they each name the practice something distinct. Based on this, we felt it was important to seek out the definitions/descriptions of practices, rather than their designated names. There does not appear to be a wide diversity in the practices that are discussed similarly, aside from minor contextual differences, and so it makes sense to try to combine practices described similarlyacross frameworks. By not focusing on titles, we are making sure to be inclusive in our review of the literature. Because of this, if a study discusses an idea extensively (even if it is named differently than in other studies, or is not included in that study's overall framework), it is still factored into our filtering process because we are looking at descriptions of CT beyond the strictly defined elements described by the authors. This allows us to narrow down the set of distinct CT practices.

\subsubsection*{CT practices/skills versus affect-based ideas (interpersonal skills and mindsets)}

A number of frameworks include affect-based ideas. Brennan \& Resnick's framework highlights various ``Perspectives,'' including ``Expressing,'' ``Connecting,'' and ``Questioning'' \cite{Brennan2012}. Berland \& Lee's work is heavily influenced by the idea of ``Distributed Computation,'' which can be associated with working in groups \cite{Berland2011}. Additionally, Barr and Stephenson outline important dispositions when engaging in CT~\cite{Barr2011}. These affect-based elements of CT are not universal in every framework, but they are somewhat common. Given the choice of whether or not to emphasize these elements in our CT framework, we have chosen to include them because we feel that they align with our teachers learning goals. As stated previously, many teachers did not have a strong basis for evaluating CT in their respective classrooms. But we know that teachers do care about these affect-based ideas. They want their students to be comfortable with computation, rather than afraid of it~\cite{Weller2019}. They have discussed the group-based nature of computational activities. Therefore, we wanted to be inclusive in regard to these affect-based ideas.

\subsubsection*{Summary: Computational thinking in computationally integrated physics classroom framework}

Our ultimate goal is to provide a CT framework for teachers that (1) helps teachers to develop a better understanding of how CT manifests in the classroom, and (2) guides teachers in their assessment strategies around CT. We also want to provide a framework that researchers can apply to similar contexts or adapt for related contexts that allows the investigation of CT in the classroom. The combination of the aforementioned goals for our framework combined with the contextual boundaries for where the framework is to be applied guides our definition of CT. The paper to this point has outlined that the following definition should in no way be interpreted as a universal definition of CT and instead as a definition that aligns with previous definitions but that has been operationalized to allow for the research of CT in the classroom setting. We align our definition of CT with aspects of Berland and Lee and with the focus on practices of Weintrop and colleagues. For our context and goals we view CT as an array of practices that students engage in as a group ($>$1 person) when constructing, adapting, and utilizing a computational model. This definition takes up the ``distributed computation'' framing of Berland and Lee by focusing on the thinking as being group oriented while also applying the idea that CT is observable in the form of the practices the group engage in. Again, this in no way implies that CT can only be engaged in a group format or that it requires interaction with a code, instead, it indicates that this is the form of CT we are interested in investigating. The distributed computation model aligns with our previous curriculum design efforts that use socio-constructivist learning theory as a foundation~\cite{Irving2020}. Using socio-constructivism as a foundation for constructing learning activities is also advocated for in the ICSAM workshop and has resulted in computation being integrated in a group format by the teachers who attend the workshop. From both a research and assessment perspective, perceiving CT as a set of practices allows for us to focus on a form of CT that is observable in the classroom. In regards to other distinctions and decisions we made that relate to the previous frameworks:

\begin{itemize}
  \item We grouped and tiered some but not all practices. Grouping was based on both an intentional sequence of CT practices that we encourage our teachers to use and an observed cycle that students engaged in. It was also felt that a grouped structure would better illustrate the meaning of various CT terminologies for unfamiliar instructors.
  \item We included affect-based ideas because they aligned with the group orientated nature of our CT definition and with the majority of our teachers intended goals for integrating computation into their curriculum.
  \item We decided that the CT practice of modeling was inherent within the our framework because students begin class with a working but inaccurate computational model so instead we focus on the CT practices that further develop that model.
\end{itemize}

The following section on our context will help outline why these decisions were made.

\section{Our specific context: Computational Modeling in High School Physics with VPython}\label{Context}

The context of this work is computationally integrated high school physics classes in Michigan. The instructors involved in our study participated in a professional development (PD) series called Integrating Computation in Science Across Michigan (ICSAM). This PD program is an NSF-funded project where teachers learn to program and to teach computational modeling to their students. It begins with a five-day summer workshop that is followed by periodic workdays every two months throughout the school year. The workshop is hosted cooperatively by the Departments of Physics \& Astronomy and Teacher Education at Michigan State University. Over the past two years, more than 35 teachers have participated in the program. At the workshop, teachers collaborate in groups and working with facilitators to design computational activities for their classrooms. At the follow-up workdays, instructors return to the university to check in and discuss challenges faced in the classroom. The workshop is split into two main themes: equity in the classroom and confidence with computing.

The programming language that teachers use for all computational modeling activities is Glowscript VPython~\cite{VPython}. This language is a specialized form of Python and the ``Visual'' 3D graphics module that can be written and executed in a browser, thus, decreasing the barrier to entry to writing scientific programs. The language is particularly well designed for modeling objects and events in a 3D environment as it was specifically designed for use in physics classrooms. For this reason, Glowscript VPython is well suited to simulate motion, collisions, and other spatial features that are commonplace in the physics classroom. Because 3D visuals are so readily available using Glowscript VPython, visual animation is emphasized as an important way of helping students learn physics. The ICSAM program employs online platforms (like Glowscript.org and Trinket.io websites) for all computational work. These platforms make it easy for teachers and students to create, save, and share their code with their collaborators.

The ICSAM model makes use of ``minimally working programs'' (MWPs) to simplify the more difficult aspects of coding, to provide scaffolding for students, and to keep the focus on physics even in during computational activities~\cite{Weatherford2012}. At first, MWPs exist as a functional computer code that will run without an error and model some aspect of a physics phenomenon. This serves as a jumping off point for students. Scaffolding can be further augmented by adding comments to the code or providing a worksheet to guide students through the beginning of an activity. Minimally working programs are beneficial because students can start activities by interacting with the physics immediately, while not being slowed down with creating a computer program from scratch. Usually, MWPs will start off as a simulation that runs without an error but somehow models the physics incorrectly. As a simple example, students may be given a model that does not have any value for the net force yet, and students will have to add an equation for the net force that references the pre-existing code. In some cases, teachers may have provided a MWP that has a simple error embedded to provide students experience with debugging. By the end of most activities, students will have modified the MWPs to implement physics equations, model phenomena, and change values in their code to rapidly test and predict the outcomes of different physical scenarios.

Numerous college-level curricula in the past have discussed the beneficial use of VPython and MWPs in physics classrooms. For example, the Matter and Interactions curriculum, developed by Sherwood and Chabay~\cite{Sherwood2001}, relies on the use of VPython for modeling canonical physics situations. Similarly, the P-cubed curriculum at Michigan State University utilizes VPython~\cite{Irving2020,Irving2017}. Thus, there is a building consensus that these types of computational classroom interventions are a good way of teaching physics at the introductory level, however, there is no explicit research of the benefits of using minimally working programs in regards to CT or physics. While ``CS Unplugged'' activities have been discussed as a way of engaging students with CT, we mostly avoided this topic due to the dissimilarities to our context. Furthermore, even before participating in the ICSAM PD series, some high school teachers have already adapted college-level physics problems to the high school context. Some of the teachers joined our program by way of a workshop held at the annual AAPT section meeting (MI-AAPT), others teachers were self-motivated to implement computation in their classes. Therefore, previous evidence from conferences and word of mouth motivated the inception of the ICSAM program.

The main motivation of the physicists involved in the ICSAM workshop is to facilitate instructors integrating computation into their curriculum through boosting their confidence with computation and assisting them in curriculum design around computation. Workshop activities are designed to help instructors practice their computing skills and design computational activities for their classes. While these activities are underway, organizers monitor the participants and assist them when necessary. One example these activities is ``Ice Station McMurdo,'' a projectile motion computational physics problem for introductory mechanics students at Michigan State. Teachers work in groups to recognize how the code is working and solve the problem. This helps teachers (especially those with little coding experience) to be more comfortable with writing and evaluating code. Towards the end of the workshop, instructors are asked to design their own computation-based physics problem(s), with the help of workshop facilitators. This is also particularly helpful for teachers, because it allows them to ask questions about anything from specific syntax to general coding principles like loops or conditional logic.

A key distinction that separates the ICSAM approach to computational integration is that curricular design is solely at the teacher's discretion. There is no set curriculum that the workshop intends to establish in every participant's classroom. In other words, the ICSAM PD experience is not a prescriptive intervention for teachers to instantly implement in their classrooms. Rather, each instructor decides how thoroughly computation will be integrated in their respective environment. We feel that this approach is necessary because instructors tend to have different levels of programming and teaching experience, different intentions for implementing computation, and different learning goals for their classrooms. For someone with little or no experience coding, seeing a fully working script can be intimidating. For teachers in this position, it is essential that we do not exacerbate this by moving too fast. In addition, while we do not want computational integration to be too daunting for newcomers, we also do not want it to be too simplistic for those experienced with computation. Therefore, we take a teacher-centered curriculum approach, and we support the teachers with the decisions they make around their curriculum. By using this approach, each individual teacher's confidence and overall comfort with learning to code are the primary concern, and this facilitates successful workshop activities.

Alongside activities in which participants practice coding, there is also a brief presentation of the CT practices in the literature. We feel that this is necessary because it provides a clearer picture of why computational integration is so important in STEM. High school physics instructors that are beginning to integrate computation into their classroom often do not know how to identify CT practices or articulate learning goals around their computational activities. The framework presented here intends to provide instructors and researchers with a means of identifying and classifying CT in the classroom. It will allow them to evaluate effectively if students are engaging in these practices while they work through computational activities. In addition, it provides guidance for computational activities in the high school physics context. For teachers who are new to designing computational activities, this framework is a jumping off point, which establishes specific themes students should be working with. Thus, it provides instructors with recognizable goals around their computational activities. In the next section, we will outline the design of our framework, and describe these specific actions in detail.

\begin{figure*}[th]
    \includegraphics[width=\linewidth]{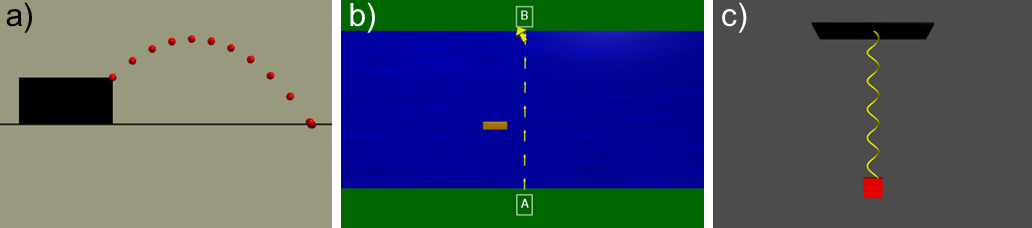}
    \caption{Outputs displayed in Michael's solutions for computational models of (a) projectile, (b) river crossing, and (c) spring energy problems.}
    \label{ProjectileRiverSpring_Fig}
\end{figure*}

Another focus of the ICSAM workshop is to promote equity around computational work in the classroom. Because working in groups is messaged heavily in the workshop, equitable participation is an important factor to keep in mind for early learners of computing. To study equity in the classroom, teachers are given experience with EQUIP (Equity QUantified In Participation) \cite{EQUIP}, a software designed to synthesize data around equity in classroom activities. Workshop attendees are trained to use EQUIP to create a student roster, choose the social markers and aspects of classroom discourse they want to track, and collect and analyze data to evaluate. This allows teachers to make sure that their classroom activities (especially group-based activities) are equitable. This is a healthy collaboration because these teachers are in the process of designing new curriculum that is centered on group-based computational activities, and this provides them with a means of evaluating the equity of these activities as early as their first implementation~\cite{Shah2020}.

\subsection{Classroom 1: Michael's AP Physics Classroom}

After teachers completed the professional development workshop over the summer, we followed a subset of the teachers to study the implementation of computation in their classrooms. In this report, video data are presented from two classrooms. We will use evidence from these two classrooms to provide proof of these CT practices within our context. The first classroom, taught by Michael, was an advanced placement (AP) physics C mechanics course for seniors. In this classroom, computational activities took place every Friday for the entirety of the school year. Students were assigned to groups of three to five students, and the group all shared one desktop computer to build computational models on the Glowscript platform. Typically, Michael's MWPs started with a model that would run without error but would not model some aspect of the physics correctly. These activities were often used as supplements to other lab or lecture activities that students worked with earlier in the week. Michael's activity prompts gave students guidance on how to complete the activities, without directly instructing them how to modify the code (e.g., ``you'll have to write lines of code to designate the forces on the block; Fgrav and Fspring'').

The three activities examined from Michael's classroom were a projectile problem, a river crossing problem, and a spring energy problem (see Fig.\ \ref{ProjectileRiverSpring_Fig}). In the projectile problem, depicted in Fig.\ \ref{ProjectileRiverSpring_Fig}(a), the students are tasked with changing the initial conditions of the model (cliff height, initial ball height, initial velocity, and angle), modifying a while loop to make the program stop when the ball hits the ground, and displaying the final results (horizontal distance, max height, and final velocity) with a print statement. This computational activity was used to model an experimental projectile motion lab that the students did earlier in the week.

For the river crossing problem, shown in Fig.\ \ref{ProjectileRiverSpring_Fig}(b), the students first calculate (on a whiteboard) the angle that a boat must travel at to reach the other side of a river straight across from where it started. Next, the students work with a computational model to simulate the scenario that they calculated on the whiteboard. To complete the computational portion, the students have to change the boat's angle and modify the boat's velocity by adding the velocity of the river's current to the velocity of the boat.

Finally, Michael's spring energy problem features a red block hanging on a yellow spring by a black support. The students have 3 primary tasks to complete this activity. First, the students need to get the block to bounce up and down by adding equations for the forces of gravity and the spring. Second, the group is tasked with adding graphs for gravitational, elastic, and total energy values; they are given a working graph of kinetic energy by Michael in the MWP. Lastly, if the students make it this far, they are prompted to add a damping force to their computational model. Overall, this activity was one of the more involved activities of Michael's classroom, which makes sense as it was positioned near the end of the students' second semester of working with computational models.

\subsection{Classroom 2: Liam's Physical Science 2 Classroom}

\begin{figure*}[th]
    \includegraphics[width=\linewidth]{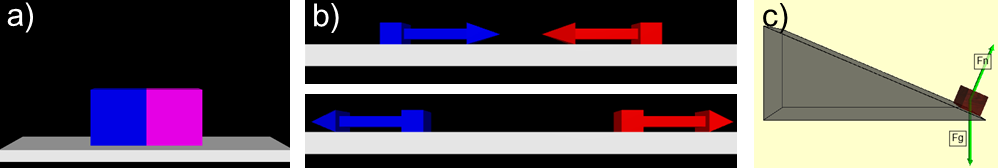}
    \caption{Outputs displayed in Liam's solutions for computational models of (a) colliding crates, (b) momentum conservation (top panel for before the collision and bottom panel for after the collision), and (c) block on a ramp activities.}
    \label{CrateMomentumRamp_Fig}
\end{figure*}

The second classroom examined in this work was Liam's physical science 2 advanced classroom for sophomore students. In this classroom, computational activities were scattered into the curriculum about once every month, giving the students about 4 total experiences with computational modeling per semester. Students were assigned to groups of three to five students, and they all shared one school-provided laptop per group. Students also worked at desks that had whiteboard tops to facilitate easy writing and sharing of ideas. Liam's computational activities were treated as introductory experiences to programming, rather than a reinforcement of physics concepts or labs experienced earlier in the course. This meant that students were usually learning how to interpret code, modify simple aspects of a computational model, and test/predict the outcome of scenarios with their simulation. Typically, Liam's students were given fully-working codes, and the worksheets provided to students asked questions that needed to be answered directly by interpreting the code. First, students were given a day to simply change code in Glowscript to spell their name by positioning objects and changing their attributes (e.g.\ size, position, angle). Our analysis focuses on three activities, following the aforementioned introductory activity from Liam's classroom: 1D motion colliding crates, 1D momentum conservation, and a block on a ramp. Figure \ref{CrateMomentumRamp_Fig} displays the output of computational models for activities from Liam's classroom.

After their first introductory experience with computation, students in Liam's classroom work with a computational model of two crates moving toward each other. The final output of the colliding crates activity is displayed in Fig.\ \ref{CrateMomentumRamp_Fig}(a). In the MWP, Liam provides his students with the correct code to model 1D motion of one crate moving across a floor object. Students are tasked with changing some parameters of the model (e.g., ``Give the crate a different constant velocity moving to the right.''), interpreting the code (e.g., ``Which lines of code make the crates move?''), adding to the model (e.g., ``Create a second crate with a different size, color, and position), and getting the simulation to stop when the two crates meet in the center (e.g., ``Figure out how to stop the program once the crates collide.''). This activity provides students with a more guided, in-depth activity to work with computational modeling and thinking practices.

The second computational activity analyzed from Liam's classroom was the momentum conservation activity (see Fig.\ \ref{CrateMomentumRamp_Fig}(b)). To give students experience with building on previous models, the computational model of momentum conservation used a code similar to the previous colliding crates activity. Liam provided his students with a worksheet that comprised several phases: analyzing the code, playing around with the model, making predictions, modifying the code, and testing predictions. Students were tasked with explaining what different parts of the code meant to them, changing the code to observe the effect of their changes, and testing/predicting the outcome of different scenarios when plugging in specific real-world values. In this activity, Liam emphasizes the utility in being able to rapidly test different situations with a computational model. By the end of the momentum conservation activity, students used their code to model elastic collisions between two boxes, a car and a truck, and a dodgeball and a teacher.

The block on a ramp problem in Liam's class was students' final experience with computational modeling in the physical science course. Figure \ref{CrateMomentumRamp_Fig}(c) illustrates the final output of the block on a ramp computational model. The MWP given to students begins with a block sliding down a ramp with the normal force and gravity force vectors depicted by green arrows. First, students are tasked with extracting important details from the code (e.g., ``What is the angle of the ramp?''). Next, the students had to make changes to the code and observe the effect (e.g., ``What happens if you increase the angle of the ramp?''). Lastly, if the students got far enough, students were instructed to add friction to their simulation. In the end, this activity was a more involved and complicated activity than the previous two assignments.

Michael and Liam demonstrate a diversity in both approach and type of activities that they used to integrate computation into their physics classroom which highlights that high school physics can be a fruitful context for engaging students in CT. As researchers when observing the classrooms we see CT happening in these contexts, but teachers often do not know how to identify CT or to articulate learning objectives around CT. The framework presented in this paper will be used to identify CT in the classroom, to assist teachers in developing learning goals around computation integrated with physics, and to serve as a jumping off point for future research.

\section{Framework Design and Research Methodology}\label{Methods}

The process of developing our CT learning goal framework for introductory physics began with a review of the relevant literature. In total, over 30 scholarly articles were read and analyzed to extract key ideas. Next, we narrowed our review to examine works closely that were frequently referenced among the literature. This narrowing process resulted in the 7 works discussed in Section \ref{LitReview}. After choosing these prominent CT studies as a basis for our framework, several researchers within our group read the papers and discussed the key ideas to identify major similarities and differences within the papers. This process helped us comprehend the critical aspects of CT that we should look for in our data. We tried to ask continuously how our context (e.g., subject matter, high school level, different classrooms) might influence the importance of different CT ideas. This literature review served as the foundation for the development of our physics-specific CT framework.

After the literature review, our team began developing a set of CT practices that, in theory, would be relevant for high school physics and, by extension, due to the frequent overlap in content: introductory physics. The goal of this development was to encompass the ideas that were highlighted in previous frameworks while also filtering those ideas through the context of physics and high school. This filtering process was based on the experiences of physics instructors and teachers within our research group and teacher cohort. We iterated on the framework's development over three main meetings, and each time a new draft arose as a result of feedback from peers and members within our team. Initially, our goal was to be as inclusive as possible, resulting in a large, multifaceted framework.

We iterated over the development of our framework with three groups of researchers. The order of these meetings was chosen based on the availability of the people involved. First, we began by reviewing our draft of CT practices internally within our research group to determine CT practices that we felt did not align with the context of a physics classroom at the introductory or high school level. At that time, the framework included 16 ideas: algorithm building, iterative problem solving, abstraction, debugging, decomposition, transferring problems to code form, generalization, modularity, groupwork, data, planning, modeling, programmer logic, parallelization, systems thinking, and extracting physical insight. After reviewing these practices with our team, we found that some were not fit for our context because they were either too complex (e.g., modularity, systems thinking, parallelization) or they were too broad to be considered exclusively related to CT (e.g., planning, iterative problem solving). The research team also suggested that categories be used to group similar practices together. After making changes based on the previous thoughts, our second draft was then presented to a larger group of physics education researchers (many of whom had integrated computation into their classrooms). Here, we learned that the usability of our framework could be improved by using concise language in our definitions and providing a wealth of examples to explain the variety of CT practices. After this, a third draft was then reviewed by external computer scientists, leading to the final set of practices in our CT framework. The computer scientists helped us realize that our CT practices should be focused on moments when students were working with computers/programming, so as to not conflate CT practices from this framework with general problem solving practices. The list of practices that were omitted from our framework will be discussed later (see Section \ref{Discussion}.B). We ended with a total of twelve practices in our framework, after deciding to omit or merge four of the original sixteen. This filtering process was the first phase of our research process. Because so many frameworks existed previously, we did not want to start from scratch but also did not assume that the practices and definitions produced through the filtering process were finalized or the full spectrum of CT practices that could occur in our context.

In order to understand how the initial filtered framework would apply to our context, we gathered audio and video data from numerous high school classrooms around the state of Michigan. Videos were gathered over the course of a single academic year by researchers visiting the classrooms of teachers participating in our professional development series. To gather the data, a camera on a tripod was positioned to record students' discussions, body language, and equipment setups (e.g., table positioning, whiteboards, computers, etc.). In addition, a remote microphone was placed in the middle of student groups to record their conversations with auditory clarity. Overall, this research report examines 6 different computational activities from 2 different classrooms (3 activities from each classroom), but the project as a whole gathered data from 15 teachers' classrooms totaling about 170 hours of in-class data. We focused on the activities from Michael's and Liam's (pseudonyms) classrooms and chose specific activities from their classrooms for further analysis because they featured a large amount of interaction between group members.

After developing the initial filtered framework, we conducted video analysis to study students engaging with CT practices. We used the video analysis techniques of Scherr as a model for our video analysis~\cite{Scherr2008}. For all video data presented, teachers selected their own pseudonyms. In order to iterate on the initial framework, we gathered evidence that outlined the existence proofs of many of the initial practices within our context while also allowing for the emergence of new CT practices or the iteration of our initial framework of practices. A mixed \emph{a priori} (i.e., pre-determined) and \emph{a posteri} (i.e., generative) coding scheme was employed to analyze video data \cite{Otero2009}. The initial filtered framework acted as a guide but the emphasis was on discovering how the CT practices manifested in the classroom. Due to differences in context, we did not want to discard practices that did not fit previous descriptions. However, previous descriptions of CT practices vary significantly and make other frameworks difficult to apply in practice. Accordingly, previous CT frameworks helped provide an initial idea of what CT looks like.

Keeping this in mind, we looked at the video data and conducted a thematic analysis \cite{Boyatzis1998, Braun2006}. The initial phase of our in-class analysis involved examining the videos and describing student behavior as they worked on the computational activities. Behaviors that seemed similar in nature where grouped together and given an initial description of the commonalities in the instances and the differences. An example of this process comes from one of the first practices that we introduce in the subsequent sections which is ``decomposing.'' Initially, a selection of behaviors that frequently occurred at the beginning of the activity were based around students deciding to focus on one piece of code as the center of their attention. Students would interpret and play with the code in order to figure out what needed to be edited. Initially, behaviors like breaking the code into segments and deciding which segment to focus on were grouped together has they fit a theme of trying to understand or gain insight about the code and task at hand. But further examination resulted in our coding for two different themes with this group of behaviors as there seemed to be an important distinction between these behaviors. The breaking down of the code into sections was determined to be distinct from deciding on which of those sections was the most important to focus on. Once these two themes had been identified, we returned to the filtered framework and examined it for alignment. If alignment was found, then the emergent theme would be paired with the CT practice from the filtered framework. If no alignment was found, then the theme identified was written up as an emergent CT practice. After this coding process, most of the CT practices were correlated with previous frameworks, but there were also some emergent practices that we thought were necessary to include in our framework.

All video data were coded using MAXQDA mixed methods analysis software. In regards to the coding of individual CT practices in the video data, we used certain markers to indicate the beginning and end of practices that allowed them to be identifiable within the classroom context. Markers that often indicated the start or end of a CT practice were when students moved on to the next portion of their assignment (some of the assignments would have parts that are explicitly aligned to a CT practice such as creating a graph), when students had a long period of silence in their discussion (i.e., 10-15 seconds or more of not talking), or when students shifted their attention to a teacher or different group member. A complication with the individual identification of practices was that it was possible for some CT practices to co-occur with each other, especially in the case of debugging. For example, students would commonly use other practices like utilizing generalization (e.g., if it worked previously, the same resolution should work here) or decomposing (e.g., breaking down different parts of code to more precisely understand why an error is occurring) while they were also debugging. We decided that co-occurrences would be coded as such rather than trying to discern out distinguishable time differences in the separate practices. Frequent co-occurrences could be an indicator of connections between CT practices in our learning environments, as discussed in more detail in Section \ref{Discussion}. For the research presented here, we report on 12 hours of in-class data collected from two different classrooms.

The coding process was conducted by two researchers and themes went through a peer review and negotiation process before being finalized. Once the pair of researchers had constructed a framework developed out of themes that emerged from the data, it was tested for validity by giving it to a third researcher. The third researcher coded the same data set with the constructed framework, and the framework underwent a refinement process based on this review. This refinement included making definitions more concise, providing examples in the codebook for each different practice, and adding common markers for the beginnings/ends of a practice. In the end, separate raters came to agreement on more than 85\% of the video segments provided (approximately 1 hour of of the total 12 hours of data analyzed). In the following sections we present the CT practice framework with existence proof of these practices that emerged from our analysis process.

\section{Computational Thinking Framework for Introductory Physics}\label{Framework}

\begin{figure*}
    \includegraphics[width=\linewidth]{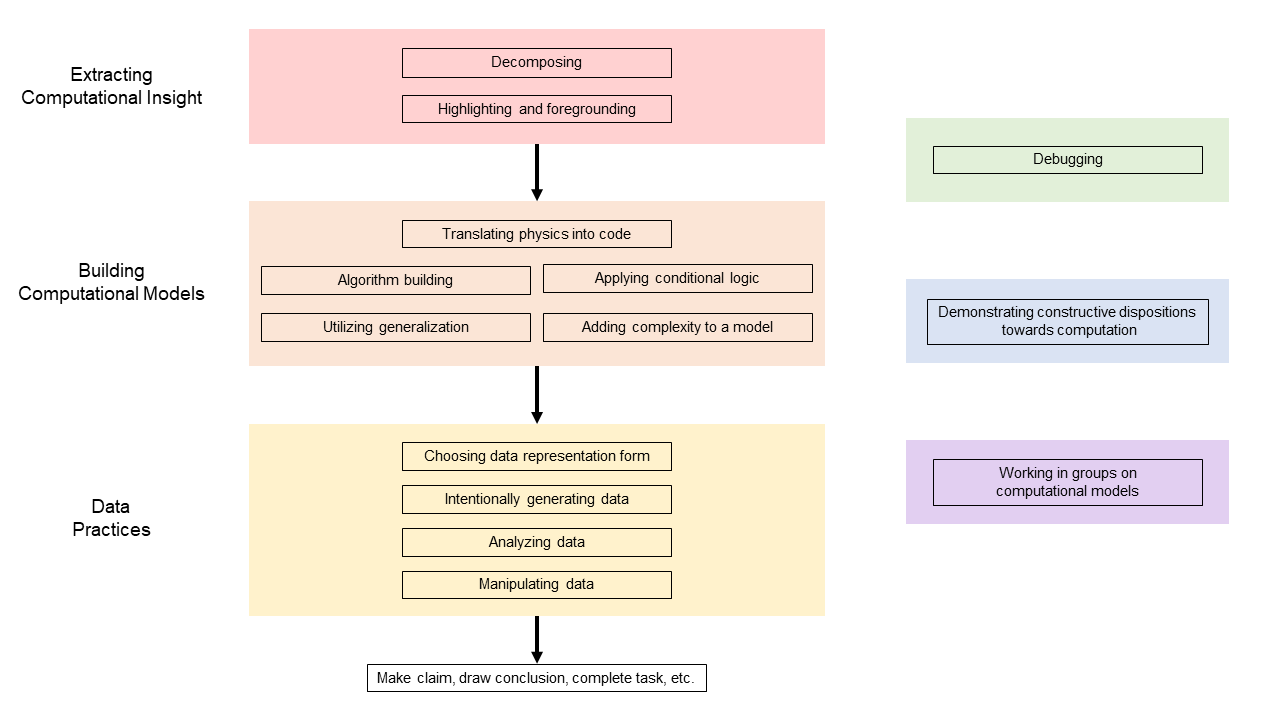}
    \caption{Outline of the practices in our computational thinking framework. Our framework has 14 practices comprising 6 distinct categories. The first three categories of practices exist as a parts of a sequence of computational thinking practices. The practices of debugging, demonstrating constructive dispositions towards computation, and working in groups on computational models may emerge throughout this sequence.}
    \label{Framework_Fig}
\end{figure*}

Our computational thinking framework contains 14 practices contained within 6 different categories. We propose that when working with computational models, students engage in the first three categories of CT practices (i.e., extracting computational insight, building computational models, and data practices) in a sequential order, as shown in Fig.\ \ref{Framework_Fig}. For the sake of clarity, we have organized and depicted the practices in this sequence, but further investigation is needed to determine if there could be other ways that students work with CT practices. First, students extract computational insight to interpret and plan their plan for the model. Then, students build and modify the computational model. Lastly, students use the model to obtain meaningful data so that they can accomplish the task at hand. Within these first three categories, there are several practices that student could engage in as they work through a computational learning activity. The other three practices (i.e., debugging, demonstrating constructive dispositions with computation, and working in groups on computational models) exist as their own distinct categories of independent CT practices. Students engage in these three CT practices at multiple stages throughout one coding activity, rather than in a sequence. As student work through this cycle of CT practices, they may or may not engage with every practice in the sequence (i.e., it is not required for students to engage in every practice) when working with computation. Our initial analysis from the classrooms indicates a sequential nature to the ordering of CT practices. More research is needed to see how these relate to each other. There are potentially cycles between practices but also between categories. 

The following hypothetical example is provided to demonstrate how students could potentially work through a computational assignment about a satellite orbiting Earth in geostationary orbit. Students often start computational assignments by interpreting the code (i.e., decomposing) and choosing one aspect to focus on (i.e., highlighting and foregrounding). In this case, they would open the minimally working program and identify the lines of code that handle the force of gravity, and they can then choose to focus on those lines as essential pieces of the code. After that, students may translate physics into code by inputting Newton's Law of Gravitation in code form. Once they have the correct code, they run the program to observe the path of the satellite's orbit (i.e., intentionally generating data). Maybe at this point, the students have completed their initial task of modeling geostationary orbit. After that, they may choose to restart the sequence to model an elliptical orbit. Throughout the process of engaging in these different practices, the students may have to debug or develop constructive dispositions as they work through the activity. In general, this is a proposal for how students could work through computational assignments, but it is not meant to be the only way they could complete an assignment. Therefore, this framework may be used to investigate the sequence of how teachers would want their students to work through computational assignments.

\subsection{Extracting Computational Insight}

\noindent\textit{Perceiving and identifying the essential components of a computational model.} \\

Extracting Computational Insight is associated with the beginning stages of working with a computational model: perceiving the model as an abstraction of a real-world physical phenomena. Abstracting is included as top-tier CT practice in a number of frameworks, including those from Shute et al.~\cite{Shute2017}, Barr \& Stephenson~\cite{Barr2011}, Rich et al.~\cite{Rich2019}, and Brennan \& Resnick~\cite{Brennan2012}. Jeanette Wing emphasizes thinking in ``multiple layers of abstraction''~\cite{Wing2006}. Grover \& Pea describe it best: ``The value of abstraction as CT's keystone (distinguishing it from other types of thinking) is undisputed.'' While there are a number of frameworks that emphasize abstraction, many of them do not specify the scale or grain-size involved. There is a larger grain size of abstraction (where students recognize that an entire model is itself an abstraction of reality), and a smaller grain size of abstraction (where students identify simplifications within a model). This is something we hope to adequately address in our framework.

As mentioned in the previous section, our context is heavily focused on modeling. Models are representations or simulations of real-world phenomena, made imperfect by assumptions and approximations of the real world (for instance, students often assume there's ``no friction'' or ``no air-resistance'' in motion problems). These assumptions and approximations establish the backbone of the model, which ultimately makes it easier to construct and improve. An example of this is a problem in which students are meant to illustrate a satellite's orbit around the Earth (a common method to learn about curved motion in opening mechanics courses is to use gravitational orbits). 
In reality, any Earth-orbiting satellite feels a somewhat sizeable gravitational force not just from the Earth, but also the moon and the sun as well. However, these elements of the model are often ignored, whether they are explicitly told to ignore them (typically at a more introductory level) or they choose to ignore them (typically at higher levels, when students have more freedom in their model construction and solution development). In any case, students are meant to recognize that ignoring the gravitational effects from any other bodies in space is an assumption that simplifies the construction of the model, while sacrificing some of its accuracy. Understanding the importance of these approximations and assumptions (and identifying where they are necessary versus where they compromise the model) is critical to both computational thinking and modeling, and is thus incredibly relevant for our context. 

Extracting computational insight refers to the first of the four cyclic CT practices from our framework. Before students go about building the model (adding new lines of code or changing a minimally-working program), they first need to think about how it will be developed. They might break down the model into manageable pieces, plan the order and construction of those pieces in relation to each other, and consider small-scale abstractions for separate stages of the model. There are three practices within extracting computational insight: decomposing, modular thinking, and highlighting \& foregrounding.

\subsubsection{\normalsize\normalfont{Decomposing}}

\noindent\textit{Separating a computational model into a series of manageable tasks.} \\

\begin{table*}[t]
  \caption{CT frameworks that mention ideas related to decomposing.
  \label{Decomposing_Table}}

  \begin{ruledtabular}
    \begin{tabular}{p{1.5in} p{1.5in} p{3.9in}}
      \textbf{Framework} & \textbf{CT Element} & \textbf{Definition} \\
      \hline
      Barr \& Stephenson, 2011 & Decomposing problems & Breaking problems down into smaller parts that may be more easily solved. \\

      AAPT, 2016 & Subdividing a model & Logically subdivide a computational model into a set of manageable computational tasks, and organize their code accordingly. \\

      Shute et al., 2017 & Decomposition & Dissect a complex problem/system into manageable parts. The divided parts are not random pieces, but functional elements that collectively comprise the whole system/problem. \\

      Rich et al., 2020 & Decomposition & Breaking apart a complex problem or situation to make it more manageable. \\
    \end{tabular}
  \end{ruledtabular}
\end{table*}

Many frameworks highlight the practice of breaking down large problems, complex systems, or multifaceted computational models into smaller pieces. The development of a computational model can be a multi-step process and might incorporate many different concepts at one time. As a result, decomposition is one of the most universal practices included in most CT frameworks. The definitions of decomposing from various sources can be seen in the Table \ref{Decomposing_Table}. Grover \& Pea include structured problem decomposition in their list of practices that are ``widely accepted as comprising CT,'' but they do not ever explicitly provide a definition for this practice~\cite{Grover2013}. Barr \& Stephenson reference problem decomposition not just in computer science, but in mathematics (e.g., following the order of operations), science (e.g., carrying out a species classification), and language arts (e.g., producing an outline)~\cite{Barr2011}. Needless to say, the general notion of reducing complex systems or problems to their simpler components is well-received as a core element of computational thinking. 

 All of the definitions from the sources include similar wording of breaking down, subdividing, dissecting, and breaking apart. In turn, the practice of decomposing can be thought of as making multiple smaller pieces from a larger entity. In addition, all of the frameworks in Table \ref{Decomposing_Table} agree that a primary purpose for decomposing is to improve the manageability of a task or to make it more easily solvable. As a result, it is important to include the idea of manageability in our definition. The AAPT framework~\cite{AAPT2016} uses the language of ``subdividing a model into a set of manageable computational tasks'' rather than ``decomposing problems'' or ``problem decomposition.'' This definition from AAPT is more relevant to our context, which puts a heavy emphasis on working with computational models. Lastly, Shute et al.~\cite{Shute2017} emphasize the connectedness of the decomposed pieces that make up an entire model or solution (e.g., ``the divided parts are...functional elements that collectively comprise the whole system''). This is the only definition that underscores the inter-related nature of decomposed pieces. Aside from minor language distinctions, each framework appears to generally agree on the meaning of decomposing as a CT practice.

Based on the frameworks reviewed, we wanted to be sure that the improved manageability of the model was contained within our definition. Furthermore, because our context heavily emphasizes computational modeling, we felt it was important to view decomposing through the lens of model construction, rather than problem solving. Accordingly, our definition uses the language pertaining to a computational model rather than a complex system or problem. Examples of decomposing are evident whenever there exists a computational model with multiple individual steps or tasks required to establish a complete, working version. For instance, to model the motion of a ball rolling on a table and then falling off the edge, students must consider the change in net force when the ball and the table are no longer in contact. To model the motion in a 3D visual environment (such as VPython), students often use loops to iteratively update the position of an object. In turn, students who are decomposing this model would think about the entirety of the motion as multiple stages, and there might be a single loop for each stage of the motion. One loop would update the ball's position while it is on the table, and another loop would update the ball's position after it leaves the table. In this hypothetical example, the students are decomposing their computational model into two separate while loops to model two different stages of a ball's motion on and off a table. Therefore, decomposing the model allows students to take complicated motion with many stages and separate it into multiple, more manageable pieces. 

\begin{figure}
    \includegraphics[width=\linewidth]{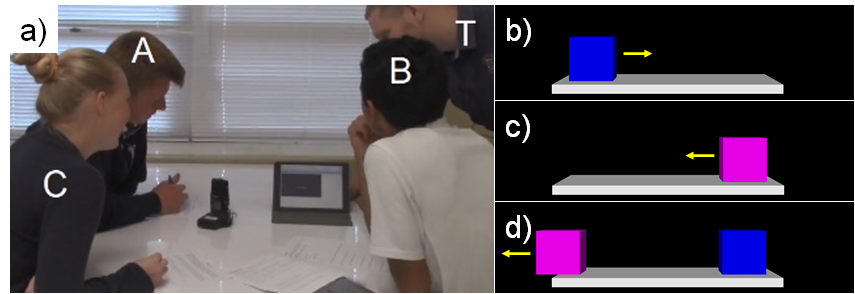}
    \caption{(a) A group of three students (pseudonyms A, B, and C) discuss their computational model with the teacher (T). The students' model displays (b) first a blue box moving to the right, (c) then the blue box stopping on the right side of the platform and a pink box appearing after the blue box completes its motion, and (d) lastly the pink box continuing its motion infinitely to the left of the platform. Yellow arrows in (b), (c), and (d) have been added to illustrate the direction of motion for each crate at different instances.}
    \label{Decomposing_Fig}
\end{figure}

In the following 23 second episode, the students demonstrate decomposition while troubleshooting some issues in their program that models two colliding crates. The group is trying to make the two crates move toward each other simultaneously such that the crates stop moving when they reach each other in the center of a platform (see Fig.\ \ref{Decomposing_Fig}). Ultimately, this program will be used to model an elastic collision between the two objects, but first, the students are simply trying to get both cubes to move toward each other at the same time and stop when they meet. When the students run their program, it initially displays a blue crate moving to the right (Fig.\ \ref{Decomposing_Fig}b). Then, when the blue crate reaches the right side of the platform, it stops its motion, and a pink crate appears at the position where the blue crate stopped (Fig.\ \ref{Decomposing_Fig}c). Subsequently, the pink crate moves to the left and continues its motion infinitely, even after it extends beyond the platform object (Fig.\ \ref{Decomposing_Fig}d). The students call their teacher over to gain some guidance on how to proceed. \\

\begin{hangparas}{1em}{1}
	B: So here’s where we are. [Student B runs the program to show their teacher the visual output of their program.]
\end{hangparas}
\begin{hangparas}{1em}{1}
	T: You have box 1 that flies across, and then box 2 comes into play and flies across.
\end{hangparas}
\begin{hangparas}{1em}{1}
	A: And, off... [The animation shows the second (pink) box moving infinitely to the left beyond the platform.]
\end{hangparas}
\begin{hangparas}{1em}{1}
	B: And then it doesn’t stop.
\end{hangparas}
\begin{hangparas}{1em}{1}
	T: No... Oh, it doesn’t stop!
\end{hangparas}
\begin{hangparas}{1em}{1}
	B: Nope, we don’t really know why.
\end{hangparas}
\begin{hangparas}{1em}{1}
	T: Man, that’s...
\end{hangparas}
\begin{hangparas}{1em}{1}
	B: There’s some problems. So, there’s a delay, that’s problem number 1. That is another problem. [Student B points at the second (pink) box still moving farther off-screen.]
\end{hangparas}
\begin{hangparas}{1em}{1}
	T: Yeah, yeah, so which problem do you want to fix first?
\end{hangparas}
\begin{hangparas}{1em}{1}
	B: The time delay.
\end{hangparas}
\begin{hangparas}{1em}{1}
	T: The time delay.
\end{hangparas}
\begin{hangparas}{1em}{1}
	A: Yeah. \\
\end{hangparas}

This is an example of decomposing problems because the group is dividing one overall task (i.e., making both crates appear at the same time and simultaneously move toward each other) into two smaller tasks (i.e., fixing the time delay before the pink crate appears and stopping the motion of the pink crate as it moves off the platform). In this case, the students identify multiple issues with their simulation that should be dealt with. They choose to address the time delay first, and this serves as a chance for the group to re-focus their efforts toward a more manageable task. It should be noted that decomposing does not always occur in the context of troubleshooting or debugging. The interaction and co-occurrence of coded video examples will be discussed more in Section \ref{Discussion}.

The transcript provided above acts as an existence proof for the presence of decomposition occurring when students work through computational modeling activities in this context. Another easily conceivable example of decomposition could emerge when students are first trying to interpret and break down the different lines of code in a minimally working program. Decomposing may occur at several different grain sizes such as decomposition at a large overall model level or at a smaller individual line of code level. The differing grain sizes for CT practices will be addressed in Section \ref{Discussion} to follow. While the example provided here occurs with the teacher being present, student B actually demonstrates decomposition without any specific prompting from his instructor. Accordingly, this example provides some evidence that the teacher's presence could lead to students more easily engaging in the practice of decomposing. Overall, this is a case of students separating their computational model into a series of manageable tasks.

\subsubsection{\normalsize\normalfont{Highlighting and Foregrounding}}

\noindent\textit{Perceiving the most important features of a model to enhance understanding, focus on essential aspects of code, and recognize unexpected behaviors.} \\


\begin{table*}[t]
  \caption{CT frameworks that mention ideas related to highlighting and foregrounding.
  \label{HighlightingForegrounding_Table}}

  \begin{ruledtabular}
    \begin{tabular}{p{1.5in} p{1.5in} p{3.9in}}
      \textbf{Framework} & \textbf{CT Element} & \textbf{Definition} \\
      \hline
      Weintrop et al., 2016 & Creating computational\phantom{xx} abstractions & The ability to conceptualize and then represent an idea or a process in more general terms by foregrounding the important aspects of the idea while backgrounding less important features. \\

      Shute et al., 2017 & Pattern recognition & Identify patterns/rules underlying the data/information structure. \\
    \end{tabular}
  \end{ruledtabular}
\end{table*}

Shute et al.'s~\cite{Shute2017} ``abstraction'' closely resembles our ``Extracting Computational Insight.'' Because of this similarity, we feel it is worth looking into the sub-practices within. In Shute et al.'s framework~\cite{Shute2017}, the practice of ``abstraction'' contains a sub-practice entitled ``pattern recognition,'' which is defined as, ``Identify(ing) patterns/rules underlying the data/information structure.'' However, there is little detail beyond this definition. Weintrop et al.'s framework ~\cite{Weintrop2016} mentions the practice of ``creating computational abstractions,'' which is described as ``The ability to conceptualize and then represent an idea or a process in more general terms by foregrounding the important aspects of the idea while backgrounding less important features'' \cite{Weintrop2016}. There is typically little or no specificity as to the grain-size of this practice, which makes it slightly problematic. In other words, we have no way of knowing whether these definitions (or those from other frameworks) are meant to apply on large scale (computational models/problems being abstractions themselves) or on a small-scale (abstracting while considering individual steps of a model's construction), and this is the main motivation for the inclusion of this practice. Highlighting and foregrounding refers explicitly to small-scale abstraction. While abstraction is thoroughly discussed in the literature, small-scale abstraction is a gray area. Many frameworks that discuss abstraction do so broadly, or without specifying any grain size. Therefore, much of our incentive for placing this practice in our context is instructor-motivated. We know that teachers affiliated with our ICSAM workshop care about how their students interpret models and problems seen in their lectures, and therefore we have included it in our framework.

Students first break a model down into its individual components (decomposing). They are encouraged to organize these individual components/tasks into connected but independently testable modules. Students then need to reflect on what assumptions/approximations go into the specific task of the module at hand. An example of this is a projectile motion problem, in which students are asked to model the motion of some object, possibly dropped from the top of a high structure or thrown in the air from the ground. Students in our context will typically be given an incomplete model – a minimally-working program – that partially models the motion. Perhaps the MWP begins with the object sitting still, and an initial velocity of \texttt{object.v} {$\mathtt{= (0,0,0)}$} is present. In this case, students might highlight that they need to change the object's velocity in order to make the ball move. They might background ideas such as air resistance's effects on the motion of the object; instead, they'll focus on simply getting the ball from point A to point B. In the big picture, the entire model is an abstraction of a real-life physical phenomenon. But, on the small-scale, there are abstractions within the model itself. The model could be more realistic if there were fewer assumptions and approximations (i.e, if air resistance was included in the model). \\

In the following instance, we present a case where, after decomposing the code, the students direct their efforts to a central goal through the practice of highlighting and foregrounding. This example takes place near the beginning of a computational activity where students are modeling the motion of a block hanging from a vertical spring. After the students dissect and interpret the different parts of the code, they decide to focus on correctly modeling the graphs of different forms of energy, instead of correctly modeling the different forces in their model. The conversation in the following transcript takes place over about 40 seconds. \\

\begin{hangparas}{1em}{1}
	C: Alright, so we need to add the forces of gravity and the spring, and we need to graph the energies. \\
\end{hangparas}
\begin{hangparas}{1em}{1}
	A: So which one should we do first? \\
\end{hangparas}
\begin{hangparas}{1em}{1}
	B: He already gave us the code for the kinetic energy graph, so maybe we should start with the graphs. \\
\end{hangparas}
\begin{hangparas}{1em}{1}
	C: Okay, so just copy this line? [Student C points to the line of code for the kinetic energy graph.] \\
\end{hangparas}

This example is a demonstration of highlighting and foregrounding because the students have perceived the different features of the model (i.e., decomposing), and then they decided to focus on coding the different graphs into their model, rather than programming in the different forces (i.e., highlighting and foregrounding). Ultimately, this leads to the students experiencing difficulty because the graphs will not appear correctly unless the forces have been correctly coded. Additionally, the students could be highlighting and foregrounding this aspect of the assignment simply because their teacher had already done part of the work for them. Ideally, students would be able to engage in a practice like this on their own. However, this example still clearly shows the students highlighting and foregrounding one specific feature of the model over another. This practice is essential to working through coding exercises because when working with computation, typically focusing on one aspect at a time is an effective approach. This practice is subtle and difficult to notice, but it is important for computational thinking. As a result, we hope to investigate this practice more in the future. In many ways, highlighting and foregrounding is a form of planning that the group negotiates among all of its members.

Students might engage in highlighting and foregrounding at multiple points throughout a computational modeling activity. Often, this practice tends to follow decomposition in the early stages of an activity because after students have broken down and interpreted the code, it naturally follows that they will choose a particular piece of the model to focus on after that. However, highlighting and foregrounding also occurs after students complete various portions of the assignment that their teacher provides. For instance, after successfully adding code to add a spring force to their model, the students might go on to focus on graphing all of the energies in their model. They could also choose to focus on different aspects of the code, such as adding friction or changing values in their model to observe the relationships between variables in the model. Most often, highlighting and foregrounding exists as students focus on one part of the code or choose where to go next with their model. Highlighting and foregrounding typically occurs before students go on to build or modify their computational models. Hence, highlighting and foregrounding is a natural practice that occurs commonly in computational modeling activities.


\subsection{Building Computational Models}

\noindent\textit{Using a computer to create an abstract representation of a physical system or phenomenon.} \\


Computational modeling is ubiquitous. Mentioned first in Wing's 2006 article ``Computational Thinking,''~\cite{Wing2006} almost every framework highlights the use of models or simulations in early education, with varying levels of emphasis. In Weintrop et al.'s~\cite{Weintrop2016} taxonomy, ``modeling and simulation practices'' is a top-level category that includes five  distinct practices. Shute et al.'s~\cite{Shute2017} framework incorporates modeling under the larger umbrella-idea of abstraction, and Grover \& Pea~\cite{Grover2013} include models and simulations as forms of ``abstractions and pattern generalizations.'' Additionally, both Berland \& Lee's~\cite{Berland2011} and Barr \& Stephenson's~\cite{Barr2011} studies refer to simulations, which they seem to associate with modeling. Needless to say, computational modeling is a widely accepted CT practice that is discussed in multiple ways by previous literature sources.

Although modeling is generally accepted to be an element of CT, different frameworks prioritize modeling quite differently. For example, Weintrop et al.~\cite{Weintrop2016} call models ``simplifications of reality that foreground certain features of a phenomenon while approximating or ignoring other features.'' Interestingly, Weintrop and colleagues highlight modeling as a large and encompassing umbrella term. On the other hand, Shute et al.~\cite{Shute2017} and Barr \& Stephenson~\cite{Barr2011} organize abstraction as the top-level term that encompasses modeling. Shute et al.~\cite{Shute2017} define modeling as ``representing how a system operates, and/or how a system will function in the future.'' Additionally, Berland \& Lee ~\cite{Berland2011} and Barr \& Stephenson ~\cite{Barr2011} reference ``simulation'' instead of ``modeling.'' Berland \& Lee~\cite{Berland2011} define simulation as ``the enactment of algorithms or plans in order to test the likely outcome.'' Barr \& Stephenson~\cite{Barr2011} provide the example of ``algorithm animation and parameter sweeping'' for simulation in a computer science course. These two definitions appear to focus on modeling specifically from the standpoint of running an algorithm, while the other definitions are more broadly associated with modeling in general. Another noteworthy idea is the difference between the NGSS idea of ``developing and using models''~\cite{K12Framework2012} and computational modeling specifically. Of course, modeling can be done without a computer. In fact, the NGSS directly includes computer simulations as one of many forms of modeling: ``A practice of both science and engineering is to use and construct models as helpful tools for representing ideas and explanations. These tools include diagrams, drawings, physical replicas, mathematical representations, analogies, and computer simulations''~\cite{K12Framework2012}. We use modeling to reference the last idea specifically. For teachers administering computationally integrated physics problems, computer simulations will be the main modeling tool for students, and so it is necessary to specify this in our framework.

As stated previously in the section ``Extracting Computational Insight,'' we argue that models themselves are abstractions. Weintrop et al.'s language of ``foregrounding certain features'' and ``approximating or ignoring other features'' is comprehensive, as these phrases are closely aligned with both models and abstractions. Therefore, our definition of ``Building Computational Models'' includes the idea that these models are abstract representations of physical phenomena that students experience inside and outside of the classroom. The recognition of models as abstract simplifications of real-world phenomena is essential, especially for a context that emphasizes the use of modeling techniques so heavily. When working through a computational physics problem, students represent many different phenomena with computational models, including free-falling objects, collisions, and rotational motion. Models have limitations (students typically ignore air resistance, friction, etc.), and when engaging in this practice, students should assess the validity of their model when they are testing outcomes or making predictions. Unlike other forms of modeling (such as drawings or physical replicas) students can use computational modeling to test multiple scenarios and evaluate several outcomes quickly. An important note is that this practice is associated with \textit{building} computational models, and as such, all of the practices contained in this category pertain to students designing, enacting, and modifying their models. Each sub-practice within is associated with some sort of tool or technique that students have at their disposal when actually interacting with a computational model. There are five sub-practices contained within building computational models: translating physics into code, algorithm building, applying conditional logic, utilizing generalization, and adding complexity to a model.

\subsubsection{\normalsize\normalfont{Translating Physics Into Code}}

\noindent\textit{Adapting the features of an analytical model to a computational environment.} \\

\begin{table*}[t]
  \caption{CT frameworks that mention ideas related to translating physics into code.
  \label{TranslatingPhysicsIntoCode_Table}}

  \begin{ruledtabular}
    \begin{tabular}{p{1.5in} p{1.5in} p{3.9in}}
      \textbf{Framework} & \textbf{CT Element} & \textbf{Definition} \\
      \hline
      Weintrop et al., 2016 & Preparing problems for\phantom{xxx} computational solutions & Reframing problems so that existing computational tools -- be they physical devices or software packages -- can be utilized. \\

      AAPT, 2016 & Translating a model into\phantom{xx} code & Translate a theoretical or algorithmic model into code that enables computation. \\
    \end{tabular}
  \end{ruledtabular}
\end{table*}

Literary support for including this CT practice came predominantly from two sources: Weintrop et al.\ and AAPT. Weintrop et al.'s framework contains the practice ``Preparing Problems for Computational Solutions'' within the ``Computational Problem Solving Practices'' category. This practice describes reframing problems ``so that existing computational tools -- be they physical devices or software packages -- can be utilized''~\cite{Weintrop2016}. AAPT's recommendations include a computational physics skill called ``Translate a Model Into Code''~\cite{AAPT2016}. This multifaceted action describes how students should ``translate a theoretical or algorithmic model into code that enables computation,'' which includes constructing readable code, using language documentation, and applying physics knowledge to make decisions~\cite{AAPT2016}. The two ideas are not a perfect match. For instance, Weintrop et al.\ focus on problems being translated into code, while AAPT refers to models and algorithms. Regardless, we believe that the notion of translating analytical (e.g., written) elements of a problem into code is something to value, especially in our context where students are typically expected to read and learn with written prompts before modeling physics with the computer.

Translating physics into code can manifest itself in a huge variety of different ways in our context. When students program physics equations in code form, they are directly translating physics concepts into something that can be understood by a computer. An example of this is the commonly-used Euler update equation, which states that a new value is equal to the old value plus a change. This is typically seen when students need to update the position of a moving object. This example is one of the more literal interpretations of this practice, but more subtle examples exist as well. When students are looking for a model that adequately portrays some physical phenomena or are using a simulation to analyze some unknown relationship, they must consider specifically how the computer is going to output the desired results. For example, if students are looking to model perpendicular and parallel forces on a satellite in a curved orbit, then they might apply arrow objects to the orbiting satellite that represent the various forces. Because  the computer does not understand commands that point the arrows or scale the arrows based on the physics, students need to translate those elements for the computer. They can determine the direction of the arrows by using the position vectors of the satellite and the mass it is orbiting, and they can adjust the size of the arrows based on the size of the force. In a MWP-based context, much of this practice stems from students consciously deciding that physical relationships are not immediately recognizable to the computer. In other words, one cannot simply tell the computer to ``exert a force'' or ``make these objects collide.'' Rather, one must program the actual changes in momentum by redefining the vectors after each iteration through an animation loop. The more students begin to recognize this, the more fluent they will become as translators between physics and computation.

Below, we provide an example of translating a physics into code when students are trying to modify their code to calculate the elastic potential energy of a vertical spring-mass system. Their teacher provided them the correct lines of code to calculate and graph kinetic energy of the system, and next the students are tasked with correctly modeling the graph for the elastic potential energy. The following transcript takes place over about 65 seconds.  \\

\begin{hangparas}{1em}{1}
	A: So I think we need to have, like, the formula for kinetic energy. Oh wait, we already have kinetic energy.
\end{hangparas}
\begin{hangparas}{1em}{1}
	B: Spring energy then?
\end{hangparas}
\begin{hangparas}{1em}{1}
	A: Yeah.
\end{hangparas}
\begin{hangparas}{1em}{1}
	B: Elastic energy equals 0.5 \textit{k}...
\end{hangparas}
\begin{hangparas}{1em}{1}
	C: Times \textit{k}...
\end{hangparas}
\begin{hangparas}{1em}{1}
	A: Oh, I forgot that I had to do that. [Student A adds an asterisk between 0.5 and \textit{k} in their computer code.]
\end{hangparas}
\begin{hangparas}{1em}{1}
	B: Okay now times... um...
\end{hangparas}
\begin{hangparas}{1em}{1}
	A: \textit{X} squared, right?
\end{hangparas}
\begin{hangparas}{1em}{1}
	C: Yeah, but we need to use what the code uses instead of \textit{x}.
\end{hangparas}
\begin{hangparas}{1em}{1}
	B: Yeah, so, times `spring displacement'...
\end{hangparas}
\begin{hangparas}{1em}{1}
	C: Times, again...
\end{hangparas}
\begin{hangparas}{1em}{1}
	A: Oops! Sorry... [Student A adds an asterisk between \textit{k} and `spring displacement' in the computer code.]
\end{hangparas}
\begin{hangparas}{1em}{1}
	B: And then put two asterisks and 2.
\end{hangparas}
\begin{hangparas}{1em}{1}
	A: Because it's \textit{x} squared?
\end{hangparas}
\begin{hangparas}{1em}{1}
	B: Yeah, that's right. \\
\end{hangparas}

The previous example is a straightforward case of students translating the equation for elastic potential energy into code form. First, the students identify the equation that they are trying to translate, and then they slowly talk out the equation in code form ($ElasticEnergy=0.5*k*x**2$). This example is specific to computational modeling for two reasons. First, multiple times, student A forgets that she needs to include an asterisk every time for multiplication between variables. A similar instance occurs when student B instructs his teammate to use two asterisks for carrying out exponential calculations, because Python uses a double asterisk to carry out exponential functions. This idea of requiring a specific symbol for mathematical operations is standard for many calculators, spreadsheet programs (e.g., Microsoft Excel), and programming languages (e.g., MATLAB). Second, student C acknowledges that they cannot just use \textit{x} as the spring's displacement, and instead, they need to use whatever variable is appropriate for this specific code. By some clever activity design, the MWP provided by their teacher explicitly defines a variable named `spring displacement,' and the students use that in their code instead of \textit{x}. Therefore, our context, specifically the computational platform, requires students to think carefully about mathematical symbols and variable names when translating physics into code.

In general, translating physics into code was one of the most commonly occurring practices observed in video data. This is somewhat expected given the nature of integrating computation with physics using a minimally working program approach. When using such programs, teachers are able to eliminate more complicated, CS-focused ideas for students while instead prompting students to engage more physics-related computational tasks. Also, this CT practice tended to occur at a small grain size that did not frequently overlap with other practices. That is to say, when translating physics into code, students were usually only thinking and discussing on the level of an individual line or only a few lines of code. Ultimately, this practice is specific to the introductory physics context because it involves bridging the gap between conceptual physics and computational content.

\subsubsection{\normalsize\normalfont{Algorithm Building}}

\noindent\textit{Planning and constructing a series of ordered steps to model a physical phenomenon or a decomposed piece of the model.} \\

\begin{table*}[t]
  \caption{CT frameworks that mention ideas related to algorithm building.
  \label{AlgorithmBuilding_Table}}

  \begin{ruledtabular}
    \begin{tabular}{p{1.5in} p{1.5in} p{3.9in}}
      \textbf{Framework} & \textbf{CT Element} & \textbf{Definition} \\
      \hline
      Barr \& Stephenson, 2011 & Abstraction & Use procedures to encapsulate a set of often repeated commands that perform a function. \\

      Berland \& Lee, 2011 & Algorithm building & In its simple form, it is planning of  actions for events that are taking place; in its complex form, it is planning for unknown events. \\

      Brennan \& Resnick, 2012 & Sequences & A series of individual steps or instructions that can be executed by the computer. \\

      Shute et al., 2017 & Algorithms & Design logical and ordered instructions for rendering a solution to a problem. The instructions can be carried out by a human or computer. \\
    \end{tabular}
  \end{ruledtabular}
\end{table*}


Algorithms are commonly referred to when discussing computational thinking. Berland \& Lee call algorithms ``data recipes''~\cite{Berland2011}. Barr \& Stephenson highlight that studying classic algorithms is an important focus, in addition to implementing algorithms~\cite{Barr2011}. Shute et al.\ include algorithms as a top-tier practice, with four sub-practices underneath it: algorithm design, parallelism, efficiency, and automation~\cite{Shute2017}. Brennan \& Resnick include the concept of ``sequences,'' which are described as ``a series of individual steps or instructions that can be executed by the computer...like a recipe''~\cite{Brennan2012}. Additionally, outside of the frameworks covered in this literature survey, Grover \& Pea include ``algorithmic notions of flow control'' as one of the most basic elements of CT~\cite{Grover2013}. Accordingly, algorithms were discussed extensively in the majority of papers reviewed in our study.

Overall, we feel that the Shute et al.\ framework gives the most comprehensive coverage with respect to what this practice looks like in the classroom~\cite{Shute2017}. The four sub-elements included in their overall practice of algorithms emphasize key components of building an algorithm. Students must go through some sort of design phase, where they might plan and construct a series of ordered steps for the model. From there, parallelism, efficiency, and automation describe how algorithms are effectively understood and optimized. In an effort to simplify the algorithm, students should try to carry out many steps in parallel if possible. An example of this might be defining two similar variables in one line of code, rather than writing a new line with similar code for each variable. Additionally, to make the algorithm more efficient, students should practice constructing algorithms that have as little redundancy as possible. Lastly, automation deals with the applicability of the model to a range of scenarios, and how easily it can work without the need for user input or influence as it executes~\cite{Shute2017}. Thus, our definition of algorithm building is inspired by the work of Shute and colleagues.

This practice is unique because it occurs at multiple grain sizes. At the largest grain size, the entire computer program can be considered a type of algorithm because it is a series of sequential steps that models a physics scenario. Therefore, one could argue that for this context, an entire MWP is an algorithm. It might consist of ordered steps such as (1) import packages, (2) define objects, (3) define initial conditions and constants, (4) construct a loop to model motion. In this way, students see the whole script as one large algorithm, and each decomposed stage or piece as one step of the algorithm. However, there is also a smaller grain size, such as the algorithm of the animation loop itself. Loops in this context often have multiple steps and calculations which serve the purpose of updating position, velocity, acceleration, force, momentum, or any other dynamic physical quantity. The order and efficiency of the loop is algorithmic in nature. When students consider control flow (i.e., order of individual lines of code), they are working to establish a correctly ordered algorithm, whether that algorithm is a model itself or one piece of a larger model. We most commonly focus on the example of an algorithm within a position update loop because that is where this practice is most easily observed in our context. As a result, the only algorithms that students work with in our context is an Euler-Cromer iteration loop to predict the position of an object after some time has passed. This limited view of algorithms might be different from what a computational physicist would commonly think of as an algorithm because it only covers a small range of possible situations when an algorithm would be useful. Owing to our approach and context, we exclude potential ways that students could engage in the practice of algorithm building.

The following transcript is an in-class example of students discussing the algorithmic nature of computer code with their teacher. This students are experiencing the same problem demonstrated in Fig.\ \ref{Decomposing_Fig}(a), (b), and (c). At this moment, the students are trying to get two box objects (i.e., ``crates'') to move toward each other simultaneously in the computational model. Currently, their simulation displays one box moving to the right, and after the first box completes its motion, the second box begins moving to the left (see Fig.\ \ref{Decomposing_Fig}). The students call their teacher over for help, and the subsequent conversation ensues over about 120 seconds. \\

\begin{hangparas}{1em}{1}
	T: So you have crate 1 and crate 2...
\end{hangparas}
\begin{hangparas}{1em}{1}
	C: We got them both to move, but it does this...
\end{hangparas}
\begin{hangparas}{1em}{1}
	B: The second one moves after the first one.
\end{hangparas}
\begin{hangparas}{1em}{1}
	T: Oh yeah, so this is a little different. You're familiar with the while loop, right? So think about that. That tells you that while this is true, it's going to do this. So if we think about coding as, like, step-by-step instructions for the computer. You told it to create a box. You told it to create another box. You gave box 1 an initial velocity. Then, you said while this is true, start moving the first box. So when is box 2 going to start moving? Where in your instruction list do you have box 2 moving?
\end{hangparas}
\begin{hangparas}{1em}{1}
	A: Oh, because it's after all this? [Student A points to the while loop, which only updates the position of the first box.] So does it have to be line in the same line?
\end{hangparas}
\begin{hangparas}{1em}{1}
	B: Or do we just put it next to it?
\end{hangparas}
\begin{hangparas}{1em}{1}
	T: Okay, so right now crate 2 doesn't get a velocity until after crate 1 moves. [The teacher points to the computer to guide his students' eyes.]
\end{hangparas}
\begin{hangparas}{1em}{1}
	B: So we should put the velocity earlier in the code, like up here, for crate 2?
\end{hangparas}
\begin{hangparas}{1em}{1}
	T: Well, do you want crate 2 to start moving at the same time as crate 1? [The students all respond with affirmation.] Yeah, so when you give crate 1 a velocity, give crate 2 a velocity at the same time. [The students pause and look slightly confused.] So what I'm saying is take line 19 and put it earlier in your instructions.
\end{hangparas}
\begin{hangparas}{1em}{1}
	A: So put it first, like before this? [Student A points to above the while loop in their code.]
\end{hangparas}
\begin{hangparas}{1em}{1}
	T: Like at line 14 or 15.
\end{hangparas}
\begin{hangparas}{1em}{1}
	A: Yeah.
\end{hangparas}
\begin{hangparas}{1em}{1}
	C: So then put everything else the same?
\end{hangparas}
\begin{hangparas}{1em}{1}
	T: Sure, we can try that! That will at least get your second box moving at the same time. Basically, it will say step 1, move this box, step 2, move this other box. Instead of move this box, and then once it's done moving, start moving the other box. Think order of operations, like a step-by-step procedure. [The teacher walks away and lets the students work amongst themselves.] \\
\end{hangparas}

After this discussion, the students try to rearrange and modify their code so that both of the crates move in the desired manner. This is a demonstration of algorithm building because the students are interacting with the step-wise nature of computer programming and they are thinking about the order of commands that need to be executed to correctly model the physics of two colliding crates. The teacher sends explicit messages about the sequential nature of programming by dissecting their computer and telling them how the computer runs each command. He even describes it as a ``step-by-step procedure,'' and he relates the idea to the mathematical order of operations that students have encountered in previous classes. After troubleshooting their code for about 5 minutes, the students cannot figure out how to fix the same problem, so they call the teacher over for help again. \\

\begin{hangparas}{1em}{1}
	B: It's still moving after the first one.
\end{hangparas}
\begin{hangparas}{1em}{1}
	T: Okay, I'm going to give you a hint on something. Notice how some things are indented? What do you think the indentation means?
\end{hangparas}
\begin{hangparas}{1em}{1}
	C: Oh, yeah, huh... So, indentation means it falls under the while loop.
\end{hangparas}
\begin{hangparas}{1em}{1}
	A: Yeah, it means it does like all those things under the while loop. [Student B deletes some lines of code to combine their two separate while loops into one while loop that will update the positions of both crates at the same time.]
\end{hangparas}
\begin{hangparas}{1em}{1}
	C: Okay, try running it now. [Student B runs the program, and it displays both crates moving toward each other at the same time.]
\end{hangparas}
\begin{hangparas}{1em}{1}
	B: Ah, I see! [The three students smile and are excited about the result.] \\
\end{hangparas}

The example provided above illustrates many of the different facets of algorithm building. The idea of parallelism (i.e., carrying out steps simultaneously in a computational model) is demonstrated when the students realize that indented lines of code after the while statement will be executed together, effectively simultaneously, throughout every iteration of the animation loop. Once the students combined their two separate while loops into a single loop that updated the position of both crates, the students realized how to arrange their code to model the physics of two colliding crates correctly.

Algorithm building did not emerge in video data frequently. This result could either be because teachers did not focus on algorithm building as an expected practice or because algorithm building is a more complex idea embedded within computational models. In the former case, teachers might have low-level goals for students with regards to algorithms. In turn, if teachers are more interested in exposing students to physics ideas rather than programming concepts, one might expect for most of the algorithm building practice to be taken care of when an instructor is creating a MWP for their class. The most straightforward form of this practice occurs when students are considering the control flow of a computer program. Other variants of algorithm building related to efficiency, redundancy, generalizability, and accuracy are expected to occur, and further investigation into this area of research could be fruitful. In the end, algorithm building is an essential CT practice when working with computational models.

\subsubsection{\normalsize\normalfont{Applying Conditional Logic}}

\noindent\textit{Discussing and planning the logical sequence of events due to conditional statements in a computational model, and editing conditional statements within the model.} \\

\begin{table*}[t]
  \caption{CT frameworks that mention ideas related to applying conditional logic.
  \label{ApplyingConditionalLogic_Table}}

  \begin{ruledtabular}
    \begin{tabular}{p{1.5in} p{1.5in} p{3.9in}}
      \textbf{Framework} & \textbf{CT Element} & \textbf{Definition} \\
      \hline

      Berland \& Lee, 2011 & Conditional logic & Conditional logic is the use of an “if-then-else” construct. \\

      Brennan \& Resnick, 2012 & Conditionals & The ability to make decisions based on certain conditions, which supports the expression of multiple outcomes. \\
    \end{tabular}
  \end{ruledtabular}
\end{table*}

Conditional statements in code are some of the most fundamental building blocks of a computational model. Berland \& Lee emphasize conditional logic, specifically if/then/else logical statements, as an element of CT that was evident when students worked together while playing board games~\cite{Berland2011}. Barr \& Stephenson include ``logic'' as a key element of computer science~\cite{Barr2011}, and Grover \& Pea highlight conditional logic as a base element of CT~\cite{Grover2013}. Brennan \& Resnick also include both ``loops'' and ``conditionals'' in their list of computational thinking concepts~\cite{Brennan2012}.

Most commonly, this practice exists when students are working with iterative while/for logic and if/then/else logic. Owing to the commonplace nature of looping and iteration in the context of introductory physics, we highly value the ability to construct a loop with logical statements. Another argument for its inclusion is that, like if/then/else logic, while/for logic is usually tied to the physics of the model (e.g., position updates within a while loop over time). Loops are often responsible for motion in computationally-integrated mechanics courses. This is because computers cannot recognize physical quantities such as ``velocity,'' and it is up to the coder to redefine the position over and over again using the velocity and a small time-step, and to repeat this calculation consecutively through the use of a loop (see the translating a model into code subsection). When students work with changing forces, they must also update the velocity value based on acceleration because whatever object of interest in the code is now feeling a net acceleration. When the forces on a moving object changes, this should signal that some new logic is required because the velocity needs to update differently as a new force is felt. In this way, the stopping condition in the while/for statement explicitly depends on where the physics changes in a scenario, particularly where the net force on an object is changed.

Consider the hypothetical example of a skateboarder jumping off the end of a ramp. Before launch, the skateboarder feels a normal force, whose vertical component opposes gravity, and if there is no friction or air resistance, then students should say that there is no net force, implying constant velocity motion. However, when the skateboarder leaves the ramp, suddenly there is no normal force, and the skateboarder will begin to accelerate towards the Earth because of the gravitational force. Because the net force on the skateboarder changed, the velocity update equation in the code must change as well, and this could be handled in a few ways. One solution would be to have two loops; one loop exists for motion while on the ground, and one loop models the motion after leaving the ramp. In both loops, the position is updated according to the velocity, but these velocities are different from one loop to the other (in the first loop, the velocity would be constant, and in the second loop, the velocity would be changing due to the gravitational force). Another solution method might involve only one loop, but instead includes an if/then/else statement within, which says something like ``\textit{If} the skateboarder is still on the ground, \textit{then} move with a constant velocity, \textit{else}, move according to a net gravitational force.'' In any case, the change in physics is represented by logical conditions in the computational model. Therefore, we claim that conditional logic in our context is a vital element of CT because it serves as the primary method by which students connect physics to computation.

\begin{figure}
    \includegraphics[width=2.5in]{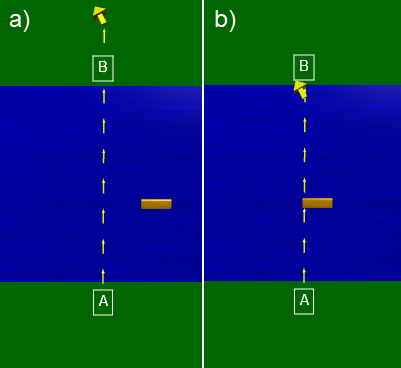}
    \caption{Computational model of a boat crossing a river. The large yellow arrow represents the boat, the smaller yellow arrows represent the boat's velocity at different time instances, and the dark yellow rectangle represents a log floating in the river (i.e., only affected by the current). Panel (a) displays the boat incorrectly continuing after it reaches the opposite shoreline, and panel (b) displays the boat correctly stopping when it reaches the opposite shoreline.}
    \label{ApplyingConditionalLogic_Fig}
\end{figure}

In the 46 second episode presented below, a group of students apply conditional logic when modeling the motion of a boat crossing a river with a nonzero current. At this point, the students have successfully added together the vectors for the river's current and the boat's motion such that the boat is travelling in a straight line across the river (i.e., the boat's horizontal velocity component is equal and opposite to the velocity component of the current vector). However, in its current state, their model shows the boat moving directly across the river, and then the boat continues its motion even after it reaches the shore (see Fig.\ \ref{ApplyingConditionalLogic_Fig}(a)). The students know that their boat should stop (i.e., the program should stop running, or the position of the boat object should stop updating) once it reaches the opposite shore labeled `B.' The students focus on the conditional statement of their while loop in an attempt to achieve this goal. \\

\begin{hangparas}{1em}{1}
	D: We should make it not go onto the grass now. [The rest of the group agrees with student D.]
\end{hangparas}
\begin{hangparas}{1em}{1}
	B: Yeah so, uh, if boat position...
\end{hangparas}
\begin{hangparas}{1em}{1}
	D: No, just while `boat.pos.y' is less than something.
\end{hangparas}
\begin{hangparas}{1em}{1}
	B: While... `boat.pos.y' is less than... [Student B is making changes to the conditional statement of the while loop within the computer code.]
\end{hangparas}
\begin{hangparas}{1em}{1}
	A: Do 120. That's where the `B' label is at.
\end{hangparas}
\begin{hangparas}{1em}{1}
	B: Oh, really? Okay.
\end{hangparas}
\begin{hangparas}{1em}{1}
	A: Or maybe, like, just before that. We'll see. [Student B runs the program, and it displays the boat moving across the river and still moving beyond the destination shore, as seen in Fig.\ \ref{ApplyingConditionalLogic_Fig}(a). Their teacher walks up as they see the outcome of their computational model.]
\end{hangparas}
\begin{hangparas}{1em}{1}
	T: How we doing? What's up?
\end{hangparas}
\begin{hangparas}{1em}{1}
	B: We just basically finished.
\end{hangparas}
\begin{hangparas}{1em}{1}
	C: We're just trying to figure out how to not go in the grass. [Student B changes the value of their while loop to a slightly different value and runs the model again.]
\end{hangparas}
\begin{hangparas}{1em}{1}
	D: I'd say we did it... Well actually, it's still barely going.
\end{hangparas}
\begin{hangparas}{1em}{1}
	B: Okay, how about 85? Let's do some 85 up in here. Run this program. [Student B changes the value in the conditional statement to 85 and runs the program again. The model shows the boat stopping as it reaches the shore on the opposite side of the river, as shown in Fig.\ \ref{ApplyingConditionalLogic_Fig}(b).] Nice! That's good. It stops like right when it gets there. \\
\end{hangparas}

Here, we observe the students applying conditional logic to get their model to stop running at the right time. They accomplish this goal by changing the conditional statement of the while loop, which animates the boat's motion depending on the its $y$-position, such that the program stops at a different time. At first, student B thinks about the dilemma with an `if' statement approach, but shortly after this, student D overrules him and decides to change something in the `while' loop. Both of these pathways are valid forms of applying conditional logic that could help the students achieve their desired result. Subsequently, student A proposes using the value of 120 in their conditional statement because that is where the `B' label is positioned (see Fig.\ \ref{ApplyingConditionalLogic_Fig}). This idea of using the relative value of one object's position as the animation requirement of another object's motion is a higher-level form of applying conditional logic than random guessing and checking. Nevertheless, the students still make slight adjustments to the conditional statement in their while loop, over multiple iterations, to get the boat to stop in the right location. Hence, the previous example demonstrates applying conditional logic because the students are editing the value in the conditional statement of the while loop of their model.

Applying conditional logic is a practice that appeared frequently in our context because every computational modeling activity that we examined contained a while loop. This practice tends to overlap with algorithm building because teachers seemed to value the enactment and consequences of while loops over simply the sequential nature of coding. Whether the students were explicitly modifying the while loop or just running the program and watching an animation, the students were exposed to the idea of conditional logic in a computer program. We observed the practice of applying conditional logic emerging in either a passive (i.e., guided, or low-level) or active (i.e., spontaneous, or high-level) form. When students were not explicitly prompted to engage with the conditions of a loop (and the lines of code contained within the loop), they did not exhibit robust, easily noticeable forms of applying conditional logic. Further investigation is required to discern the variation in this practice for introductory physics courses with computation integrated. Accordingly, when working with computational models in introductory physics, students experience a unique opportunity to practice applying conditional logic.

\subsubsection{\normalsize\normalfont{Utilizing Generalization}}

\noindent\textit{Importing previous approaches, algorithms, or specific code into a model.} \\

\begin{table*}[t]
  \caption{CT frameworks that mention ideas related to generalization.
  \label{Generlization_Table}}

  \begin{ruledtabular}
    \begin{tabular}{p{1.5in} p{1.5in} p{3.9in}}
      \textbf{Framework} & \textbf{CT Element} & \textbf{Definition} \\
      \hline
      Brennan \& Resnick, 2012 & Reusing and remixing & Building on other people’s work. \\

      Shute et al., 2017 & Generalization & Transfer CT skills to a wide range of situations/domains to solve problems effectively and efficiently. \\

      Rich et al., 2020 & Patterns & Looking for similarities between new problems and problems that have already been solved. \\
    \end{tabular}
  \end{ruledtabular}
\end{table*}

Generalization is a widely accepted idea in the realm of CT. In the context of discussing their CT practice of ``reusing \& remixing,'' Brennan \& Resnick declare, ``Building on other people’s work has been a longstanding practice in programming, and has only been amplified by network technologies that provide access to a wide range of other people’s work to reuse and remix''~\cite{Brennan2012}. Shute et al.'s framework describes the practice of generalization as ``[transferring] CT skills to a wide range of situations/domains to solve problems effectively and efficiently''~\cite{Shute2017}. Rich et al. include ``patterns'' as a practice in their study, which teachers describe as ``looking for similarities between new problems and problems that have already been solved''~\cite{Rich2019}. Although these frameworks name the practice of generalization something different, they all talk about similar ideas, so it is important to look at descriptions and definitions rather than terminology.

We chose to name this practice ``utilizing generalization,'' as opposed to just generalization, because we felt that the latter was ambiguous as to whether the practice meant utilizing generalized code or writing code in a general way. While both of these practices are important, our MWP-based context focuses more on the utilization of existing code that has been previously written. By contrast, one could envision a distinct, high-level form of building computational models in a generalized way, such that the model accommodates a wide range of circumstances or variables. Most MWP's have already made decisions on how the code is structured, and so it is difficult to emphasize construction of new code such that it may be used in future problems.  Nonetheless, for our context, the learning objective of getting students to reuse, remix, and utilize general codes/structures was highly valued. Thus, the focus is on using general code as a resource for constructing a computational model.

There is a grain size discussion with this practice as well, which refers to utilizing general code from an external model versus utilizing pre-existing code from within the current model. For example, say students have constructed a projectile motion model for a ball, given an initial velocity. Their model displays the trajectory of the ball with a Motion Map, which plants arrows that point in the direction of the velocity every few time-steps -- like a stroboscopic image with arrows to indicate velocity. The code for something like a Motion Map is typically provided for students because it is syntax-heavy. Now, say the students wanted to model a second ball with the same initial velocity, this time with air resistance taken into account. To add the second Motion Map, most students will copy the previous Motion Map line, and tweak its name and values. If students are able to recognize it, this process is much easier and less time-consuming than writing the new line from scratch. This is an example of small grain-size generalization because it references code from within the MWP itself. If instead students had copied Motion Map code from a problem they worked on a few weeks ago, that would be a larger grain-size, which deals with recognizing similarities between entirely distinct models, as opposed to pieces of the same model. The grain size variation of utilizing generalization is worthy of exploring in the introductory physics context.

In the next example, the students utilize generalization to easily and quickly create a second box object by remixing the working code of the first box object within the same program. This occurs through the simple act of copying and pasting one line of code, and then making modifications to the replicated code. The following 80 second conversation takes place after the students are tasked with creating a second crate object with some different attributes than the first one. \\

\begin{hangparas}{1em}{1}
	A: Create a second crate with a different size and color, and then place it on the far right side of the floor. [Student A reads the prompt from their activity worksheet.] Okay, crate equals box... Copy all of that. [Student A highlights the entire line of code that defines the attributes of the first box.] Copy. I'm just going to put it right here. Paste. Okay. [Student A pastes the line of code directly below the original copied line.] So that'll put it...
\end{hangparas}
\begin{hangparas}{1em}{1}
	B: It's going to put a cube in the same place as the last one.
\end{hangparas}
\begin{hangparas}{1em}{1}
	A: Same exact spot, same exact color, same everything. So um...
\end{hangparas}
\begin{hangparas}{1em}{1}
	C: Well, we need to change the size and color.
\end{hangparas}
\begin{hangparas}{1em}{1}
	A: So let's put the color back to red. Nah, let's do magenta. Magenta actually works, I tried it last time. [Student A changes the code to make the second crate magenta instead of blue.]
\end{hangparas}
\begin{hangparas}{1em}{1}
	B: So magenta and then change the size. Let's just make it 30. [Student A changes the code to make the second crate larger.]
\end{hangparas}
\begin{hangparas}{1em}{1}
	A: Crate equals... Okay, so let's change the position vector to... So, negative 20 puts it on the same axis as the first crate, so let's try positive 50. [Student A changes the code to make the second crate's \textit{x}-position start at positive 50 instead of negative 20.]
\end{hangparas}
\begin{hangparas}{1em}{1}
	B: Let's give this a shot. Boom! [Student A runs the program, and it displays two crates of different size, color, and \textit{x}-position.]
\end{hangparas}
\begin{hangparas}{1em}{1}
	A: Wow, that was super easy. Different size, different color, and it's on the far right side. We did it! \\
\end{hangparas}

This example candidly demonstrates utilizing generalization because the students are copying (i.e., importing) a line of code from elsewhere in their program, and then they are modifying that line to easily create another box object in their model. After reading the prompt from the worksheet, student A begins by simply copying the code for the first crate and pasting it directly below the original line; this is a form of reusing code. Next, the group acknowledges that the cube will be located at the same position with the same characteristics as the original cube. Consequently, they start the process of remixing the code by changing its color, size, and initial position. The students are able to rapidly create another crate because they are skilled at using this code in a generalized way. By the end of the interaction, student A admits with glee that the entire process was ``super easy'' because they had experience with this CT practice. Thus, the previous example is one representative instance where students utilize generalization to quickly build or modify parts of a computational model.

The way a teacher designs their computational modeling activities may be a significant factor that leads to students engaging with utilizing generalization. The importance of this practice was especially messaged by one instructor. He believed this practice was one of the major benefits of exposure to CT, and in turn, he designed his activities to engage students with the idea of generalization. In the end, utilizing generalization is an important practice for students to learn when exposed to CT.

\subsubsection{\normalsize\normalfont{Adding Complexity to a Model}}

\noindent\textit{Iteratively adding complexity to a model of a physical phenomenon based on the output of the current model.} \\

\begin{table*}[t]
  \caption{CT frameworks that mention ideas related to adding complexity to a model.
  \label{IterativeModelDevelopment_Table}}

  \begin{ruledtabular}
    \begin{tabular}{p{1.5in} p{1.5in} p{3.9in}}
      \textbf{Framework} & \textbf{CT Element} & \textbf{Definition} \\
      \hline
      Brennan \& Resnick, 2012 & Being incremental and\phantom{xxx} iterative & Iterative cycles of imagining and building -- developing a little bit, then trying it out, and then developing further, based on [students'] experiences and new ideas. \\

      Shute et al., 2017 & Iteration & Repeat design processes to refine solutions, until the ideal result is achieved. \\
    \end{tabular}
  \end{ruledtabular}
\end{table*}

Another CT idea that is particularly relevant in our context is iterating on a model or solution. Much of the motivation for including this practice stems from Weintrop et al.'s practices of ``assessing computational models'' and ``constructing computational models'' practices, located within the ``Modeling and Simulation Practices'' category~\cite{Weintrop2016}. The first practice emphasizes assessing a model's validity based on assumptions and approximations, and the second practice encourages students to ``implement new model behaviors,'' which can be done by ``extending an existing model''~\cite{Weintrop2016}. This language resonates with the idea on iterating on the design of MWPs, where the models are typically constructed for students, and they only have to add simple features with significant scaffolding. As another example, Shute et al.\ include the practice of ``iteration,'' which is to ``repeat design processes to refine solutions, until the ideal result is achieved''~\cite{Shute2017}. Combining the ideas of repetition and improving/assessing a model leads to our definition for adding complexity to a model.

Our context is heavily influenced by MWPs, and we believe that this has a strong influence on the relevance of adding complexity to a model. This practice is linked to MWPs because students are often encouraged to run their code a number of times throughout a class session. Students will run the code first to see what the MWP has provided them already, and then when changes are made, they might run the code again and again, to see if the changes were implemented correctly. By the time the model is near completion, students should ask themselves, ``How can this model be improved?'' Accordingly, students make adjustments that add even more complexity to the model. Examples of these additions include color-coding arrows or objects, adding titles, x and y labels, and legends to a graph, and incorporating air resistance in a projectile motion model. Sometimes, these improvements are merely cosmetic, and they focus on making the model easier to comprehend. Other changes may be extensions of the physics, such as incorporating air-resistance or friction when it was not originally present. The major takeaway is that this process is iterative. Students can run the code, decide if an extension is called for, and then make the necessary edits. After a number of iterations, the model is hopefully in a state in which the relevant physics is easy to perceive, and the results are presented in an efficient, interpretable way.


The following example of adding complexity to a model emerges when students are working through an activity that tasks them with modeling the motion and energy of a vertical spring-mass system. Eventually, once the students model the correct graphs for all of energies successfully, the students continue to read the next task on their worksheet. \\



\begin{hangparas}{1em}{1}
	D: In a real situation, the spring would run out of energy. As the spring bounces up and down some of the energy will be lost to the surroundings as thermal energy. Add to your forces a new force called `F damp.'  [Student D reads the final prompt on their worksheet aloud to the rest of his group.] I don't know why it's called \mbox{`F damp.'}
\end{hangparas}
\begin{hangparas}{1em}{1}
	C: Because it's damping, like dampening vibrations.
\end{hangparas}
\begin{hangparas}{1em}{1}
	D: Okay, sure. This force will need to slow the spring on each successive bounce so that it eventually comes to a stop. The `F damp' should be proportional to and opposite the velocity. [Student D continues to read the prompt from the worksheet.] Okay, so, thermal energy is some force times distance, right?
\end{hangparas}
\begin{hangparas}{1em}{1}
	C: Yes, because it's work. [The students go on to discuss their ideas about how to proceed.] \\
\end{hangparas}

The above example demonstrates the students being tasked with adding a damping force to their model. Even though this case was specifically prompted by the activity's design, the students are still gaining experience with adding new physics to their model. This new feature will lead to a much different outcome than was shown by their previous model, and by building this prompt into the assignment, this instructor is asking his students to comprehend that a computational model can always be further developed to more closely match reality. After the discussion above, the students continue to engage in other CT practices while trying to achieve their end goal of adding a damping force. After about 11 minutes and 30 seconds, the students get a hint from their teacher on how to proceed. The next transcript shows the students coming to a resolution and completing this saga of adding complexity to a model. \\

\begin{hangparas}{1em}{1}
	D: So for `F damp,' put in negative `cube velocity,' and then we also have to add in the damping fraction. [Student A changes the line of code corresponding to `F damp.']
\end{hangparas}
\begin{hangparas}{1em}{1}
	A: This is some very thick air. [The other students in the group laugh at student A's joke.]
\end{hangparas}
\begin{hangparas}{1em}{1}
	D: Yeah, now in `F net,' do all of that minus `F damp.' [Student A makes the change that student D is describing.] Wouldn't it be minus `F damp?'
\end{hangparas}
\begin{hangparas}{1em}{1}
	A: It's already negative in the line above.
\end{hangparas}
\begin{hangparas}{1em}{1}
	D: Oh, okay. [Student A runs the program, and it displays the mass oscillating up and down on the spring with the damping force correctly added.] Did we just get it? [The students continue to watch the simulation.] \\
\end{hangparas}

At this point, the students have successfully added a damping phenomenon to their computational model. They do this by translating the equation for the damping force into the code, and then adding the damping force to the net force equation. In this case, we see that adding complexity to a model emerges at a very large grain size over a gradual series of interactions between computational thinkers. Over the entire class session, the students successfully added new calculations and graphs of the energies, as well as adding complexity to the physics underlying the model (i.e., adding damping). The idea of making assumptions as a way of simplifying models is paramount to engaging with CT in a physics context.

Video data around adding complexity to a model was scarce because this practice tends to occur near the end of modeling activities. Adding complexity to a model occurs at a larger grain size than most other practices in our framework, and we leave our description of the practice open-ended to include other possible forms of adding complexity to a model. For example, at an advanced level, students should be focusing on improving the modeling of physics in their simulation, rather than simply adding new visual elements to their code. In addition, expanding on a computational model is a complicated process that requires experience and confidence with the computational medium. Because computationally integrated physics courses focus less on providing students experience with programming, students might not have the self-efficacy that allows for effortless communication and advanced ideas around model development. On the other hand, this lessened focus on computer science ideas would be expected to enable students to think more readily about the physics modeled by a simulation. Therefore, adding complexity to a model is a high-level CT practice that can be demonstrated by adding new physical features to a working computational model. \\

\subsection{Debugging}

\noindent\textit{Remedying unexpected behaviors or error messages encountered when working with a computational model.} \\

\begin{table*}[t]
  \caption{CT frameworks that mention ideas related to debugging.
  \label{Debugging_Table}}

  \begin{ruledtabular}
    \begin{tabular}{p{1.5in} p{1.5in} p{3.9in}}
      \textbf{Framework} & \textbf{CT Element} & \textbf{Definition} \\
      \hline

      Berland \& Lee, 2011 & Debugging & Determining problems in order to fix rules that are malfunctioning. \\

      Brennan \& Resnick, 2012 & Testing and debugging & Developing strategies for dealing with – and anticipating – problems. \\

      Weintrop et al., 2016 & Troubleshooting and\phantom{xxxxx} debugging & Figuring out why something is not working or behaving as expected. \\

      AAPT, 2016 &  Debug, test, and validate\phantom{x} code & Resolving error messages and other incorrect behavior. \\

      Shute et al., 2017 & Debugging & Detecting/identifying errors, and then fixing the errors when a solution does not work as it should. \\

      Rich et al., 2019 & Debugging & Systematically finding and correcting problems and errors. \\
    \end{tabular}
  \end{ruledtabular}
\end{table*}

Debugging is a practice with an overwhelming amount of literary support. Table \ref{Debugging_Table} shows all of the definitions for debugging provided in major CT works reviewed. Nearly every framework included in this study makes mention of some sort of troubleshooting phenomenon associated with CT. Brennan \& Resnick's ``Testing and Debugging'' practice highlights that students should be creating strategies for correcting and anticipating errors in their computational endeavors \cite{Brennan2012}. Similarly, Weintrop et al.'s ``troubleshooting and debugging'' practice calls debugging the computer-science form of troubleshooting, and offers possible strategies for debugging, including ``clearly identifying the issue, systematically testing the system to isolate the source of the error, and reproducing the problem so that potential solutions can be tested reliably'' \cite{Weintrop2016}. One of the AAPT's recommended computational physics skills is to ``[d]ebug, test, and validate code,'' which is unique in that it does not imply an error message, and instead can be something as simple as ``asking whether the results are physically plausible'' \cite{AAPT2016}. Shute et al. include a straightforward definition of debugging in their framework: ``Detect and identify errors, and then fix the errors, when a solution does not work as it should'' \cite{Shute2017}. In addition to these four frameworks, a number of others provide support for the idea of debugging as it pertains to CT, including Berland \& Lee \cite{Berland2011} and Rich et al.~\cite{Rich2019}.

Although many frameworks contain a straightforward presentation of debugging, there is still room for interpretation. One example of this is whether or not debugging explicitly includes an error message. For almost any coding software, if there is an error in the code (unrecognized or ill-formatted code, typos, etc.), the code will not be able to fully run. Instead, it will run up to where the error is encountered, and print an error statement. Typically, this error statement includes a location (line number) and a basic error classification (e.g., type error, operational error, syntax error, etc.). Many frameworks include language along the idea of remedying ``errors,'' without specifying if these errors are uniquely code-breaking error messages or, more broadly speaking, any sort of general issue encountered in the process of developing a computational model. We have chosen to take the broader approach, and include any sort of computational hiccup as something to be remedied through a process of debugging. We feel that the very nature of MWP's and the types of computational activities in this context will yield a number of unexpected behaviors that do not give an explicit error message.

Another example is specifically where debugging occurs, and how it relates to other CT practices. Say a student runs a code and finds an unexpected behavior. After careful analysis, the student decides to try swapping the order of two lines of code, and it corrects the unexpected behavior. We call the process by which the student came to that decision debugging. It might have included consulting a fellow student or professor, reviewing a previously completed assignment that contains a similar computational layout, or more carefully reviewing the physics to see if there was an incorrect order of calculations. Based on our definitions of utilizing generalization and algorithm building, one could argue that the student might have engaged in these practices while debugging. The student might have referenced a previously-completed assignment (one that was generally similar to the current assignment) to fix this bug, which is a form of utilizing generalization. Or, perhaps they looked more carefully at the ordered steps of the physical solution, and realized that there was a control flow issue in the code's algorithm, a form of algorithm building. This example highlights that debugging is ubiquitous, and can occur alongside any CT practices. One could argue that CT practices are carried out in one of two ways: (1) to construct something brand new, or (2) to edit/Debug something that already exists. In light of this, we have placed debugging within its own category, and highlighted the fact that it can occur at any point in the CT cycle.

A third example is the differing grain-sizes of debugging, especially as they pertain to our context, which uses MWP's. When looking specifically at unexpected behaviors (``bugs'' that don't have a code-breaking error message), one could argue that the entirety of the process by which an MWP is completed can be thought of as a debugging process, in the sense that students take an investigative approach to finding the source of the model's imperfections, and then remedy said imperfections. However, throughout this approach, students might run into other errors, either created by attempted solutions or discovered at a later stage of the model development. In fact, MWP's are often constructed in certain ways to plant specific bugs that the instructor wants students to practice resolving. These planted bugs can yield error messages that break the code, or simply produce an incorrect model/unexpected behavior. Thus, within the larger debugging process (fixing the MWP to accurately model the desired phenomena), there are smaller-scale bugs, either created by the students' attempted solutions or planted purposely within the MWP. The ways in which debugging as a CT practice relates to grain-size and MWP's need to be further investigated.

When applying debugging to our context specifically, we say that the practice encapsulates everything from the moment an unexpected behavior is identified to when a remedy has been put in place. This not only includes realizing that an error exists, but also identifying its location and its source, and ultimately deciding on the best solution for the issue. When it comes to specific error messages, students need practice reading, interpreting, and understanding the computer's outputted error messages. Over time, students will likely have developed a catalog of the different types of computational errors, and have ideas on how to resolve them. As for the non-error-message unexpected behaviors, the process is a bit more organic. Firstly, students have to recognize that an issue is present, which is not always readily obvious, because there is no error message and the code runs through to the end. Students ultimately have to consider the physics to make a decision on whether or not the model is correct. In addition, once students decide that there is an issue with the model, they have to figure out for themselves where the issue is and why it is happening. Again, this relates specifically to the actual physics. If a certain object is moving incorrectly, it is likely because the physics that the computer is interpreting is incorrect, leading to unexpected results. One could argue that the best approach to diagnosing these types of unspecified issues is to question what physical phenomena \textit{should} be accounting for the unexpected motion, and reviewing how that physics is implemented in the code. This is a large part of why debugging is so crucial for our context: it is often the implementation of the relevant physics that students are evaluating when they encounter unexpected behavior that needs to be remedied, which makes this process particularly noteworthy for our context.

\begin{figure}
    \includegraphics[width=2.5in]{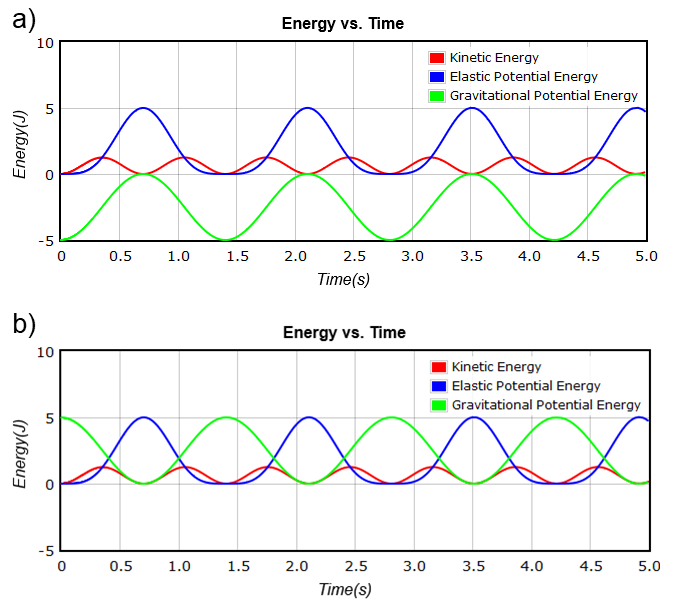}
    \caption{(a) Incorrect graph of gravitational potential energy in a vertical spring-mass system simulation. (b) Correct graph of gravitational potential energy in a vertical spring-mass system simulation.}
    \label{Debugging_Fig}
\end{figure}

The following example of debugging features a computational model of a hanging spring-mass system to explore the concept of energy conservation. After the students successfully get their spring to oscillate, they are tasked with producing a line graph of the kinetic, elastic potential, gravitational potential, and total energies of the system. Their teacher already provided the correct code in the minimally working program to graph the kinetic energy, so the students needed to utilize the same syntax for the other forms of energy. The students (pseudonyms A, B, C) have just added two lines of code to produce a line graph of the system's elastic potential energy and gravitational potential energy versus time. The group runs their program to see if they correctly programmed the physics, and the program displays the graph in Fig.\ \ref{Debugging_Fig}(a). \\

\begin{hangparas}{1em}{1}
	A: Oh no! I'm sorry... I don't know! [Student A is distressed because of the negative gravitational potential energy in their graph.]
\end{hangparas}
\begin{hangparas}{1em}{1}
	C: Oh gosh! Energy can't be negative.
\end{hangparas}
\begin{hangparas}{1em}{1}
	A: And it's still in phase...
\end{hangparas}
\begin{hangparas}{1em}{1}
	C: Wait... It needs to be flipped because actually it's just upside down.
\end{hangparas}
\begin{hangparas}{1em}{1}
	A: Yeah, it's a sign error!
\end{hangparas}
\begin{hangparas}{1em}{1}
	B: No, even if you flipped it, the gravitational potential energy wouldn't start at the right spot. It starts high... It should start at zero, right?
\end{hangparas}
\begin{hangparas}{1em}{1}
	C: No, it shouldn't. Gravitational is the most at the beginning because it's all the way wound up.
\end{hangparas}
\begin{hangparas}{1em}{1}
	A: Okay, let's check this... [Student A changes the code by adding a negative in front of the equation for gravitational potential energy, they run the program, and the graph in Fig.\ \ref{Debugging_Fig}(b) is displayed.]
\end{hangparas}
\begin{hangparas}{1em}{1}
	B: I guess if g was negative, it would have worked. \\
\end{hangparas}

The previous transcript provides a case where the program runs without encountering a fatal error, but the students still identify an unexpected behavior in the gravitational potential energy in Fig.\ \ref{Debugging_Fig}(a). First, the students express frustration through exclamations of distress (``Oh no!'' and ``Oh gosh!") when initially encountering the problem. In this way, the group is analyzing the data produced by their model to inform them that they need to debug the physics behind the gravitational potential energy. The students know from class that energy is a scalar quantity and that it should not be negative in this case. Initially, they are using the equation $U_g=m*g*(spring~displacement)$, where $U_g$ is the gravitational potential energy, $m$ is the mass of the hanging cube, $g$ is the gravitational constant, and $spring~displacement$ is the position of the cube minus its resting position. This equation could either be corrected by multiplying by a negative (either in front of the $U_g$ line or by making the value of $g$ negative) or by changing the way that the vector for $spring~displacement$ is calculated. The students choose to simply add a negative in front of the line of code for $U_g$, which is the simplest way to make their model behave in the way they expect it to, even though the equation is not necessarily correct for calculating gravitational potential energy. While student A is making changes to the code, student B and C discuss what they expect the value of $U_g$ to be at the beginning of their model. Here, we observe that debugging can be a group-oriented process of everyone contributing their ideas about the code and the physics underlying the model. After making the change to the code, the students successfully produce the correct graph for gravitational potential energy.

Overall, debugging was one of the most frequently emerging CT practices in the videos analyzed. Typically, debugging tended to follow after the data analysis practice because the easiest way for students to identify errors or observe unexpected behaviors was by running the program. Moreover, troubleshooting attempts did not always end with students being successful. However, as students became more familiar with the coding environment, they could more readily determine how to resolve common issues. For example, students would regularly encounter an error where the computer program cannot add a vector and scalar quantity. In many cases, this error can easily be remedied by using a magnitude command (e.g., $KE=0.5*mass*mag(velocity)^2$, where $KE$ is kinetic energy) to make vector quantities become scalars. Of course, the solution might depend on the specific problem, but for the observed data in this study, using the magnitude command resolved all scalar and vector issues. After a few experiences with coding in their physics classes, students would readily try using this strategy to resolve the vector/scalar addition error. One might expect for debugging to be one of the most recurrent CT practices because it applies when either the physics (leading to unexpected behaviors in the model) or the computer code (leading to fatal errors in the computer program) are incorrect. We also note that debugging can occur in both a systematic (i.e., planned) and non-systematic (i.e., random or guess-and-check) way. While we value the systematic form of debugging as a higher level variant of this CT practice, debugging that is haphazard can still be an effective approach to troubleshooting computational models~\cite{Obsniuk2015}. Nevertheless, further investigation is needed to explore the variation of frequently occurring CT practices. Therefore, debugging is an important, repeated CT practice in our context. \\

\subsection{Data Practices}

\noindent\textit{Creating or collecting data to be represented in a model. Processing data to prepare it for analysis. Analyzing data to make claims and draw conclusions. Visualizing data by producing tables, plots, or animations.} \\

Data Practices as a whole are discussed extensively in the literature. Brennan \& Resnick state that the computational concept of data involves ``storing, retrieving, and updating values''~\cite{Brennan2012}. Interestingly, Shute et al. place ``Data collection and analysis'' within the larger ``abstraction'' practice, and they define it as ``collecting the most relevant and important information from multiple sources and understand the relationships among multilayered datasets''~\cite{Shute2017}. Barr \& Stephenson's framework includes three data-related concepts/capabilities: collection, analysis, and representation~\cite{Barr2011}. AAPT's recommendations include two data-related ``technical computing skills,'' which are to ``Process Data'' and ``Represent Data Visually''~\cite{AAPT2016}. Weintrop et al.'s framework includes a five-piece category of data practices, which includes practices such as collecting, creating, manipulating, analyzing, and visualizing data.

Data practices serve as the final piece of our computational thinking sequence. By this point, students have worked through the ``Building Computational Models'' CT practices, and have some form of a working finished product. When an MWP is completed, it has the capability of producing meaningful data. It is important to note that the way we have used Glowscript VPython and the way that are teachers use it limits our ability to produce data. This is because we do not ask students or teachers to use arrays to store data, as might be typical in other coding environments. Instead, variables store data at a specific instance of the computational model, and when we use Glowscript VPython, much of the data is in the form of a visual output (rather than numerical values). Nonetheless, data practices are associated with how this data is presented, interpreted, and analyzed. The practices within involve modifying a model's output so that it can be used to make scientific claims backed by evidence in the data.

\subsubsection{\normalsize\normalfont{Intentionally Generating Data}}

\noindent\textit{Producing some form of data through the enactment of a computational model.} \\

\begin{table*}[t]
  \caption{CT frameworks that mention ideas related to intentionally generating data.
  \label{IntentionallyGeneratingData_Table}}

  \begin{ruledtabular}
    \begin{tabular}{p{1.5in} p{1.5in} p{3.9in}}
      \textbf{Framework} & \textbf{CT Element} & \textbf{Definition} \\
      \hline

      Barr \& Stephenson, 2011 & Data collection & Collect data from an experiment. \\

      Weintrop et al., 2016 & Creating data & Define computational procedures and run simulations that create data they can use to advance their understanding of the topic under investigation. \\

      Shute et al., 2017 & Data collection and\phantom{xxxxxx} analysis & Collect the most relevant and important information from multiple sources. \\
    \end{tabular}
  \end{ruledtabular}
\end{table*}

Intentional data generation happens once students have a functional computational model. Generating data is particularly unique in our context. According to Weintrop et al., data can be collected (``through observation or measurement'') or created (when ``investigating phenomena that cannot be easily observed or measured or that are more theoretical in nature'')~\cite{Weintrop2016}. Barr \& Stephenson include data collection their framework~\cite{Barr2011}. Shute et al.\ title their data-related piece ``data collection and analysis,'' and emphasize collecting ``important and relevant information from multiple sources''~\cite{Shute2017}. Of the frameworks reviewed, Weintrop et al.'s section on Data is far more extensive than any other.

In our context of MWP-based computationally integrated physics courses, generating data is hard to do consciously. When the code is ran, data is generated, but the question becomes whether or not the student is consciously considering the run of the code as a data generation technique. For example, say students are given a MWP that is meant to model a child swinging on pendulum-like swing set, and to start, the MWP assigns no initial speed or forces on the child (aside from gravity and the associated normal force). Students may run a code like this first, just to see what it does and gain some insight into what pieces of the model are missing. Technically, this initial run generates data; the computer may be storing the child's position, speed, and other dynamical quantities of interest. However, students are not generating this data thinking that the model is in a complete state, and they can use the generated data to make a scientific claim. The students are well aware that they are not out of the ``building computational models'' phase, and the majority of the work before analysis is not yet complete. Additionally, the ``collecting'' data element is difficult to witness in this context. Students are not typically running real-world experiments, and using the computer as a mechanism to record raw data. Instead, everything is in the ``creating'' data realm; students are given some sort of MWP that creates a unique set of data that pertains to the hypothetical situation being modeled.

The following example from in-class data shows students intentionally generating data in order to think more deeply about what they expect their model to display. This case occurs when students are trying to graph the different forms of energy for a vertical spring-mass system. After the spring force has correctly been implemented, the students move on to the next task: adding graphs of kinetic, elastic, gravitational, and total energies. After choosing to focus on this part of the activity, the students discuss what they expect the energies to look like. The transcript below showcases a discussion that takes place over about 39 seconds. \\

\begin{hangparas}{1em}{1}
	C: Kinetic energy is already done. \\
\end{hangparas}
\begin{hangparas}{1em}{1}
	D: So, when the spring is at its lowest, it will be... \\
\end{hangparas}
\begin{hangparas}{1em}{1}
	C: The energy will be all elastic... \\
\end{hangparas}
\begin{hangparas}{1em}{1}
	D: At the end, it will be... Well actually, go back to the program... \\
\end{hangparas}
\begin{hangparas}{1em}{1}
	B: For what? Just run it again? \\
\end{hangparas}
\begin{hangparas}{1em}{1}
	D: Yeah, run it again. [Student B runs the code and it shows a vertical spring-mass system oscillating up and down. The simulation also displays a graph of the kinetic energy versus time throughout the motion.] So, that's kinetic energy... \\
\end{hangparas}
\begin{hangparas}{1em}{1}
	B: So, it's just saying when it's at the bottom, it has zero kinetic energy, which makes sense. \\
\end{hangparas}
\begin{hangparas}{1em}{1}
	A: Yeah, so what I expect to see for elastic potential is like this... [Student A points her finger in a sinusoidal fashion that is out-of-phase with the graph of the kinetic energy.] You know what I mean? \\
\end{hangparas}
\begin{hangparas}{1em}{1}
	C: Oh yeah. So how do we graph that? \\
\end{hangparas}

This is a case of students intentionally generating data because the students choose to run their program for a specific purpose. They are trying to visualize and to predict what they expect to see for the graphs of different energies in the model. There are many instances in these types of activities where students simply run a computer simulation to gather information and to decide what to do next. Whether students are running code to remind themselves of where they are at, to hash out what they expect to see, or to see if their changes achieved a desired outcome, these are all cases of intentional data generation.

We distinguish the practice of intentionally generating data from other previous definitions for data collection, data creation, and data generation by specifying that there must be a purpose behind why students are generating data. If students specify a reason for running their program, then they are engaging in this practice with purpose, rather than passively generating data by clicking the ``Run'' button on their computer interface. This intentionality is what explicitly indicates this idea to be a \textit{thinking} practice, rather than a straightforward action where students are passively following a set of instructions. When data generation is inherent (i.e., passive), students are still engaging in the action of generating data, but true computational thinking requires an intention when generating data. Although not shown in this example, commonly the visualization of the model can serve as a viable form of data, as opposed to more traditional forms of data like graphs and numerical values. Consequently, data generation might look different in programming contexts that are less visually-dependent. Therefore, intentionally generating data is an important practice for students to gain experience with in computational modeling activities.


\subsubsection{\normalsize\normalfont{Choosing Data Representation Form}}

\noindent\textit{Implementing the best approach, technique, or tool to convey the results of a computational model.} \\

\begin{table*}[t]
  \caption{CT frameworks that mention ideas related to choosing data representation form.
  \label{ChoosingDataRepresentationForm_Table}}

  \begin{ruledtabular}
    \begin{tabular}{p{1.5in} p{1.5in} p{3.9in}}
      \textbf{Framework} & \textbf{CT Element} & \textbf{Definition} \\
      \hline

      Barr \& Stephenson, 2011 & Data representation & Summarize data from an experiment. \\

      Weintrop et al., 2016 & Visualizing data & Use computational tools to produce visualizations that convey information gathered during analysis. \\

      AAPT, 2016 & Represent data visually & Students should be able to produce static visualizations (i.e., plots) of data because plots are fundamental to facilitating analysis and communication of data. \\
    \end{tabular}
  \end{ruledtabular}
\end{table*}

This practice most closely resembles the data-visualization pieces of various CT frameworks. Weintrop et al. emphasizes the importance of communicating results, which is enhanced through the use of computational tools~\cite{Weintrop2016}. Barr \& Stephenson's ``data representation'' in science is put simply as ``summarizing data from an experiment''~\cite{Barr2011}. AAPT's ``Represent Data Visually'' states that ``students should be able to produce static visualizations (i.e., plots) of data because plots are fundamental to facilitating analysis and communication of data''~\cite{AAPT2016}. Since there are so many methods of producing visualizations with different IDEs (Integrated Development Environments), this practice is bound to occur in almost any computationally-integrated science course.

The key discussion point around this practice is whether or not the 3D visualizations produced in Glowscript VPython are considered ``Representation Forms.'' With the practice of Generating Data listed above, we argued that students who exhibited a passive/unconscious action that pertains to the practice does not mean that the student is engaging in computational thinking, and the same argument can be made here. When students run an MWP, it will almost always produce a visualization. These visualizations are an integral part of computationally-integrated physics courses, because they allow the students to visualize motion of objects, usually represented by spheres or boxes in an endless black background. However, this practice places an emphasis on the communicative ability of data visualizations. Students engage in this CT practice when they consider how to make data visualizations more effective at conveying scientific results. If a student simply runs the code and a visualization of the objects is displayed, we do not consider this to be engaging in CT unless the student is assessing the visualization for any need of improvement.

AAPT mentions the explicit use of plots, which we find to be a noteworthy example. Consider a model in which students display a satellite moving in a circular orbit. If the students are asked how they might prove that orbit is perfectly circular, one way they could do this is by plotting the radius of the orbit over time, and seeing that it is a straight horizontal line. Students engage in CT when they ultimately decide that visualizing the data (radius vs. time) via a graph is an effective way of demonstrating evidence for the claim that the satellite is in a circular orbit. This is one technique; another technique could be printing the value of the radius for each iteration, and seeing that it does not change. It is up to the students to decide the most illustrative and effective route for presenting their data.

The next example is an illustration of students choosing a data representation form for the quantities in their computational model. After successfully translating the equation for Hooke's Law into their model, the simulation displays a vertical spring mass oscillating with simple harmonic motion. At this point, the students continue to add other forms of energy to the graph provided in the minimally working program. Their teacher designed the program to model a graph of the kinetic energy versus time correctly, and now the students have to model the elastic, gravitational, and total energies. The following conversation takes place over 25 seconds. \\

\begin{hangparas}{1em}{1}
	C: Alright, so right now we have energy versus time. [The program displays a cube oscillating up and down on a vertical spring and the corresponding kinetic energy versus time graph.]
\end{hangparas}
\begin{hangparas}{1em}{1}
	B: It just keeps doing the same thing. There's no friction or anything. Are we going to have to add that later?
\end{hangparas}
\begin{hangparas}{1em}{1}
	D: Yeah, that's what we do at the bottom. [Student D is refers to the last question on their worksheet, which tasks them with adding a damping force to their computational model.] Right now we have to graph all of the energies. Kinetic energy is already done. We need the gravitational potential, the elastic potential, and the total energy.
\end{hangparas}
\begin{hangparas}{1em}{1}
	B: We need to graph those all on the same graph?
\end{hangparas}
\begin{hangparas}{1em}{1}
	D: Yeah. \\
\end{hangparas}

This example shows the students choosing a data representation form because they realize that the energy versus time data can be effectively visualized with a line graph. Additionally, they ask if the data should all be displayed on the same graph, which is one choice of data representation that will affect the clarity of their results. Although the previous example does exhibit students engaging in this practice, it is a more passive form than it would be if the students had actively made some of these decisions around data representation themselves. Instead, the teacher encourages students to engage with this practice by designing the activity to require students to graph all of the energies. This result might be the case because graphing is a difficult technique to carry out in Glowscript VPython, and this might have been these students' first experience with graphing in a computational environment. Therefore, choosing a data representation form can exist in a simple passive form, but one could also envision a more complicated and purposeful variant of this practice.

\subsubsection{\normalsize\normalfont{Manipulating Data}}

\noindent\textit{Preparing and reshaping data for further investigation by processing, organizing, and cleaning the dataset.} \\

\begin{table*}[t]
  \caption{CT frameworks that mention ideas related to manipulating data.
  \label{ManipulatingData_Table}}

  \begin{ruledtabular}
    \begin{tabular}{p{1.5in} p{1.5in} p{3.9in}}
      \textbf{Framework} & \textbf{CT Element} & \textbf{Definition} \\
      \hline
      Weintrop et al., 2016 & Manipulating data & Manipulate datasets with computational tools, reshaping the dataset to be in a desired or useful configuration so that it can support further investigation. \\

      AAPT, 2016 & Process data & Students should be able to use computers to process data, which includes reducing, fitting, filtering, and/or averaging data, and computing uncertainties from measurements. \\
    \end{tabular}
  \end{ruledtabular}
\end{table*}

Data manipulation is associated with the handling and managing of data in preparation for some sort of analysis stage. A number of studies reference this action, the most notable of which are AAPT and Weintrop et al.. AAPT's technical computing skill ``Process Data'' describes ``reducing, fitting, filtering, and/or averaging data, and computing uncertainties from measurements''~\cite{AAPT2016}. Weintrop et al.'s framework directly describes the practice: ``Data manipulation includes sorting, filtering, cleaning, normalizing, and joining disparate datasets''~\cite{Weintrop2016}. Both sources illustrate some sort of data cleaning/preparation that can be done before the relationships within the data are analyzed.

When looking at this practice in our context, we believe that it closely resembles the agreed upon definitions from Weintrop et al.\ and AAPT. Students have a working model of some kind, which produces data in a meaningful way. This data can be anything from lists of position coordinates for some object to force arrows on an object. A good example of this is the latter. Using our satellite-in-orbit example, let's say students wanted to turn the motion into an elliptical orbit, and are now tasked with placing force arrows on the satellite. They need to model the size and direction of various forces, including the net force, gravitational force, and parallel and perpendicular forces associated with curved motion. By the time they establish these quantities and they update correctly, the arrows still might not be recognizable, because they are not scaled properly. The directions of the arrows may be correct, and they may have the correct sizes relative to each other, but they are difficult to see, so students can simply multiply the size of the arrows by a scale. Increasing the size of these arrows (which represent the force vectors on the satellite) does not alter any physical meaning from the model, unless the arrows are scaled in dramatically different ways; they simply allow someone viewing the orbit to recognize the relationships between the forces more easily. In this case, the data that has been produced (the force arrow sizes and directions) needed to be manipulated (size was scaled to larger values), so that the data could be more effectively analyzed. Data manipulation should not alter the physics or the relationships within the data, but rather make these things more apparent and easier to analyze.

We were not able to find a presentable example of manipulating data within our context. The closest case we have from our research is students changing the value of a damping coefficient variable to make the physics of damping more pronounced, but this manipulation actually changes the physics at play. In turn, changing the value of a parameter that significantly alters the physics is not an ideal instance of manipulating data. Instead, we could envision a situation where the computational model displays an arrow for some vector quantity (e.g., force, velocity, magnetic field, etc.). Take for example the case of modeling the electric field around a uniformly charged wire. In this case, students might use a scaling factor to show the electric field arrows on a similar size scale as the rest of the objects in their program. The arrows could appear to be much larger than the wire cylinder or charged particle objects because of the way the program interprets different quantities in the model. To make the model easier to understand, students might choose to use a scaling factor to shrink the size of the arrow so that they on a similar scale as the rest of the objects. In this hypothetical case, the students are not changing any values of the physical scenario, instead they are only manipulating data to make the model easier to analyze.

\subsubsection{\normalsize\normalfont{Analyzing Data}}

\noindent\textit{Extracting meaning from a dataset.} \\

\begin{table*}[t]
  \caption{CT frameworks that mention ideas related to analyzing data.
  \label{AnalyzingData_Table}}

  \begin{ruledtabular}
    \begin{tabular}{p{1.5in} p{1.5in} p{3.9in}}
      \textbf{Framework} & \textbf{CT Element} & \textbf{Definition} \\
      \hline
      Barr \& Stephenson, 2012 & Data analysis & Analyze data from an experiment. \\

      Weintrop et al., 2016 & Analyzing data & Analyze a given set of data and make claims and draw conclusions based on the finding from their analysis. \\

      Shute, 2017 & Data collection and\phantom{xxxxxx} analysis & Understand the relationships among multilayered datasets. \\
    \end{tabular}
  \end{ruledtabular}
\end{table*}

Data analysis is the climax of the modeling process, because it is the stage in which students will make scientific claims, based on what the data is telling them. Fundamentally, meaning is extracted from a dataset when claims are made based on relationships within the data. Shute et al. focus on the understanding the ``relationships among multilayered datasets''~\cite{Shute2017}. Barr \& Stephenson's framework includes ``Data Analysis,'' and examples of this practice include language like ``identify trends in data'' and ``identify patterns''~\cite{Barr2011}. Weintrop et al. includes its own data analysis practice, strategies for which are given by ``looking for patterns or anomalies, defining rules to categorize data, and identifying trends and correlations''~\cite{Weintrop2016}. Lastly, there is an emphasis on making claims based on the analysis.

In our context, analyzing data describes the analysis of a finished MWP that produces meaningful data. When the MWP is in a completed stage, it is capable of producing data for which analysis could reasonably provide enough information to make a claim based on the data. For example, if students have finished their MWP from the previous example regarding the satellite in circular orbit, then they have a finished model that produces meaningful data in the form of position, velocity, acceleration, momentum, force, etc., all of which can be used in an analysis stage. As stated above, students might need to analyze the data to prove that the orbit is indeed circular. After they have chosen a representation form and displayed the information in the selected way, they must then analyze it. They might simply observe the constant radius over time, or they could go further with their analysis, possibly fitting the curve and printing the slope of the fitted line, which should be zero. In any case, students begin with a claim (in this case, the fact that the orbit is circular), and the proceed to with analysis that ultimately proves that hypothesis. If done thoroughly, the data analysis stage provides evidence of whether or not the scientific hypothesis was true or false.

\begin{figure}[htbp]
    \includegraphics[width=2.5in]{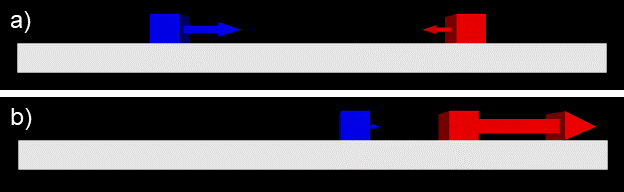}
    \caption{Computational model of two cubes colliding before and after an elastic collision. The blue cube represents a semi truck and the red cube represents a red Ferrari that is speeding toward the truck, with arrows representing each object's velocity. Panels (a) and (b) are what the simulation displays before and after the collision, respectively. }
    \label{AnalyzingData_Fig}
\end{figure}

In the following video example, a group of students is trying model the situation of a head-on collision between a red Ferrari (velocity = 30 m/s) and a blue semi truck (velocity = -50 m/s). The truck has a mass 4 times that of the Ferrari. At this moment, their computational model displays two cube objects undergoing an elastic collision conservation of momentum correctly added in (see Fig.\ \ref{AnalyzingData_Fig}). \\

\begin{hangparas}{1em}{1}
	A: Alright, let's try this again. [Student A runs the code and it displays the situation depicted in Fig.\ \ref{AnalyzingData_Fig}.]
\end{hangparas}
\begin{hangparas}{1em}{1}
	D: Okay, that seems reasonable.
\end{hangparas}
\begin{hangparas}{1em}{1}
	A: So blue barely moved after the collision, and red completely shot off.
\end{hangparas}
\begin{hangparas}{1em}{1}
	D: So the red Ferrari has a higher final velocity because it has less mass?
\end{hangparas}
\begin{hangparas}{1em}{1}
	A: Yeah, I think that's right. \\
\end{hangparas}

In the previous existence proof, we observe the students analyzing the visual data provided by their model in order to draw conclusions about momentum conservation. After entering the parameters for the real-world situation given by their teacher, the students generate data and extract meaning from the output of their simulation. There are many outcomes that they could focus on, but the one they end up realizing is that, after the collision, the lower mass object (i.e., the red Ferrari) will have a greater velocity in the end. While this case does illustrate students analyzing data, they could have gone further by using their theoretical physics knowledge (e.g., equations for momentum conservation) to verify the results of their computational model. Thus, analyzing data can occur at different levels when students are working with computational models.

Analyzing data tended to occur often. Some of the markers for students engaging in this practice include students checking to see if their model is behaving as expected, testing a scenario to understand the outcome of a physical situation, or inputting a physics equation to investigate the relationship between variables in a model. All of these cases could be considered analyzing data.

\subsection{Demonstrating Constructive Dispositions Towards Computation}

\noindent\textit{Recognizing, persevering, and overcoming the trial-and-error nature of computation.} \\

\begin{table*}[t]
  \caption{CT frameworks that mention ideas related to demonstrating constructive dispositions towards computation.
  \label{Dispositions_Table}}

  \begin{ruledtabular}
    \begin{tabular}{p{1.5in} p{1.5in} p{3.9in}}
      \textbf{Framework} & \textbf{CT Element} & \textbf{Definition} \\
      \hline
      Barr \& Stephenson, 2012 & CT dispositions & Persistence in working with difficult problems. \\
    \end{tabular}
  \end{ruledtabular}
\end{table*}

In our previous work on learning goals around computation, ICSAM teachers indicated a main learning goal and motivation for integrating computation into their physics curriculum was to impact students' affect around computation \cite{Weller2019}. Our teachers explicitly state as a goal that they are trying to counteract any fear or intimidation that computation might bring forward when students are faced with it in their future. Within the computational thinking frameworks that we reviewed this goal is discussed as ``dispositions'' in the work of Barr and Stephenson \cite{Barr2011} and by ``persevering'' in Brennan and Resnick \cite{Brennan2012}. We believe that the goal that the teachers are proposing can also be related to the concept of ``grit'', which has been defined by Duckworth \cite{Duckworth2009} in the context of computation as the ``ability to sustain long-term interest in an effort to complete an ongoing task.'' The concept of ``resilience'' also seems applicable as it is described as having optimism to continue in the face of experienced failures \cite{Miller2021}. Grit and resilience are concepts that have been also related to mindset, which is essentially a self perspective on one's ability to change qualities about oneself through effort \cite{Dweck2007}. This connected tree of concepts highlights the complexity associated with understanding how to treat the affective impact teachers want to mediate in their classrooms but also highlights the importance of their instinct that this should be a goal for their attempts to integrate computation in their classrooms. For our framework, we are going to focus on the concept of computational thinking dispositions because there is previous work that has laid the groundwork connecting computational thinking to student affective impacts~\cite{Perez2018}.

Perez builds off previous recommendations by the International Society for Technology in Education (ISTE) and the Computer Science Teachers Association (CSTA) that encourage a focus not just on the practices of computational thinking but also the dispositions students develop towards computation~\cite{Perez2018}. A disposition is described as how a learner's inclination (tendency toward a particular way of thinking or acting), sensitivity (attentiveness to opportunities to engage the particular way of thinking or acting), and ability (ability to produce thought or action after inclination and sensitivity) interact in their thoughts when faced with a context. Demonstrating positive dispositions is highlighted as being a key component of facilitating the transferability of computational thinking practices from one discipline to another~\cite{Perez2018,Brennan2012,K12Framework2012}. ISTE and CSTA initially identify 5 different computational thinking dispositions: (1) confidence in dealing with complexity; (2) persistence in working with difficult problems; (3) tolerance for ambiguity; (4) the ability to deal with open-ended problems; and (5) the ability to communicate and work with others to achieve a common goal or solution~\cite{Barr2011}. Perez refined and detailed these dispositions into 3 dispositions: (1) tolerance for ambiguity; (2) persistence on difficult problems; and (3) collaboration with others to achieve a common goal~\cite{Perez2018}. We will talk in more detail about collaboration as a disposition and computational thinking practice in the next section and instead focus on tolerance for ambiguity and persistence on difficult problems here.

Both tolerance for ambiguity and persistence on difficult problems are typically framed from the perspective of a high tolerance versus developing tolerance~\cite{Perez2018}. A student with a high tolerance for ambiguity is someone who frames an ambiguous situation as an opportunity to discover something new, whereas, a student who is in the process of developing a tolerance for ambiguity is more likely to have negative reactions to situations where they are faced with ambiguous stimuli. Similarly, a student who is disposed to being persistent with difficult problems is likely to continue to engage purposefully in a challenging problem even when experiencing difficulty, obstacles, or failure until they achieve a resolution~\cite{DiCerbo2016}. A student who is demonstrating a disposition to persisting with difficult problems will persist with many attempts at solving the problem but might not achieve a resolution. Low persistence is characterized by students who avoid challenging problem solving situations~\cite{Perez2018}. Demonstrating a disposition towards persisting with problem solving and being unafraid of tasks with levels of ambiguity aligns well with our teachers intentions of students having a positive experience with computation so that they are not afraid of a  future learning situation involving computation. From an observable practice perspective, we have to reframe development that often relates to change over an extended period of time and instead focus on demonstrations of constructive dispositions. Tolerance for ambiguity was difficult to discern with specific examples in the data set. Although students expressed frustration with the computational activities in general, this frustration never manifested as direct comments about the ambiguity of the problems. We would argue that students are demonstrating a tolerance for ambiguity just by engaging with the minimally working program activities, which often have a high level of ambiguity associated with them. We believe that specific markers for tolerance for ambiguity could be found in in-class computational activity data but that this might be something that can be assessed more appropriately with a survey or interviews. Oppositely, we did find markers for persistence with difficult problems in our data set as evidenced by the next episode.

The following video example displays students demonstrating constructive dispositions towards persisting with computational problems when working through the colliding crates activity in Liam's classroom (same group as shown in Fig.\ \ref{Groupwork_Fig}(b)). In this situation, the students simulation runs, but it is not behaving as they expected (i.e., the students are engaging in debugging). We will see student B providing positive reinforcement when his teammates suggest an idea. The following transcript takes place over 76 seconds. \\

\begin{hangparas}{1em}{1}
	A: Hm... That's not what I expected to happen. [The program displays one box moving across a platform, and then another box appearing once the first box stops.]
\end{hangparas}
\begin{hangparas}{1em}{1}
	B: Not at all. Are they moving at the same rate? Rate 50, rate 50... I don't know. I'm stumped.
\end{hangparas}
\begin{hangparas}{1em}{1}
	A: I don't know what to change. Maybe change that to a negative 35? [Student A points to a line of code on their computer.]
\end{hangparas}
\begin{hangparas}{1em}{1}
	B: Oh, this one right here? I wonder what that will do. [Student B runs the program and the model remains unchanged.]
\end{hangparas}
\begin{hangparas}{1em}{1}
	C: Oh, maybe we should subtract `dt' in the `t=t+dt' line.
\end{hangparas}
\begin{hangparas}{1em}{1}
	B: So make that negative, you're saying?
\end{hangparas}
\begin{hangparas}{1em}{1}
	C: Yeah, try that.
\end{hangparas}
\begin{hangparas}{1em}{1}
	B: Oh, okay. Let me just put this back to zero. [Student B undoes the first change suggested by student A.] So `t=t-dt'?
\end{hangparas}
\begin{hangparas}{1em}{1}
	C: Yeah, try that.
\end{hangparas}
\begin{hangparas}{1em}{1}
	B: That's actually a really good idea! We can try it, maybe! \\
\end{hangparas}

In the previous example, we observe the students making multiple attempts to address an unexpected behavior in their model. Throughout the interaction, student B remains positive and encouraging, and this attitude causes his group mates to feel comfortable suggesting their ideas, even if they may not work. Student B makes multiple utterances that indicate he is maintaining a persistence with solving the problem (e.g., ``I wonder what that will do'' or ``That's actually a really good idea''). Even though the students are stuck on this part of the assignment, they keep a positive attitude and continue to try making changes to their model. This could be contrasted with a group that has a negative attitude toward computational material where students are resigned to feeling powerless when working through early computational experiences. 
Typically, debugging is one of the most frustrating but also one of the more prevalent parts of coding. For this reason, evidence of dispositions does emerge when students are debugging, however, in the in-class data this evidence is often difficult to assign to one disposition or attribute of the students' interactions. This indicates that a deeper exploration of the markers of dispositions needs to be completed while also exploring alternative ways of assessing development in this area. As the teachers indicated, demonstrating constructive dispositions towards computation is an important goal because it teaches students how to persevere through adversity when learning through activities that have computation in them. These dispositions need to be explored beyond in-class data.

\subsection{Working in Groups on Computational Models}

\noindent\textit{Engaging as a member of a team to gain understanding, develop creativity, and complete the task at hand.} \\

\begin{table*}[t]
  \caption{CT frameworks that mention ideas related to working in groups on computational models.
  \label{ComputationalGroupwork_Table}}

  \begin{ruledtabular}
    \begin{tabular}{p{1.5in} p{1.5in} p{3.9in}}
      \textbf{Framework} & \textbf{CT Element} & \textbf{Definition} \\
      \hline
      Berland \& Lee, 2011 & Distributed computation & Distributed computation is an inherently social aspect of computational thinking, in which different pieces of information or logic are contributed by different players over just a few seconds during the process of debugging, simulation, or algorithm building. \\

      Brennan \& Resnick, 2012 & Connecting & Creativity and learning are deeply social practices, and so designing computational media with Scratch is unsurprisingly enriched by interactions with others. \\

      Barr \& Stephenson, 2012 & CT dispositions & Setting aside differences to work with others to achieve a common goal or solution. Knowing one's strengths and weaknesses when working with others. \\

    \end{tabular}
  \end{ruledtabular}
\end{table*}

Working in groups seems an odd fit to be classified as a computational thinking practice. Groupwork is in no way exclusive to programming or in this case programming in a physics classroom nor is it easy to classify as a ``thinking'' practice. However, several of the frameworks that we have drawn upon highlight the social component of computational thinking and provide multiple perspectives of the importance of its inclusion when speaking about computational thinking practices. Berland and Lee highlight that practices such as debugging, simulation, and algorithm building are enhanced when approached using a distributed computation model \cite{Berland2011}. Barr and Stephenson discuss the need for the development in the practices of negotiation and consensus building in order for successful group programming to occur \cite{Barr2011}. Brennan and Resnick argue that computation inherently involves learning and creativity and describe these practices as ``deeply social practices'' that are enriched by collaboration with others \cite{Brennan2012}. Finally Korkmaz makes a strong argument that cooperativity is a ``decisive'' skill of computational thinking \cite{Korkmaz2017}. Korkmaz's perspective is based in the context of learning computation and how group based learning is often cited as more efficient. Korkmaz also goes on to make the point that groups of people working together is inevitable in the 21st century and therefore cooperativity is important to learning and participating in computational thinking.

Korkmaz's broad reasoning would seem to align with the majority of our teachers who often placed an emphasis on the importance of group work and designed activities and learning environments that were collaborative in nature. However, teachers never specifically describe group work as a computational thinking practice or even a learning goal for their introduction of computation into their curriculum. It could be argued that the teachers value group work around computation inherently because they are integrating using a group based approach. Although they did not set specific group work based learning goals, they did expect some sort of development in that area.

The inherent nature of group work in our context was problematic from a research perspective initially, as the question we asked ourselves was if we were interested in exploring group work what exactly would we be looking for to code. Initially, it fell under a similar category of modeling -- because the activities are modeling based then the students could be perceived as automatically modeling by engaging with the activity and so we would be coding the whole of the video for modeling. The same would be true for group work, because they are working in groups, we would end up coding the whole of the video for group work. An alternative approach was taken for coding for group work. Instead of the broad notion of group work, the focus would be on aspects of group work that could be relatively unique to the context of working on computational models of physical phenomena. For example, the positioning of the computer that is the focus of the coding could result in both inclusive and exclusionary group dynamics. Calling back to Perez's \cite{Perez2018} work and the aforementioned collaborative CT disposition, another relatively unique computational collaborative practice is the splitting of stages of a computational model into individual tasks. The splitting of a group project into individual tasks is not a unique branding of group work, however, there might be some uniqueness in the task being a computational code that relies on the interconnected bits to run successfully. Due to the complexity of collaboration and its presence in several of the frameworks, it felt important to illustrate how it might manifest in the classroom, but not explore the unique aspects of the intersection of computation and group work in a complete way as it was not the focus of the paper.

\begin{figure}[htbp]
    \includegraphics[width=\linewidth]{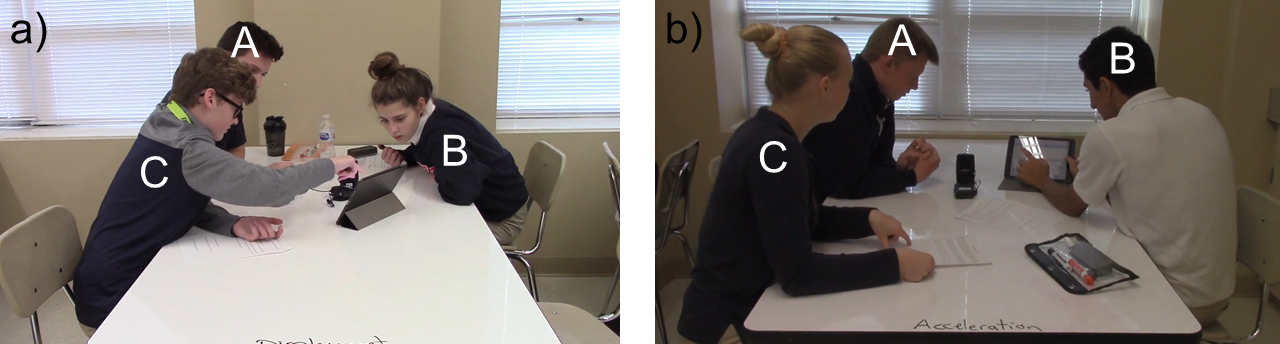}
    \caption{Images of two contrasting groups working on the ``Colliding Crates'' activity from Liam's classroom. The left panel (a) shows students A and C being able to easily see the computer, while student B has to lean in significantly to see the device. The right panel (b) shows students A, B, and C all with an easy line of sight to the computer so that they can easily collaborate and observe the results of their code.}
    \label{Groupwork_Fig}
\end{figure}

Approaching our coding from this perspective resulted in the emergence of several instances of disunity and collaboration within groups based on practices that had elements unique to the context of computation integrated into a physics classroom. For the following existence proof, we will provide two examples from Liam's classroom. The first example will demonstrate a team of students working in a way that is not conducive to collaborative group work. In contrast, the second example will exhibit students working through the same assignment with collaborative group work practices. See Fig.\ \ref{Groupwork_Fig} for images of the two contrasting groups. From a cursory observation of these images, group (a) clearly could arrange themselves better to encourage good group work such that student B does not need to lean in significantly to see what her other group mates are doing with the computer. On the other hand, group (b) is sitting in a configuration that enables its members to all easily see the computer and discuss the results of the model. The decision to force students to interact and perceive the code using one device is relatively unique to the integration approach taken by our teachers and results in the ability for the groups to be set up physically to be collaborative or resisting to the computational activities. The transcripts in the following paragraphs will provide further evidence to support our claims around collaborative and independent approaches to the computational activities.

In the next example, we are going to focus on discussing disunity as it is easier to observe and to illustrate by examples in our observed  groups. The presence of the type of disunity described below in the majority of the groups we recorded is illustrative of the need to develop learning goals around group work in computation. This case should serve as a counter-example for what good group work looks like. The following video example from Liam's colliding crates activity demonstrates disunity among a group that is working through one of their first computational experiences. One student (student C) has been told by his instructor that he is not allowed to touch the keyboard because he has a significant amount of experience with programming comparing to his group mates. The following conversation takes place over 122 seconds. \\

\begin{hangparas}{1em}{1}
	C: What if we just alternated these commands? So like crate1, crate2, while, while, rate, rate, you know? Just like alternate them...
\end{hangparas}
\begin{hangparas}{1em}{1}
	A: No, no, I don't know. [Student A speaks in a defeated tone.]
\end{hangparas}
\begin{hangparas}{1em}{1}
	B: Yeah, I don't really understand. I feel like I'm going to mess it up. You can just do it. I don't want to mess it up because I don't understand this.
\end{hangparas}
\begin{hangparas}{1em}{1}
	C: I don't really understand what we can do here. It's making me really mad that I can't touch the computer.
\end{hangparas}
\begin{hangparas}{1em}{1}
	B: Wait, so if they have to move at the same time... I don't know, I'm like the worst person to ask about this stuff. I don't get it. [The group members sit silently for about 30 seconds.]
\end{hangparas}
\begin{hangparas}{1em}{1}
	A: [Student C] is about to have an aneurysm over here. [Student A uses Student C's real name in the video clip.] He's so mad.
\end{hangparas}
\begin{hangparas}{1em}{1}
	C: I want to touch the keyboard, but I'm not allowed to.
\end{hangparas}
\begin{hangparas}{1em}{1}
	B: Why not?
\end{hangparas}
\begin{hangparas}{1em}{1}
	C: Because he told me I can't. [Student C refers to their teacher with this statement.]
\end{hangparas}
\begin{hangparas}{1em}{1}
	B: Why?
\end{hangparas}
\begin{hangparas}{1em}{1}
	C: I don't know. Because he thinks I would do everything if I did. \\
\end{hangparas}

The above transcript helps to give insight around the group dynamics of this team. At the beginning of the conversation, student C is trying to guide his group to the correct solution. Nevertheless, students A and B have such a low self-efficacy with this subject, they are unable to make sense of student C's ideas. Furthermore, student C does not make any attempts to encourage his fellow team members to try diving more deeply into the coding. Instead, the students just express their discomfort throughout the entire interaction. Clearly, student C is frustrated because their teacher, Liam, made up a rule that student C should not touch the keyboard while working with his group. Liam put this guideline in place so that the students would have the opportunity to physically do some of the coding while also allowing for the development of more group-oriented skills like communication and collaboration. Unfortunately, this approach leads to the group getting frustrated and not making good progress on the assignment. It seems that the students do not yet believe in the recommendations of their teacher. The above example from in-class data demonstrates a snapshot of disunity and a break down in collaboration when working with the computational activities.

While the previous case served as a counterexample of students effectively collaborating, the next case will show some aspects of positive collaboration when working with computation. Liam's students are trying to fix a bug in their code, and this group tends to exhibit positive group interactions like the students sharing the computer amongst each other and keeping a positive attitude. This episode, which shows a discussion that takes place over about 95 seconds, was captured the same day as the previous example (so working on the same activity but in a different group). \\

\begin{hangparas}{1em}{1}
	C: Okay, what's with this floor line? [Student C points to the line of code relating to the floor object in the model.] Let's try indenting that line. Because the box has the size and color underneath it.
\end{hangparas}
\begin{hangparas}{1em}{1}
	B: Here, you can do it. I don't understand what you mean. [Student C passes the computer over to student B.]
\end{hangparas}
\begin{hangparas}{1em}{1}
	C: Like this, so that they're indented. [Student C makes the change to the code and then shows it to her groupmates.] Because when we were using the objects to make our names, all of the size and colors were indented below the first lines.
\end{hangparas}
\begin{hangparas}{1em}{1}
	B: Okay. [Student B runs the program to observe their results, and the program experiences an error immediately upon running.] Unexpected error still...
\end{hangparas}
\begin{hangparas}{1em}{1}
	A: Can I see that real quick? [Student B passes the computer to student A. Student A begins comparing their computer code to the code provided by their teacher on the assignment hand-out.] We're missing some stuff. We're missing `t=t+dt,' and it's indented. It's the last line.
\end{hangparas}
\begin{hangparas}{1em}{1}
	B: `t=t+dt'? No, we have it. It's there, but it's not indented.
\end{hangparas}
\begin{hangparas}{1em}{1}
	A: Oh, we do?
\end{hangparas}
\begin{hangparas}{1em}{1}
	B: Yeah, we do have it. See right there? [Student B points to the line in their code.] We do have it, but it's not indented. So it should be indented, you think? On a new line?
\end{hangparas}
\begin{hangparas}{1em}{1}
	A: Yup. [Student B takes the computer back and makes the change to their code.]
\end{hangparas}
\begin{hangparas}{1em}{1}
	B: What are the chances this will work? [Student B runs the program, and it successfully runs without an error.] Hey! It actually worked. That was a good idea. Good stuff! \\
\end{hangparas}

\begin{figure}[htbp]
    \includegraphics[width=\linewidth]{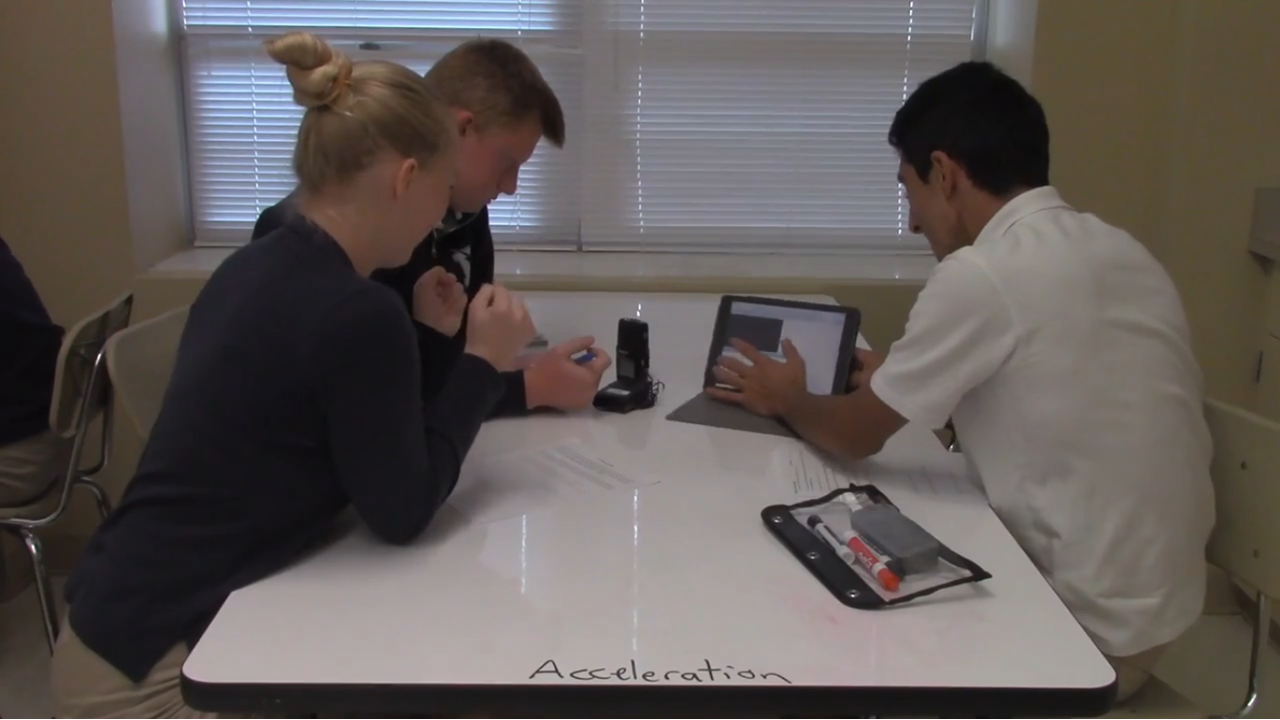}
    \caption{Images of a group of students experiencing success while working through a computational activity.}
    \label{Success_Fig}
\end{figure}

The above situation depicts positive collaboration in a number of ways. First, the students are all sharing resources like the computer and the assignment hand-out (the tablet is moving back and forward between students). They use these resources to collaborate and to try new ideas, rather than simply giving up or thinking that they are powerless. Moreover, the students are showing that they are encouraging and feeling positive about the computation. When student B says that student A's suggestion was a good idea, he reinforces the idea that this is a collaborative space where everyone's ideas are welcome. Overall, student B is encouraging and inquisitive when his group members suggest a new idea, and this is especially important if he is the student who is currently making changes to the computational model. Figure \ref{Success_Fig} displays the students celebrating after they have successfully debugged their code. The student on the left has her hands raised and everyone is smiling, indicating that they are pleased with their computational model. All of the actions discussed here can be indicators of positive collaboration.

The previous two examples indicate the ability to identify episodes of collaboration and disunity. The episodes also highlight that having the work focused around one device or having the physical coding being dominated by one group member can cause inequities in who gets to join in on the activity. These episodes are likely scratching the surface of the ways in which group work can transpire in the high school context when computation is integrated. There are more codes to identify and a more detailed code book to be developed to successfully answer research questions focused on group dynamics.

\section{Application of Framework for Video Analysis and Future Research Opportunities}\label{Results}

A consequence of our search for existence proofs of practices within these learning environments was that we also were able to investigate the emergence, frequency, and relationships between CT practices. While this paper's intention was to focus on providing existence proofs of these practices in action, the framework allows us to see relationships between curricular design choices and the frequency of codes for the teachers' classrooms we studied. Tables \ref{Michael_Code_Table} and \ref{Liam_Code_Table} display the results of our video analysis of Michael's and Liam's classrooms respectively. It is important to note than students in groups labeled ``1'' and ``2'' were kept consistent throughout the study.

\begin{table*}[t]
  \caption{Summary of codes emerging in the analysis of Michael's classroom.\footnote{P1=Projectile activity, group 1; P2=Projectile activity, group 2; R1=River crossing activity, group 1; R2=River crossing activity, group 2; S1=Spring energy activity, group 1; S2=Spring energy activity, group 2. }
  \label{Michael_Code_Table}}

  \begin{ruledtabular}
    \begin{tabular}{|p{2.8in} | p{0.4in} | p{0.4in} | p{0.4in} | p{0.4in} | p{0.4in} | p{0.4in} |}
      \textbf{Practice} & \textbf{P1} & \textbf{P2} & \textbf{R1} & \textbf{R2} & \textbf{S1} & \textbf{S2} \\
      \hline
      Decomposing & & & \phantom{x}2 & \phantom{x}1 & \phantom{x}2 & \phantom{x}1 \\
      \hline
      Highlighting and foregrounding & & & \phantom{x}2 & \phantom{x}3 & \phantom{x}5 & \phantom{x}4 \\
      \hline
      Translating physics into code & & & \phantom{x}2 & & \phantom{x}6 & \phantom{x}4 \\
      \hline
      Algorithm building & \phantom{x}2 & & \phantom{x}5 & \phantom{x}3 & \phantom{x}1 & \\
      \hline
      Applying conditional logic & \phantom{x}1 & \phantom{x}1 & \phantom{x}1 & \phantom{x}1 & \phantom{x}2 & \\
      \hline
      Utilizing generalization & & & & & \phantom{x}1 & \phantom{x}2 \\
      \hline
      Adding complexity to a model & & & & & \phantom{x}2 & \\
      \hline
      Debugging & \phantom{x}2 & \phantom{x}3 & \phantom{x}4 & \phantom{x}6 & \phantom{x}8 & \phantom{x}6 \\
      \hline
      Intentionally generating data & & & & & \phantom{x}1 & \\
      \hline
      Choosing data representation form & & & & & \phantom{x}2 & \\
      \hline
      Manipulating data & & & & & \phantom{x}2 & \\
      \hline
      Analyzing data & \phantom{x}1 & \phantom{x}1 & & & \phantom{x}7 & \\
      \hline
      Demonstrating constructive dispositions & \phantom{x}2 & & & \phantom{x}2 & & \\
      \hline
      Working in groups & & \phantom{x}1 & & \phantom{x}1 & \phantom{x}1 & \\
    \end{tabular}
  \end{ruledtabular}
\end{table*}

\begin{table*}[t]
  \caption{Summary of codes emerging in the analysis of Liam's classroom.\footnote{C1=Colliding crates activity, group 1; C2=Colliding crates activity, group 2; M1=Momentum activity, group 1; M2=Momentum activity, group 2; B1=Box on ramp activity, group 1; B2=Box on ramp activity, group 2. }
  \label{Liam_Code_Table}}

  \begin{ruledtabular}
    \begin{tabular}{|p{2.8in} | p{0.4in} | p{0.4in} | p{0.4in} | p{0.4in} | p{0.4in} | p{0.4in} |}
      \textbf{Practice} & \textbf{C1} & \textbf{C2} & \textbf{M1} & \textbf{M2} & \textbf{B1} & \textbf{B2} \\
      \hline
      Decomposing & & & \phantom{x}4 & \phantom{x}3 & \phantom{x}5 & \phantom{x}1 \\
      \hline
      Highlighting and foregrounding & \phantom{x}2 & & & & \phantom{x}2 & \\
      \hline
      Translating physics into code & & & \phantom{x}2 & \phantom{x}2 & \phantom{x}1 & \\
      \hline
      Algorithm building & \phantom{x}8 & \phantom{x}5 & & & & \\
      \hline
      Applying conditional logic & & \phantom{x}4 & \phantom{x}1 & & & \\
      \hline
      Utilizing generalization & \phantom{x}4 & \phantom{x}1 & & \phantom{x}1 & & \\
      \hline
      Adding complexity to a model & & & & & & \\
      \hline
      Debugging & \phantom{x}9 & & \phantom{x}2 & \phantom{x}1 & & \\
      \hline
      Intentionally generating data & & & & & & \\
      \hline
      Choosing data representation form & & & & & & \\
      \hline
      Manipulating data & \phantom{x}6 & & & & & \\
      \hline
      Analyzing data & \phantom{x}3 & & & & \phantom{x}1 & \\
      \hline
      Demonstrating constructive dispositions & \phantom{x}3 & & & & \phantom{x}1 & \\
      \hline
      Working in groups & \phantom{x}2 & & & & \phantom{x}1 & \\
    \end{tabular}
  \end{ruledtabular}
\end{table*}

As demonstrated by video evidence in the previous section, the effects of group composition and messaging from the teacher are most evident in Liam's classroom. The first group of Liam's classroom was encouraging, positive, and open-minded with respect to computation. This group also had more CT practices occurring which might have been due to the more open and trustful group environment they constructed. Moreover, Liam often met with these students during class and primed them for hardship by telling them things like ``I'm not sure if this change to the code will work, but you can try yourself just to see. If it doesn't work, you can just undo your change.'' While this utterance might be well understood by an experienced programmer, this kind of messaging is extremely important for students to have a positive early experience with coding. In contrast, the second group of Liam's classroom often had students demonstrating signs of negative computational dispositions and unproductive group dynamics. For instance, students in the second group often felt lost by coding, and in turn, they felt powerless to make changes in the code. One of the students in this group was highly confident in coding, which  led to some high-level CT practices being discussed by the group. However, even after the high-confidence student shared his ideas, the other group members were unable to comprehend the changes made by the more advanced student. This difference observed between the two groups in Liam's classroom serves as an initial data point that shows how group composition could impact different levels of engagement with CT practices. In this way, teachers should be especially particular when mentioning ideas around persevering and encouraging good teamwork skills when working with computational content.

Broadly, we observed that activity design and the teacher's messaging around CT practices can lead to different results with respect to the practices that students experience. This supports the underlying idea that the context and curricular design choices teachers make can have important implications for engagement in computational thinking. This top-level analysis provides evidence that when we apply the framework to different classrooms, it can yield different results. Across different teachers/classrooms, computational activities within one classroom, and groups within one activity we observe differences in CT practices that emerge. As a result, we demonstrate that the framework presented in this paper can be used to look at finer details that lead to different CT practices emerging. While we believe this to be true, we still do not know enough to precisely explain the differences, but this tool can be used to show that there are differences. Future investigations employing this framework can further explore questions such as what curricular design features matter for CT practices.

Along with initial insights on the impact of activity design and group composition our initial analysis also revealed several other areas of research that need to be investigated further. An aspect of CT practice that our initial analysis hinted at but did not provide enough evidence to make a complete argument for is that there may be sub-themes (i.e., a discernible variation) in the way that a CT practice will be applied in a context. An example of this would be decomposing different parts of computer code to make it easier to interpret versus decomposing the tasks required to complete an assignment provided by one's teacher. In this example, we may expect for the decomposition to look quite different. When decomposing code, students will work on improving their coding skills rather than focusing on other scientific skills. On the other hand, when decomposing a task, students are working on their scientific reasoning and teamwork skills. Along the same theme is the idea of levels of practice, that more complex versions of a CT practice might be able to occur depending on what the activity is demanding and what scaffolds are being provided. A possible example of levels from our data is  in the case of algorithm building. At a simple level, when students understand the step-wise nature of computer programming, they are engaging in the practice of algorithm building. However, one could envision a higher quality form of algorithm building where students are building an algorithm that would work for a range of circumstances (similar to the idea of writing code in a way that it may be generalized to a large range of situations). Further investigation of both levels and variation of CT practices would add complexity to our framework but also inform curriculum design around CT practices in a physics context in order to successful scaffold student learning of CT practices. This analysis should serve as a proof of principle for the utility of this framework. Therefore, if an educator or researcher wants to examine differences in computational implementations, this framework could be used to address those areas.

\subsection{Opportunities for Future Research}

Differences were observed in the CT practices that emerged in Michael's versus Liam's classrooms. While both classrooms were tagged with a similar frequency of codes (approximately 5-10 codes per activity per group over an hour long class session), we observe different CT practices for the two different classrooms. Both classes frequently dealt with decomposition because the students would often interpret and break down computer code when first encountering a minimally working program. Similarly, debugging occurred often, which was expected for introductory programming experiences. However, students from all groups in Michael's classroom engaged in debugging with every computational activity, and this could potentially be a result of Michael's curricular design choices. In Michael's activities, he would often have students build on the models and change the code significantly, which may result in a greater frequency of debugging. On the other hand, Liam sometimes would have students use fully working codes to have students engage in more of demonstration-style coding activities. These different design choices could lead to an increased presence of debugging. As another example, Michael's classroom typically worked with practices in the category of building computational models, whereas Liam's classroom often gained experience with data practices. This result makes sense when looking at the teachers' activity design choices. Michael's assignments often focused more on creating a model and inputting physics concepts into the code. Accordingly, students often translated physics into code and iterated on their model's development. On the other hand, Liam's activities focused more on \textit{using} computational models, rather than building models. In turn, Liam's students would often work with data practices because their assignments would ask them to predict and to test outcomes of fully working computational models. These initial extractions from our analysis of Michael's and Liam's classrooms demonstrate the applicability of using this framework to investigate the patterns of emergence of CT practices in different classroom environments with different curricular design choices.

Within one classroom, we observe different CT practices emerging over time. In other words, students engage in different CT practices as their experiences evolve over many computational learning activities. For example, in Michael's classroom, the amount of practices present appears roughly to increase as students progress through the computational unit. In the first activity analyzed, the projectile problem (P1 and P2 in Table \ref{Michael_Code_Table}), we see a high frequency of codes for CT practices. It could be that students do not engage in many CT practices because they are learning the basics of coding. By contrast, by the time students are working on the spring energy problem (S1 and S2 in Table \ref{Michael_Code_Table}, they seem to be engaging in many more practices. It is possible that as students gain more experience with computation, they are able to engage in more computational thinking. This trend could also be explained by other factors, but such a claim would require further investigation. Regardless, this framework provides educators and researchers with the tools needed to closely examine the emergence of CT practices within different computational activities.

In a similar way, the CT framework presented in this paper also yielded different results for differing groups within the same computational activity. For instance, the two different groups of Michael's spring energy problem (S1 and S2 in Table \ref{Michael_Code_Table}) engaged in CT practices with differing frequency. We observe many more CT practices in the first group than in the second group. It is possible that these differences emerge because of the different students involved in each group. In group S1, the students had developed high confidence with coding and physics in general, while group S2 exhibited low confidence with programming. As another example, clear differences are noticeable in Liam's B1 and B2 groups (see Table \ref{Liam_Code_Table}). For this box on a ramp activity, the students in group B2 were trying to figure out the trigonometry before engaging with computational content. Such a result could be used to make claims about how a teacher should design activities as reinforcement exercise rather than a learning activity for exploring new conceptual ideas. As a result, our CT framework might be used to help teachers design activities and student groups to best fit their specific learning goals. Thus, this framework can help point out trends in how CT practices emerge for different student groups.

\section{Discussion}\label{Discussion}

The framework described above bridges the gap between previous CT literature and the application of CT frameworks to a specific context. It demonstrates an approach to taking CT frameworks and applying them a specific context but also puts forward a framework for a commonly used approach to integrating computation. We propose that the application of this framework to more in-class data will allow us to answer many open questions around the integration of computation into the high school and introductory physics classroom space. Through the construction of our framework, we observe similarities and differences in the CT practices between the two classrooms examined in this report. Broadly, we see that activity format, computational platform, pedagogical approach, and emphasis in the professional development workshop all influence teachers' choices around the approach they are taking to integrating computation in their teaching environment. With that being said, there is still more to be understood around how MWPs versus other programming approaches can affect CT practices. This supports the argument that we emphasized throughout this paper that CT frameworks have to account for the importance of context for them to applicable to your teaching environment.

Students clearly engaged in practices differently when the teacher was present. They made decisions based on what design choices their teacher made. Moreover, when the teacher could give them some guidance, students tended to engage in CT practices at a higher level. We did not choose to exclude teacher-implemented practices from our analysis. However, more research is needed to investigate how the teacher's presence may influence students' computational thinking.

The CT practices presented here seem to follow a cycle starting with extracting computational insight, building computational models, and then data practices. The other three categories of practices (i.e., debugging, working in groups, and demonstrating constructive dispositions towards computation) appear to emerge throughout the entire class session. There is a logic behind these findings because when students first receive a MWP, they need to do some comprehending of the code (i.e., extracting computational insight), then they can modify the code (i.e., building computational models), and then they can use the model for some purpose (i.e., data practices). While this was a general pattern that emerged in our data, it is possible for these practices to happen in a different order based on how a teacher chooses to integrate computation within their course.

\subsection{Practices omitted from this framework}

In general the decision to exclude a practice often fell down to a sequence of questions. Did it emerge from our coding? If the answer to this was ``no'', then we would ask ourselves and our teachers theoretically could you describe what this practice would look like in the classroom based on our experiences integrating computation? If the answer to this was ``no'', then we omit the practice with the caveat that it might occur in a different data set of classrooms. If the answer was yes then the final question we asked ourselves was whether the practice was too universal to have unique CT features. For this last question we use group work as an example. Group work would be considered somewhat of a desired and universal practice in STEM and beyond, but was included in the framework due to there being enough unique contextual features to discuss the CT version of this practice such as the sharing of the device that ran the code. The following paragraphs discuss the practices that were not kept as part of the framework.

Parallelism has been discussed by a limited number of sources previously. In Shute et al.'s framework, the practice of parallelism is contained within the category of algorithms, and it is defined by them as ``[carrying] out a certain number of steps at a time''~\cite{Shute2017}. From Barr \& Stephenson's framework, it is defined as ``dividing up data or a task in such a way that it can be processed in parallel''~\cite{Barr2011}. We also interpret this to generally refers to multitasking, or working on multiple tasks in parallel (at the same time) instead of in series (one after the other). After an internal research meeting in which we asked fellow researchers for good examples of this practice, we found that the idea of parallelism was not present in our context due to the level of programming that students were working with at the time. The simplicity of the computer programs does not require  Students should learn to process tasks in parallel, but we argue that it is more of a general skill that improves efficiency in a number of different fields. Therefore, we decided it did not fit in our CT framework.

The concept of planning is discussed briefly in Weintrop et al., as well as some other frameworks that were not included in this study. We ultimately decided that planning is a universal skill that is certainly appropriate in many fields outside of computer science, physics, and STEM in general. It did occur in the classroom although rarely and never in a robust way. Planning is a multifaceted practice and the aspect of planning that was occurring in the classroom was often coded as decomposition, the definition of which has overlap with the idea of planning. Beyond decomposing, planning rarely occurred in the groups analyzed. Thus, it was impossible to understand what might be specific to planning in a CT context that was not true of planning universally. We omitted planning from the framework.


The practice of critical thinking comes from Korkmaz \cite{Korkmaz2017}, and is defined as ``the use of cognitive skills or strategies that increase the possibility of the desired behaviors.'' Korkmaz argues that a major issue with our current education system is ``rote learning that is the result of traditional instruction.'' While we agree that critical thinking could be further emphasized, we do not feel that it is closely related to computational thinking specifically, but rather learning in general. For this reason, we did not believe that it had a place in our framework.

The paper out of AAPT refers to ``extract physical insight'' as a ``technical computing skill,'' and states that ``students should be able to extract physical insight from a computation by converting the raw output of a computation into a useful form, asking interesting questions, and using the computation to answer these questions''~\cite{AAPT2016}. We believe that there is overlap between this practice, and Shute et al.'s definition of abstraction, ``Extract(ing) the essence of a complex system.'' We argue that this practice might be absorbed into abstraction within our framework, and in turn, we have decided to omit it from our framework.

\subsection{Recommendations for Teachers Integrating Computation into their Curriculum}

After developing the set of learning goals presented in this report, we think it is important to provide some insights for practitioners trying to implement computational interventions in their classrooms. While these recommendations are not meant to be prescriptive, we do hope that they communicate areas worthy of investigation in the future. Foremost, we noticed that student buy-in (i.e., their willingness to participate) for computational activities was not consistent. In other words, some students enjoyed computational activities and participated actively while other students vehemently did not want to participate in learning about computation. For this reason, it is especially important for instructors to think carefully about their messaging around computational ideas. For example, in the previous groupwork transcript for group (a), one student was told not to touch the computer because the teacher thought that particular student would do all the coding if he could. This strategy could be effective if the students understood and believed in their teacher's motives. However, because the students do not appreciate their teacher's approach in this instance, they instead end up complaining and not making good progress on the assignment. We believe this type of angst could be alleviated if the teacher made his reasoning more clear to students. Rather than simply commanding one student to sit back, Liam could have used these computational activities as a chance for his students to learn about important teamwork skills.

The importance of clear messaging is not only relevant to groupwork practices. When students experience frustration with debugging, building models, or general coding, their negative experiences could be ameliorated by their teacher being more explicit about what problems the students might face in their early computational experiences. Accordingly, it is important for instructors to give clear messaging around working in groups on computational models. If students do not appreciate the reasoning that their teacher uses for the role they are being asked to play, then it could lead to students experiencing difficulty adapting to computation in their physics curriculum. Therefore, teachers should think carefully about how they are communicating their expectations around CT practices when students are first experiencing computational content. In general, we are creating this framework in order to study computation, but we are not ready to give numerous recommendations for teachers. We hope to use this framework to investigate activity design or group composition in the future.

\section{Conclusions}\label{Conclusions}

This study presents some limitations that should be carefully considered when applying this framework to other contexts. First off, our literature review did not include the entirety of papers written on the topic of computational thinking. Instead, we tried to consider sources that were particularly relevant to the context of introductory physics. Moreover, some of these sources include ideas that are akin to CT without actually being called CT, like where the AAPT source describes these ideas as computational physics skills or technical computing skills. Another limitation is that we did not include some of the more advanced ideas that were highlighted in other frameworks (e.g., systems thinking from Weintrop et al.). We chose not to keep some of these more advanced ideas because they were either not relevant or too advanced for the context that we are interested in exploring.

In conclusion, introductory physics is a discipline well-suited to explore computational thinking practices because physics is becoming increasingly computational in nature and students are being encouraged to learn programming ideas in core science disciplines. Science standards and recommendations from professional physics organizations provide motivation for teachers to integrate computation within science. With the lack of clear guidelines around computational thinking, a CT framework, like the one presented in this report, can serve as a tool for further examining computational and scientific practices in early physics courses. This framework will help teachers by giving examples of indicators of CT practices. This framework will also be useful for researchers who would also like to investigate how various factors (e.g., activity design, student group composition, messaging from the teacher) can lead to the emergence of different CT practices. Our future work hopes to address the variation observed within each CT practice to further realize the different ways these CT practices can emerge. We also plan to collect empirical survey data from teachers and students to better understand their perspectives around these types of CT classroom interventions. While there is more to be understood about CT practices in this framework and different contexts, the practices shown here are a jumping off point for more closely examining CT in introductory and high school physics.

\acknowledgments{We would like to thank the National Science Foundation (\mbox{DRL-1741575}) for funding and the teachers for graciously donating their time and effort to our research. We also wish to thank Joe Krajcik, Bob Geier, and Sue Carpenter for the use of CREATE for STEM facilities for our professional development program. Finally, we thank Paul Hamerski, Tor Ole Odden, and Devin Silvia for insightful discussions. We also acknowledge that this work would not be possible without the many students who allowed us to study their experiences with computing.}

\bibliography{CTLGFramework_BIB.bib}

\end{document}